\documentclass[preprint,aps,prd,nofootinbib]{revtex4-1}

\usepackage[T1]{fontenc} 
\usepackage{lmodern}
\usepackage{hyperref}
\usepackage{graphicx}
\usepackage{bm}
\usepackage{amsmath}
\usepackage{amssymb}
\usepackage{color}
\usepackage{placeins}
\usepackage{subfigure}

\begin{document}

\title{Neutrino flavor transformations in supernovae as a probe for nonstandard neutrino-scalar interactions}
\author{Yue Yang}
\author{James P.~Kneller}
\affiliation{Department of Physics, North Carolina State University, Raleigh, NC 27695 USA}

\email{yyang30@ncsu.edu}
\email{jpknelle@ncsu.edu}

\begin{abstract}
We explore the possibility of probing the nonstandard interactions between the neutrino and a hypothetical massive scalar or pseudoscalar via neutrino flavor transformation in supernovae. We find that in the ultrarelativistic limit, the effective interaction between the neutrinos vanishes if neutrinos are Dirac fermions but not if they are Majorana fermions. The impact of the new neutrino interaction upon the flavor transformation above the neutrinosphere is calculated in the context of the multi-angle ``neutrino bulb model''. We find that the addition of the nonstandard neutrino self-interaction (NSSI) to the ordinary V-A self-interaction between neutrinos is capable of dramatically altering the collective oscillations when its strength is comparable to the standard, V-A, interaction. The effect of flavor-preserving (FP) NSSI is generally to suppress flavor transformation, while the flavor-violating (FV) interactions are found to promote flavor transformations. If the neutrino signal from a Galactic supernova can be sufficiently well understood, supernova neutrinos can provide complimentary constraints on scalar/pseudoscalar interactions of neutrinos as well as distinguishing whether the neutrino is a Majorana or Dirac fermion.
\end{abstract}


\date{\today}

\maketitle


\section{Introduction}
\label{sec:intro}

The physical conditions found in the core of a core-collapse supernova (CCSN) provide us with an alternative and complimentary laboratory for probing the properties of the neutrino. In addition to the extreme matter density, the neutrino density in the vicinity of the proto-neutron star (PNS) is so high that neutrinos can experience coherent forward-scattering from the other neutrinos emitted from the PNS. Indeed, during some epochs of the explosion, this neutrino-neutrino self-interaction can dominate the flavor evolution. The complete description of the flavor transformation in CCSN is given in terms of Quantum Kinetic Equations \cite{1993NuPhB.406..423S,2005PhRvD..71i3004S,2013PhRvD..87k3010V,2014PhRvD..89j5004V} which are found to reduce to a Schr\"{o}dinger-like equation in the limit where the exchange of energy and momentum between neutrinos and the medium vanishes. Using Standard Model physics, the Hamiltonian $H$ that enters this equation is built out of a vacuum contribution $H_V$, a matter contribution $H_{M}$, and a self-interaction $H_{SI}$. The self-interaction makes the flavor evolution of one neutrino dependent upon the flavor evolution of every other neutrino it encounters. The full problem is currently beyond the scope of computing platforms. The current state-of-the-art model for the calculations of neutrino flavor transformation in supernovae is known as the ``neutrino bulb model'' which imposes both spherical symmetry for neutrino emissions from the neutrinosphere, and axial symmetry around every radial ray, in order to reduce the number of independent variables needed to describe the neutrino field to just three. The three degrees of freedom are typically chosen to be: the radial coordinate along a ray, the neutrino energy, and the angle of emission relative to the normal at the neutrinosphere \cite{Duan:2006an}. Multiple studies of the neutrino flavor transformation in CCSN using the bulb model have found the addition of $H_{SI}$ can leave distinct features in the neutrino spectra which vary with time and which one would hope to observe in the signal from a future Galactic supernova: for recent reviews we refer the reader to Mirizzi \emph{et al.} \cite{2017arXiv170901515H} and Horiuchi \& Kneller \cite{2016NCimR..39....1M}

The conditions found in a CCSN mean that any change to the properties of the neutrino often modify the outcome of the flavor transformation. For example, new - sterile - flavors of neutrinos have been considered on several occasions \cite{1997PhRvD..56.1704N,1999PhRvC..59.2873M,2001NuPhB.599....3P,2006PhRvD..73i3007B,2012JCAP...01..013T,2014PhRvD..90j3007W,2014PhRvD..89f1303W,2014PhRvD..90c3013E}. Authors have found that active-sterile mass-splittings of order $\sim 0.1\;{\rm eV}^2$ or greater, and mixing angles larger than $\sim 0.01^{\circ}$ can introduce new adiabtaic Mikheyev-Smirnov-Wolfenstein (MSW) \cite{Mikheyev:1985aa,Mikheyev:1986tj,1978PhRvD..17.2369W} resonances close to the PNS whose effect upon the neutrino flavor composition of the flux changes the dynamics of the explosion \cite{2012JCAP...01..013T,2014PhRvD..89f1303W} as well as the flavor evolution at larger radii and the neutrino signal \cite{2014PhRvD..90c3013E,2012JCAP...01..013T}. Similarly one can also consider new interactions of neutrinos coupled via some new field to either matter (electrons and quarks) or to other neutrinos. There are several studies of the effect of nonstandard interactions of neutrinos with charged fermions and a pair of recent reviews can be found in Miranda aand Nunokawa \cite{1367-2630-17-9-095002} and Ohlsson \cite{2013RPPh...76d4201O}. 
Again, these scenarios often lead to new resonances and flavor evolution which differs substantially from the Standard Model, V-A,  case \cite{1987PhLB..199..432V,1996PhRvD..54.4356N,1996NuPhB.482..481N,1998PhRvD..58a3012M,2002PhRvD..66a3009F,PhysRevD.76.053001,Blennow:2008er,2010PhRvD..81f3003E,Stapleford:2016jgz}. For example, it has been shown one can observe neutrino self-interaction effects in the normal mass ordering when nonstandard interactions are included that cannot occur with just Standard Model physics \cite{PhysRevD.76.053001,Blennow:2008er,2010PhRvD..81f3003E,Stapleford:2016jgz}. 
Alternatively one can also consider non-standard interactions of neutrinos among themselves - so-called non-standard self-interactions (NSSI). Compared with nonstandard interactions of neutrinos with charged fermions, the parameters of NSSI are much less constrained by terrestrial experiments \cite{Bilenky:1992xn,Bilenky:1994ma,Masso:1994ww,Bilenky:1999dn} and current constraints show that NSSI can be as large as the standard neutrino self-interaction. This provides an unique opportunity for us to take advantage of the CCSN environment as a neutrino laboratory and place complimentary constraints upon unknown interactions among neutrinos. 

The form of the NSSI is not unique. Blennow et al. \cite{Blennow:2008er} and Das et al. explored NSSI for supernova neutrino originating from a non-standard model gauge boson. This form of interaction leads to an effective neutrino-neutrino interaction Hamiltonian similar to the standard V-A except for a flavor-dependent coupling strength and flavor-violating terms \cite{Das:2017iuj}. Dighe and Sen later applied instability analysis to study the ``fast conversion'' in the presence of such a NSSI \cite{dighe2018nonstandard}. These works show clearly that the presence of NSSI can have significant influence on neutrino flavor transformation in supernovae. For example, it is pointed out the presence of NSSI can lead to flavor equilibration in both mass hierarchies \cite{Blennow:2008er}, and it can also cause collective oscillation in normal mass hierarchy if NSSI is stronger than standard V-A \cite{Das:2017iuj}.   

While the gauge boson model is well-motivated, it represents just one category of possible NSSI candidates. Another strong candidate for NSSI is a Yukawa coupling between neutrinos and nonstandard scalar or pseudoscalar fields. This type of interaction has a long history and is used in several models to explain the origin of neutrino mass. One prominent example is the ``majoron model'' by Gelmini \cite{Gelmini:1980re,Gelmini:1982rr}. Indeed, constraints on the neutrino-majoron coupling by using the neutrino signal from SN1987A have been made \cite{Kolb:1987qy,Chang:1993yp,Choi:1987sd,Kachelriess:2000qc,Tomas:2001dh,Farzan:2002wx} although these previous works did not link the neutrino-scalar coupling to neutrino flavor transformation. 

Our goal in this paper is to explore the consequence of a neutrino-scalar/pseudoscalar interaction upon the flavor transformation. Our paper is organized in the following way. In section \S\ref{sec:evolution} we write out the neutrino evolution equation and derive the single-particle effective Hamiltonian of NSSI under the mean field framework, showing the difference between the case of a Dirac neutrino and a Majorana neutrino. In section \S\ref{sec:transformations} we solve the neutrino flavor evolution equations numerically with the NSSI term added to the standard Hamiltonian, using realistic supernova profiles and spectra, and show its impact on neutrino collective oscillations at two different snapshots of a CCSN. We also make a comparison of the results by ``single-angle'' approach and ``multi-angle'' approach. In \S\ref{sec:summary} we summarize our results and conclude.

\section{The flavor evolution of supernova neutrinos}
\label{sec:evolution}
\FloatBarrier

In this section we describe the formulism of neutrino flavor transformations in the supernova environment. During a supernova explosion, the ambient region around the contracting core is an environment featuring dense matter, violent turbulence, and an intense flux of neutrinos.  What we want to compute is the flavor evolution history of the $\sim 10^{58}$ neutrinos emitted as the PNS cools. As mentioned earlier, a full treatment of neutrino flavor evolution requires solving the quantum-kinetic equations taking all refraction and scattering effects into account. This is a gigantic task in terms of computational expense. Fortunately it has been demonstrated that neutrino flavor transformations usually happens in regions relatively far from the core due to the dense matter and multiangle suppression effect \cite{2011PhRvL.107o1101C,2011PhRvL.106i1101D}, thus only the refraction effect is relevant and the Schr\"{o}dinger-like flavor evolution equation for streaming neutrinos can be applied\footnote{We also note that more recent works on ``neutrino fast conversion'' \cite{2005PhRvD..72d5003S,chakraborty2016self,sen2017supernova,izaguirre2017fast,capozzi2017fast,abbar2017fast,dasgupta2018fast,dighe2018nonstandard} indicate flavor transformations may occur close to the PNS potentially upsetting this paradigm.}. 

\subsection{The equations of flavor evolution}
The flavor evolution equation of a test neutrino propagating with momentum ${\bf{q}}$ in the supernova environment takes the following form:
\begin{equation}
i\frac{dS_{\bf{q}}}{d\tau} = H\left(\tau, \bf{q} \right)S_{\bf{q}},
\end{equation}
where $\tau$ is the ``local proper time'' \cite{Yang:2017asl} and $S_{\bf{q}}$ is the matrix encoding the evolution history of the test neutrino. In ultrarelativistic and weak gravity limit, we can replace $\tau$ with the distance $r$ from the center of the neutrinosphere\footnote{Throughout the paper we set $\hbar=c\equiv 1$.}. The probability that a neutrino in some generic initial state $\nu_{j}$ with momentum ${\bf q}$ at distance $r_0$ is later detected as state $\nu_i$ at distance $r$ is $P(\nu_j \rightarrow \nu_i) = P_{ij} = |S_{{\bf{q}};ij}(r;r_0)|^2$. Similarly, the evolution of the antineutrinos is given by an evolution matrix $\bar{S}$ which evolves according to a Hamiltonian $\bar{H}$. The total Hamiltonian can be divided into three parts as
\begin{equation}
H(r,{\bf{q}}) = H_{\rm V}(E) + H_{\rm M}(r) + H_{\rm SI}(r,{\bf{\hat{q}}})
\end{equation}
with $\bf{\hat{q}}$ indicating a unit vector in the direction of the neutrino's momentum. 
Note that the vacuum term $H_V$ is only a function of neutrino energy $E = |{\bf q}|$, while the matter term $H_M$ is only dependent on position $r$. The vacuum term and matter term are straight-forward to write out in the flavor basis for a relativistic three flavor neutrino with energy $E$:
\begin{equation}
H_{\rm V} = \frac{1}{2E}\,U_{\rm V} \left( \begin{array}{*{20}{c}} m_1^2 & 0 & 0  \\ 0 & m_2^2 & 0 \\ 0 & 0 & m_3^2\end{array} \right) U_{\rm V}^{\dagger},\quad\quad H_{\rm M} = \sqrt{2}\,G_{\rm F}\,n_e(r) \left( \begin{array}{*{20}{c}} 1 & 0 & 0   \\ 0 & 0 & 0 \\ 0 & 0 & 0 \end{array} \right).
\end{equation}
In the standard model the self-interaction term in the Hamiltonian, $H_{\rm SI}$, has a form which arises from the V-A interaction and is dependent on both the position and direction of the neutrino's momentum. The expression for the self-interaction from the V-A interaction is 
\begin{equation}
H_{\rm V - A}\left( r,{\bf{\hat{q}}} \right) = \sqrt 2 {G_{\rm F}}\int {\left( {1 - {\bf{\hat p}} \cdot {\bf{\hat q}}} \right)\left[ {{\rho}(r,{\bf{p}})\,d{n_\nu }\left( {r,{\bf{p}}} \right) - \bar \rho^ * (r,{\bf{p}})\,d{n_{\bar \nu }}\left( {r,{\bf{p}}} \right)} \right] dE_{\bf p}}.
\end{equation}
where $\rho(r,{\bf{p}})$ is the density matrix of the ambient neutrinos at position $r$ with momentum ${\bf p}$ and $dn_{\nu}(r,{\bf p})$ is the differential neutrino number density \cite{Duan:2006an}, which is the differential contribution to the neutrino number density at $r$ from those neutrinos with energy $E_{\bf p}=\left|{\bf p}\right|$ propagating in the directions between ${\bf{\hat p}}$ and ${\bf {\hat p}}+d{\bf{\hat p}}$, per unit energy (the hat indicates a unit vector). 
The quantities $\bar{\rho}(r,{\bf p})$ and $dn_{\bar{\nu}}(r,{\bf p})$ are similar in meaning but for antineutrinos. The differential contribution ${\rho}\left( r,{\bf p} \right)\,d{n_\nu}\left( {r,{\bf{p}}} \right)$ can be further decomposed into ${\rho}\left( r,{\bf{p}} \right)\,d{n_\nu }\left( {r,{\bf{p}}} \right) = \sum\limits_{\alpha  = e,\mu ,\tau } {\rho_{\underline{\alpha}}} \left( r,{\bf{p}} \right)\,dn_{\nu_{\underline{\alpha}}}\left( {r,{\bf{p}}} \right)$ by summing over the original flavor states of the neutrinos at the neutrinosphere.

\FloatBarrier
\subsection{The effective Hamiltonian of NSSI}
Let us consider the form of the additional contribution to $H_{SI}$ from a hypothetical coupling between neutrinos via a scalar or pseudoscalar interaction. Instead of asking the nature of the hypothetical scalar fields, we focus on the phenomenological consequences if such a Yukawa coupling between neutrinos and some scalar fields exists. Generally the coupling can be written as
\begin{equation}\label{lagrangian1}
- {\mathcal{L}_{{\mathop{\rm int}} }} = \frac{1}{2}{g_{\alpha \beta }}{\bar \nu _\alpha }{\nu _\beta }\phi  + \frac{i}{2}{h_{\alpha \beta }}{\bar \nu _\alpha }{\gamma ^5}{\nu _\beta }\chi, 
\end{equation}
where the $\phi$/$\chi$ is the hypothetical scalar/pseudoscalar field, and $\bf g$ and $\bf h$ are the hermitian coupling matrices\footnote{For simplicity we assume $\bf g$ and $\bf h$ are real and symmetric in the following without loss of generality.}. In many models the scalar fields are taken to be massless leading to new long range interactions, while in other models the scalar fields are massive leading to a shortening of the range of the interaction considerably. The assumed mass of the scalar/pseudoscalar field and the typical energy of the neutrinos have considerable impact upon the neutrino phenomenology. In this paper we assume the scalar/pseudoscalar field has a mass larger than the $\rm{GeV}$ scale, which is well beyond the typical energies of supernova neutrinos. This excludes many scenarios in which the neutrino-scalar field coupling could change the CCSN dynamics through ``cooling effects'' \cite{Farzan:2002wx}. This also makes it possible to adopt the ``4-fermion'' approximation, which is the basis of discussing neutrino-neutrino coherent forward scattering effect in the supernova environments. With this assumption, we can derive an effective neutrino NSSI Hamiltonian in addition to the regular V-A type neutrino self-interaction. 

Under the assumption that the mediating particles $\phi$ and $\chi$ are sufficiently massive, the effective interaction Hamiltonian can be written in a 4-fermion form
\begin{equation}
\label{4-fermion Hamiltonian}
{\mathcal{H}}_{\rm int} = - {\mathcal{L}}_{\rm int} \approx \frac{1}{{8m_\phi ^2}}{g_{\alpha \beta }}{g_{\xi \eta }}\left( {{{\bar \nu }_\alpha }{\nu _\beta }} \right)\left( {{{\bar \nu }_\xi }{\nu _\eta }} \right)
- \frac{1}{{8m_\chi ^2}}{h_{\alpha \beta }}{h_{\xi \eta }}\left( {{{\bar \nu }_\alpha }{\gamma ^5}{\nu _\beta }} \right)\left( {{{\bar \nu }_\xi }{\gamma ^5}{\nu _\eta }} \right), 
\end{equation}
where $m_{\phi}$ and $m_{\chi}$ are the rest mass of $\phi$ and $\chi$, respectively. Note that a factor of $1/2$ has been introduced to avoid double counting.
\begin{figure}[]
\centering
\includegraphics[scale=0.45]{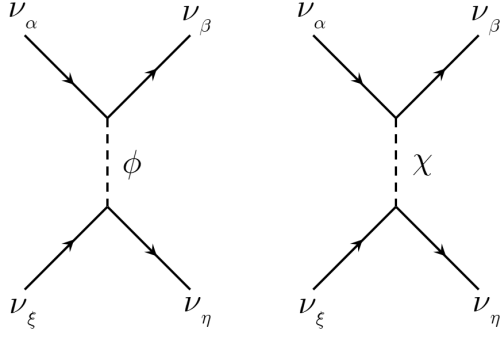}
\caption{The lowest order scalar and pseudoscalar interactions between neutrinos.}
\label{fig:interaction}
\end{figure}
Just as with the V-A self-interaction, by applying the mean field approximation we can transform the 4-neutrino operators into an effective 2-neutrino operator (see appendix \ref{sec:meanfield}). Interestingly, the resulting effective Hamiltonian holds different implications for Dirac neutrino and Majorana neutrino. For the Dirac neutrino we find
\begin{equation}
\label{4-nu scalar Dirac}
\left( {{{\bar \nu }_\alpha }{\nu _\beta }} \right)\left( {{{\bar \nu }_\xi }{\nu _\eta }} \right) \approx 
- \frac{1}{2}\left\langle {{{\bar \nu }_{\alpha L}}{\gamma ^\mu }{\nu _{\eta L}}} \right\rangle \left( {\bar \nu _{\xi R}{\gamma _\mu }\nu _{\beta R}} \right) - \frac{1}{2}\left\langle {\bar \nu _{\xi R}{\gamma _\mu }\nu _{\beta R}} \right\rangle \left( {{{\bar \nu }_{\alpha L}}{\gamma ^\mu }{\nu _{\eta L}}} \right) + \left( {\alpha \eta  \leftrightarrow \xi \beta } \right)
\end{equation}
and
\begin{equation}\label{4-nu pseudo Dirac}
\left( {{{\bar \nu }_\alpha }{\gamma ^5}{\nu _\beta }} \right)\left( {{{\bar \nu }_\xi }{\gamma ^5}{\nu _\eta }} \right) \approx 
\frac{1}{2}\left\langle {{{\bar \nu }_{\alpha L}}{\gamma ^\mu }{\nu _{\eta L}}} \right\rangle \left( {\bar \nu _{\xi R}{\gamma _\mu }\nu _{\beta R}} \right) + \frac{1}{2}\left\langle {\bar \nu _{\xi R}{\gamma _\mu }\nu _{\beta R}} \right\rangle \left( {{{\bar \nu }_{\alpha L}}{\gamma ^\mu }{\nu _{\eta L}}} \right) + \left( {\alpha \eta  \leftrightarrow \xi \beta } \right)
\end{equation}
where we have used $\left( {\alpha \eta  \leftrightarrow \xi \beta } \right)$ to denote the same terms as the earlier part of the equation but with subscripts exchanged. Thus we have decomposed the scalar/pseudoscalar coupling of neutrino fields into products of left-left coupling and right-right coupling of the vector-vector type. However, in the ultrarelativistic limit the right-handed component of neutrino fields vanishes, resulting in a zero contribution to these equations from right-handed neutrino current. So in the Dirac neutrino case, neither scalar nor pseudoscalar interactions can give observable effects in the limit of vanishing neutrino mass.

But if neutrinos are Majorana fermions we find instead
\begin{equation}\label{4-nu scalar majorana}
\left( {{{\bar \nu }_\alpha }{\nu _\beta }} \right)\left( {{{\bar \nu }_\xi }{\nu _\eta }} \right) \approx \\
- \frac{1}{2}\left\langle {{{\bar \nu }_{\alpha L}}{\gamma ^\mu }{\nu _{\eta L}}} \right\rangle \left( {\bar \nu _{\xi L}^C{\gamma _\mu }\nu _{\beta L}^C} \right) - \frac{1}{2}\left\langle {\bar \nu _{\xi L}^C{\gamma _\mu }\nu _{\beta L}^C} \right\rangle \left( {{{\bar \nu }_{\alpha L}}{\gamma ^\mu }{\nu _{\eta L}}} \right) + \left( {\alpha \eta  \leftrightarrow \xi \beta } \right)
\end{equation}
and
\begin{equation}\label{4-nu pseudo majorana}
\left( {{{\bar \nu }_\alpha }{\gamma ^5}{\nu _\beta }} \right)\left( {{{\bar \nu }_\xi }{\gamma ^5}{\nu _\eta }} \right) \approx \\
\frac{1}{2}\left\langle {{{\bar \nu }_{\alpha L}}{\gamma ^\mu }{\nu _{\eta L}}} \right\rangle \left( {\bar \nu _{\xi L}^C{\gamma _\mu }\nu _{\beta L}^C} \right) + \frac{1}{2}\left\langle {\bar \nu _{\xi L}^C{\gamma _\mu }\nu _{\beta L}^C} \right\rangle \left( {{{\bar \nu }_{\alpha L}}{\gamma ^\mu }{\nu _{\eta L}}} \right) + \left( {\alpha \eta  \leftrightarrow \xi \beta } \right).
\end{equation}
Unlike the Dirac neutrino, the charge conjugate currents of Majorana neutrino do not vanish even in the limit of zero neutrino mass. From the effective Hamiltonian operators (\ref{4-nu scalar majorana}) and (\ref{4-nu pseudo majorana}) we can derive the single-particle Hamiltonian that can be used in neutrino flavor evolution equations by evaluating the average value of neutrino currents under single-particle states. In the following derivation we consider a 2-flavor neutrino but from our result the generalization to neutrinos with more then 2 flavors is straightforward. The single-particle states for neutrino and antineutrino with momentum $\bf p$ are
\begin{equation}\label{flavor_state}
\left| {\nu \left( \bf p \right)} \right\rangle  = {a_e}\left| {{\nu _e}\left( \bf p \right)} \right\rangle  + {a_x}\left| {{\nu _x}\left( \bf p \right)} \right\rangle, \;\; \left| {\bar \nu \left( \bf p \right)} \right\rangle  = {{\bar a}_e}\left| {{{\bar \nu }_e}\left( \bf p \right)} \right\rangle  + {{\bar a}_x}\left| {{{\bar \nu }_x}\left( \bf p \right)} \right\rangle .
\end{equation} 
Evaluating the average values on the single-particle states we obtain (see appendix \ref{sec:meanfield})
\begin{equation}\label{normal_current}
\left\langle {\nu \left( \bf p \right)} \right.\left| {{{\bar \nu }_{\alpha L}}{\gamma ^\mu }{\nu _{\beta L}}} \right|\left. {\nu \left( \bf p \right)} \right\rangle  = \frac{{{p^\mu }}}{{E_{\bf p}\,V}}a_\alpha ^ * {a_\beta },\;\;\left\langle {\bar \nu \left( \bf p \right)} \right.\left| {{{\bar \nu }_{\alpha L}}{\gamma ^\mu }{\nu _{\beta L}}} \right|\left. {\bar \nu \left( \bf p \right)} \right\rangle  =  - \frac{{{p^\mu }}}{{E_{\bf p}\,V}}{\bar a_\beta} ^ * {\bar a_\alpha }
\end{equation} 
for normal currents and 
\begin{equation}\label{conjugate_current}
\left\langle {\nu \left( \bf p \right)} \right.\left| {\bar \nu _{\alpha L}^C{\gamma ^\mu }\nu _{\beta L}^C} \right|\left. {\nu \left( \bf p \right)} \right\rangle  =  - \frac{{{p^\mu }}}{{E_{\bf p}\,V}} a_\beta ^ * {a_\alpha },\;\;\left\langle {\bar \nu \left( \bf p \right)} \right.\left| {\bar \nu _{\alpha L}^C{\gamma ^\mu }\nu _{\beta L}^C} \right|\left. {\bar \nu \left( \bf p \right)} \right\rangle  = \frac{{{p^\mu }}}{E_{\bf p}\,V}\bar a_\alpha ^ * {{\bar a}_\beta }
\end{equation}
for charge conjugate currents, respectively. Here $p^{\mu} \equiv (E_{\bf p},{\bf p})$ is the 4-momentum. If we define the single-particle density matrices as \cite{Duan:2006an}
\begin{equation}\label{density matrices}
{\rho}({\bf{p}}) = \left( {\begin{array}{*{20}{c}}
{{{\left| {{a_e}} \right|}^2}}&{{a_e}a_x^ * }\\
{a_e^ * {a_x}}&{{{\left| {{a_x}} \right|}^2}}
\end{array}} \right),\:\:{{\bar \rho }}({\bf{p}}) = \left( {\begin{array}{*{20}{c}}
{{{\left| {{{\bar a}_e}} \right|}^2}}&{{{\bar a}_e}\bar a_x^ * }\\
{\bar a_e^ * {{\bar a}_x}}&{{{\left| {{{\bar a}_x}} \right|}^2}}
\end{array}} \right)
\end{equation}
for neutrinos and antineutrinos respectively, then the final single-particle effective Hamiltonian of the nonstandard neutrino self-interaction can be obtained as (see the appendix \ref{effective Hamiltonian} for details)
\begin{equation}\label{scalar}
{H_{\rm S}}\left( r,{\bf{\hat{q}}} \right) = 4\,\int {\left( {1 - {\bf{\hat p}} \cdot {\bf{\hat q}}} \right)\left\{ {{\bf{\tilde g}}\left[ {\rho^* (r,{\bf{p}})\,d{n_\nu }\left( {r,{\bf{p}}} \right) - {{\bar \rho }}(r,{\bf{p}})\,d{n_{\bar \nu }}\left( {r,{\bf{p}}} \right)} \right]{\bf{\tilde g}}} \right\} dE_{\bf p}} 
\end{equation}
for neutrino-neutrino interaction via a scalar field and similarly,
\begin{equation}\label{pseudo}
{H_{\rm P}}\left( r,{\bf{\hat{q}}} \right) = 4\,\int {\left( {1 - {\bf{\hat p}} \cdot {\bf{\hat q}}} \right)\left\{ {{\bf{\tilde h}}\left[ {\rho^* (r,{\bf{p}})\,d{n_\nu }\left( {r,{\bf{p}}} \right) - {{\bar \rho }}(r,{\bf{p}})\,d{n_{\bar \nu }}\left( {r,{\bf{p}}} \right)} \right]{\bf{\tilde h}}} \right\} dE_{\bf p}} 
\end{equation}
for neutrino-neutrino interaction through a pseudoscalar field. Here $E_{\bf p}$ is the energy of the background neutrinos with momentum ${\bf p}$, and the elements of $\bf{\tilde g}$ and $\bf{\tilde h}$ are $({\bf{\tilde g}})_{\alpha\beta}\equiv{\tilde g}_{\alpha\beta}=\frac{1}{4m_{\phi}} g_{\alpha\beta}$ and $({\bf{\tilde h}})_{\alpha\beta}\equiv{\tilde h}_{\alpha\beta}=\frac{1}{4m_{\chi}} h_{\alpha\beta}$. Note that Eqs. (\ref{scalar}) and (\ref{pseudo}) are valid for a neutrino model with arbitrary number of flavors. 

Thus we can add to the standard V-A self-interaction a new term given in Eqs. (\ref{scalar}) and/or (\ref{pseudo}) so that 
\begin{equation}
H_{\rm SI} = H_{\rm V-A} + H_{\rm S/P}.
\end{equation}
At first glance the expressions for the NSSI looks very similar to the NSSI Hamiltonian due to gauge bosons \cite{Das:2017iuj}, as both of them have a current-current nature and are modulated by the coupling matrix $\tilde{\bf g}$. However, they are distinct in that wherever the gauge boson Hamiltonian uses the density matrix $\rho_{\bf p}$ ($\bar \rho_{\bf p}^{*}$) the NSSI uses $\rho_{\bf p}^{*}$ ($\bar \rho_{\bf p}$). In addition, the NSSI mediated by a scalar or pseudoscalar field emerges only from the ``exchange terms'' of the interaction so we do not find the term ${\bf{\tilde g}}\,{\rm{Tr}}\left[ {\left( {{\rho _{\bf{p}}} - \bar \rho _{\bf{p}}^ * } \right)\,{\bf{\tilde g}}} \right]$ which appears in the gauge boson case \cite{Das:2017iuj}.
We shall see that these subtle nuances between the form of the self-interaction with the standard V-A or, indeed, any gauge-mediated boson interaction, and a scalar/pseudoscalar interaction are key for the NSSI to have distinct observable effects. 


\begin{figure}[t]
\centering
\subfigure{
\includegraphics[width=.45\textwidth]{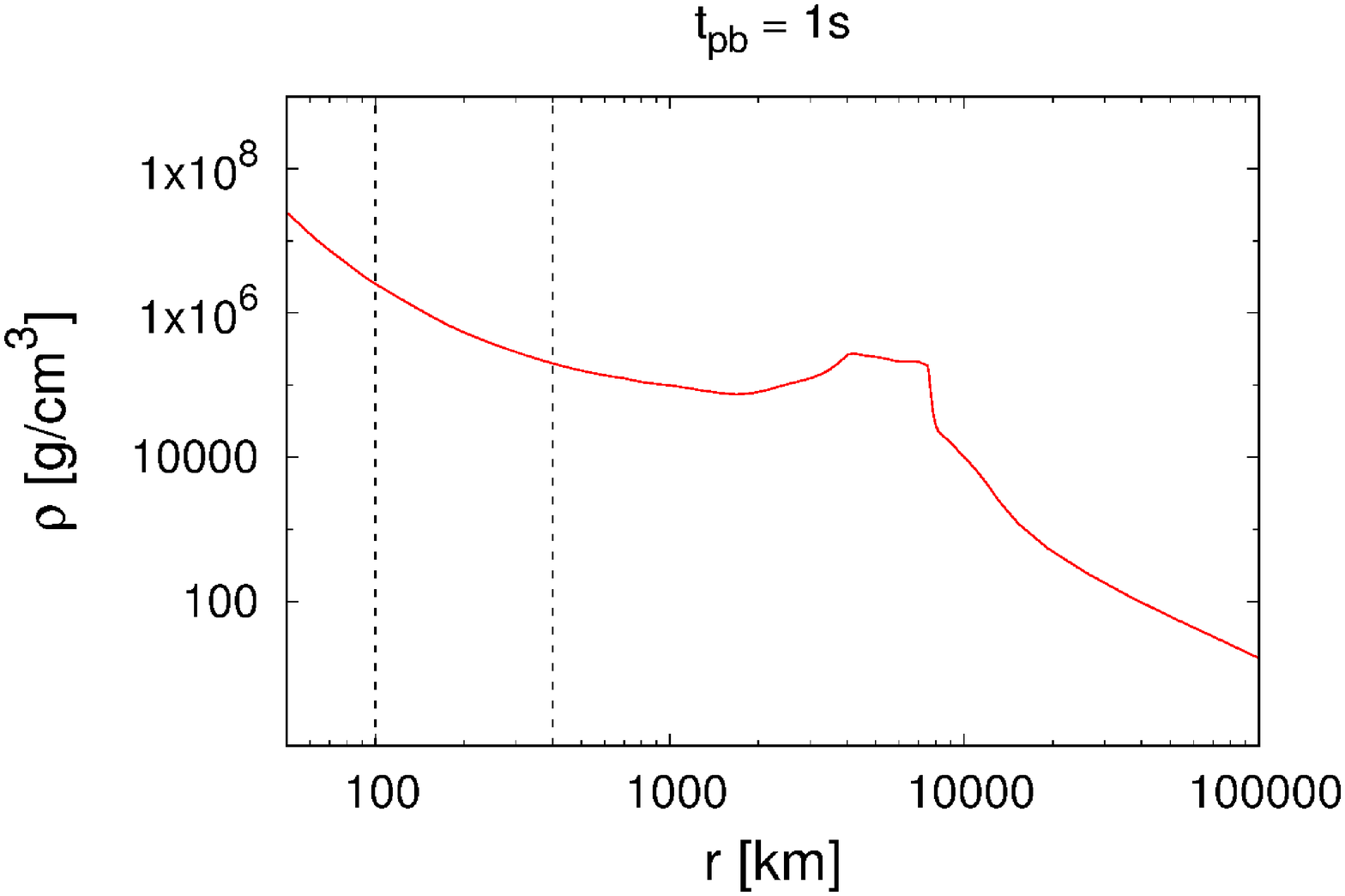}
}
\subfigure{
\includegraphics[width=.45\textwidth]{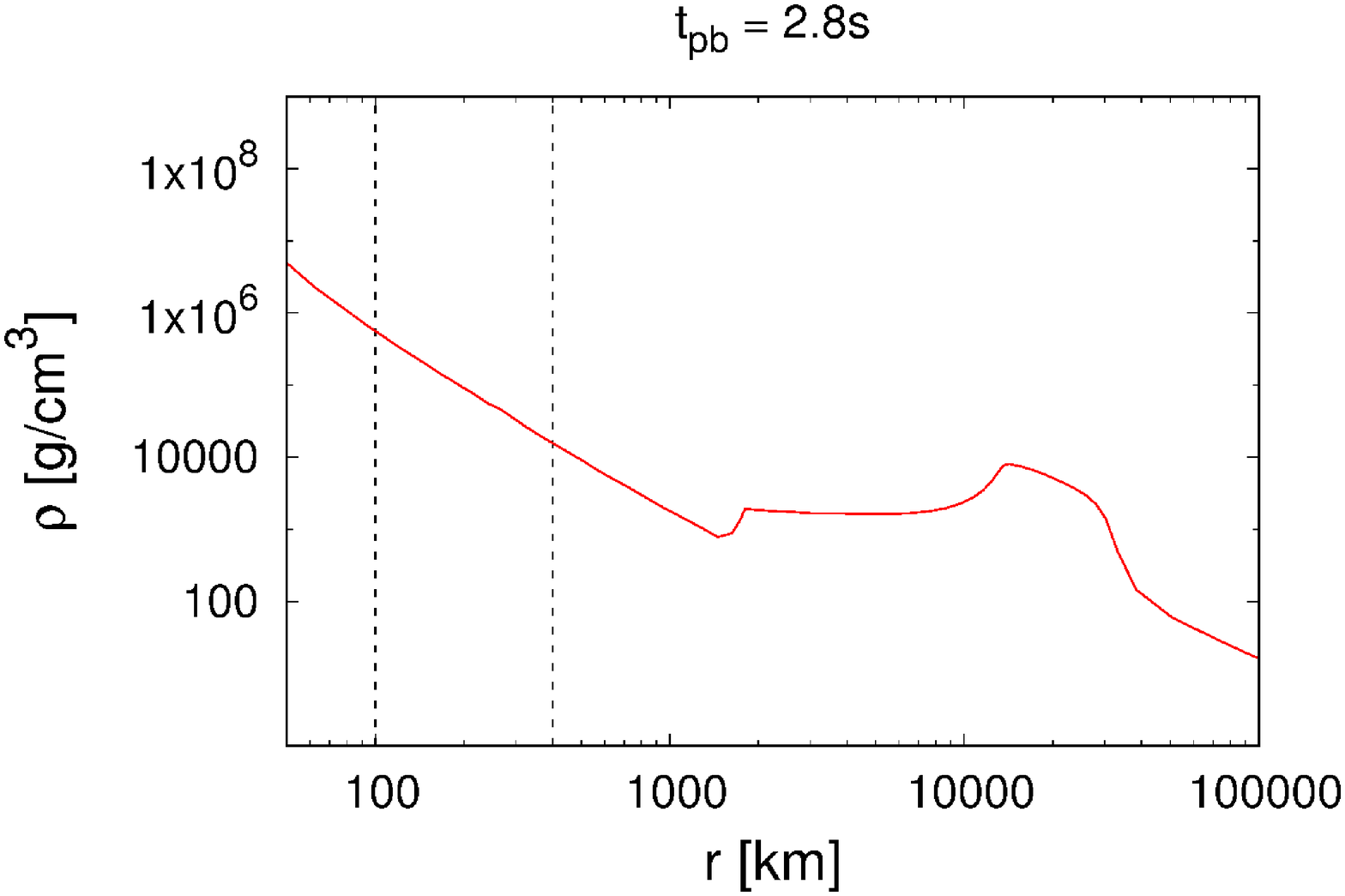}
}
\caption{The matter density profiles being used for the calculations of neutrino flavor transformation. The two dashed lines in each plot indicate the beginning and end of the calculation.}
\label{fig:profiles}
\end{figure}

\section{The effects of NSSI on neutrino flavor transformation in supernovae}
\label{sec:transformations}

Since the NSSI from scalar and pseudoscalar interactions have the same form we treat them as indistinguishable and focus on the phenomenological consequences of the scalar part of the NSSI. We define two parameters $\alpha_1$ and $\alpha_2$ so that the $\bf{\tilde{g}}$ matrix is parameterized as
\begin{equation}
{\bf{\tilde g}} = \left[\frac{{\sqrt 2 }}{4}{G_{\rm F}}\right]^{1/2}\left( {\begin{array}{*{20}{c}}
{\alpha _1}&{{\alpha _2}}&{{\alpha _2}}\\
{{\alpha _2}}&{\alpha _1}&{{\alpha _2}}\\
{{\alpha _2}}&{{\alpha _2}}&{\alpha _1}
\end{array}} \right).
\end{equation}
The parameter $\alpha_1$ indicates the strength of flavor-preserving (FP) NSSI while $\alpha_2$ indicates the strength of flavor-violating (FV) NSSI. When $\alpha_1$ or $\alpha_2$ is equal to unity it means the corresponding NSSI has an strength equal to the standard V-A interaction. For simplicity we have assumed the flavor-preserving and flavor-violating parameters are identical for all flavors but note this is a restriction that can be relaxed. 

The neutrino mixing angles and square mass differences we adopt throughout the rest of the paper are $m^2_2-m^2_1=7.59\times10^{-5}\;\text{eV}^2$, $\left| {m_3^2 - m_2^2} \right|=2.32\times10^{-3}\;\text{eV}^2$ $\theta_{12}=33.9^\circ$ $\theta_{13}=9^\circ$ and $\theta_{23}=45^\circ$ which are consistent with the Particle Data Group evaluations \cite{Olive:2016xmw}. The CP phase $\delta_{\rm CP}$ is set to zero. In the following calculations we will generally work with the inverted mass ordering (IMO) but will show some results using the normal mass ordering (NMO) and will indicate when this occurs.

\begin{table}[t]
\begin{tabular}{l*{3}{c}}
Flavor & \;Luminosity $L_{\nu,\infty}$  & \;Mean Energy $\langle E_{\nu,\infty}\rangle$ &\;rms Energy $\sqrt{ \langle E^2_{\nu,\infty}\rangle }$\\
\hline
$e$ & $4.606\times 10^{51}\;{\rm erg/s}$ & $10.24\;{\rm MeV}$ & $11.44\;{\rm MeV}$ \\
$\mu$,$\tau$ & $5.473\times 10^{51}\;{\rm erg/s}$ & $14.32\;{\rm MeV}$ & $16.78\;{\rm MeV}$ \\
$\bar{e}$ & $4.572\times 10^{51}\;{\rm erg/s}$ & $12.88\;{\rm MeV}$ & $14.51\;{\rm MeV}$ \\
$\bar{\mu}$, $\bar{\tau}$ & $5.522\times 10^{51}\;{\rm erg/s}$ & $14.42\;{\rm MeV}$ & $16.93\;{\rm MeV}$
\end{tabular}
\caption{The luminosities, mean energies, and rms energies used for the $t_{pb}=1.0{\;\rm s}$ calculation.} 
\label{tab:t=1.0s}  
\end{table}
\begin{figure}[b]
\centering
\subfigure{
\includegraphics[width=.45\textwidth]{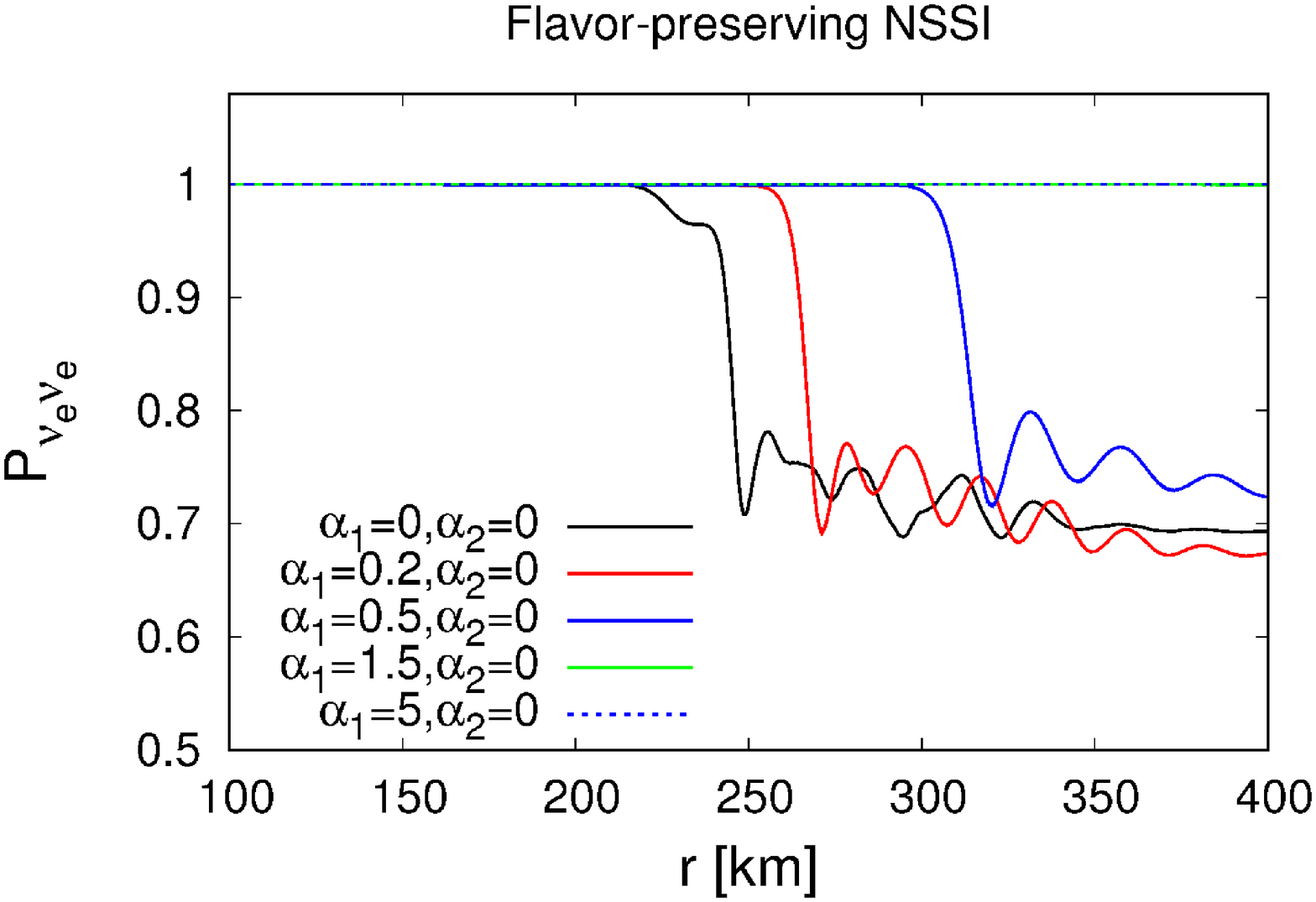}
}
\subfigure{
\includegraphics[width=.45\textwidth]{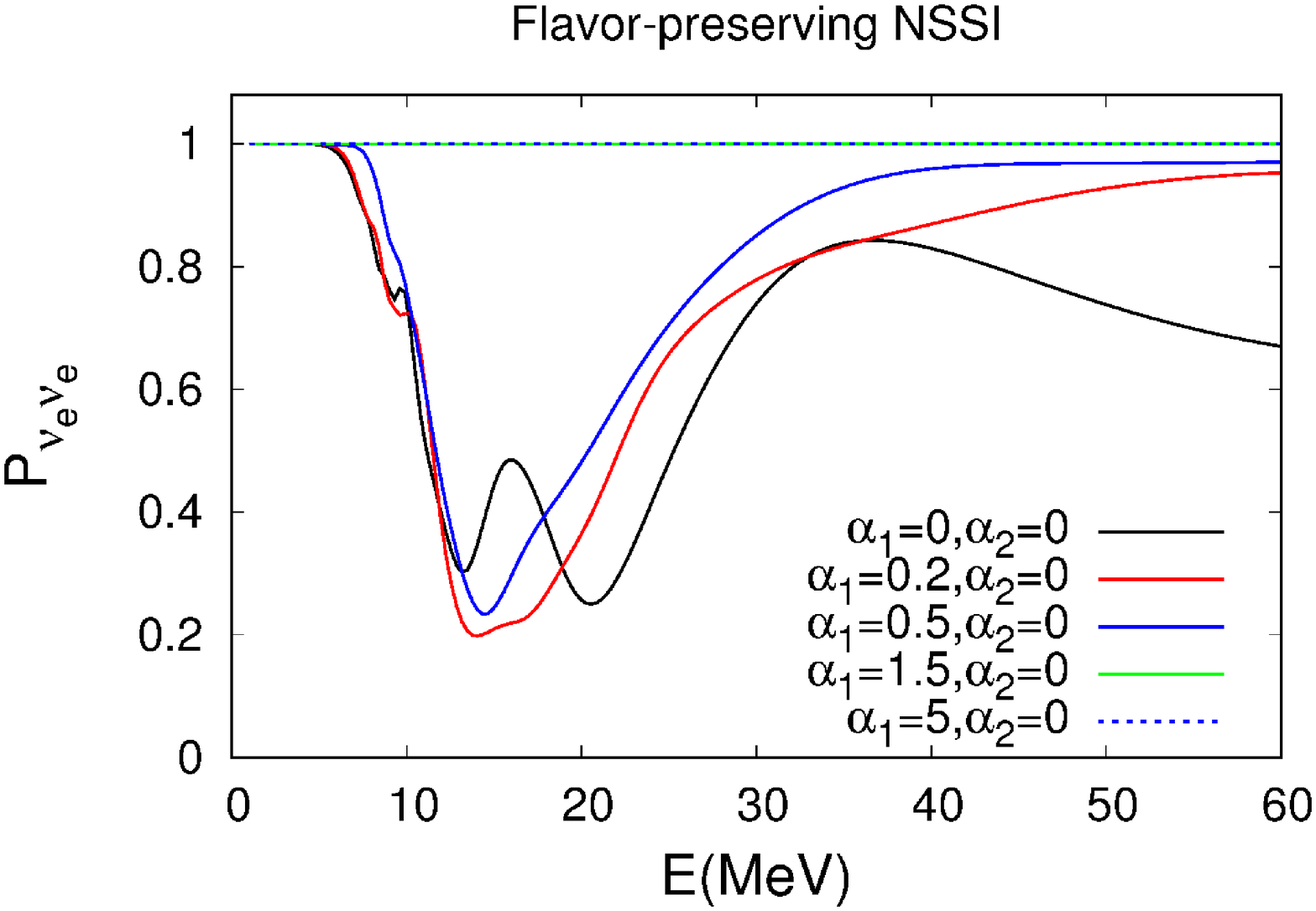}
}
\subfigure{
\includegraphics[width=.45\textwidth]{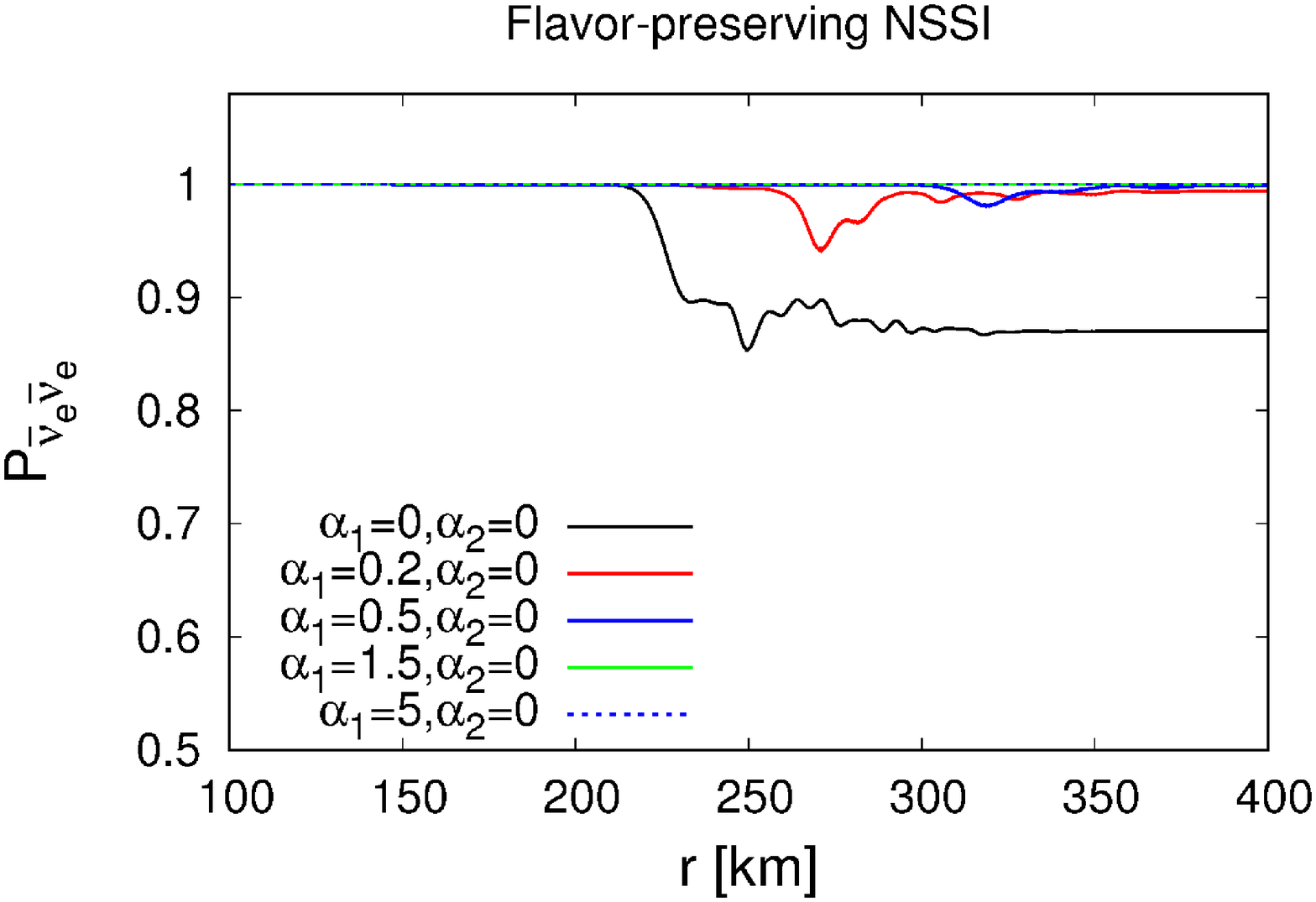}
}
\subfigure{
\includegraphics[width=.45\textwidth]{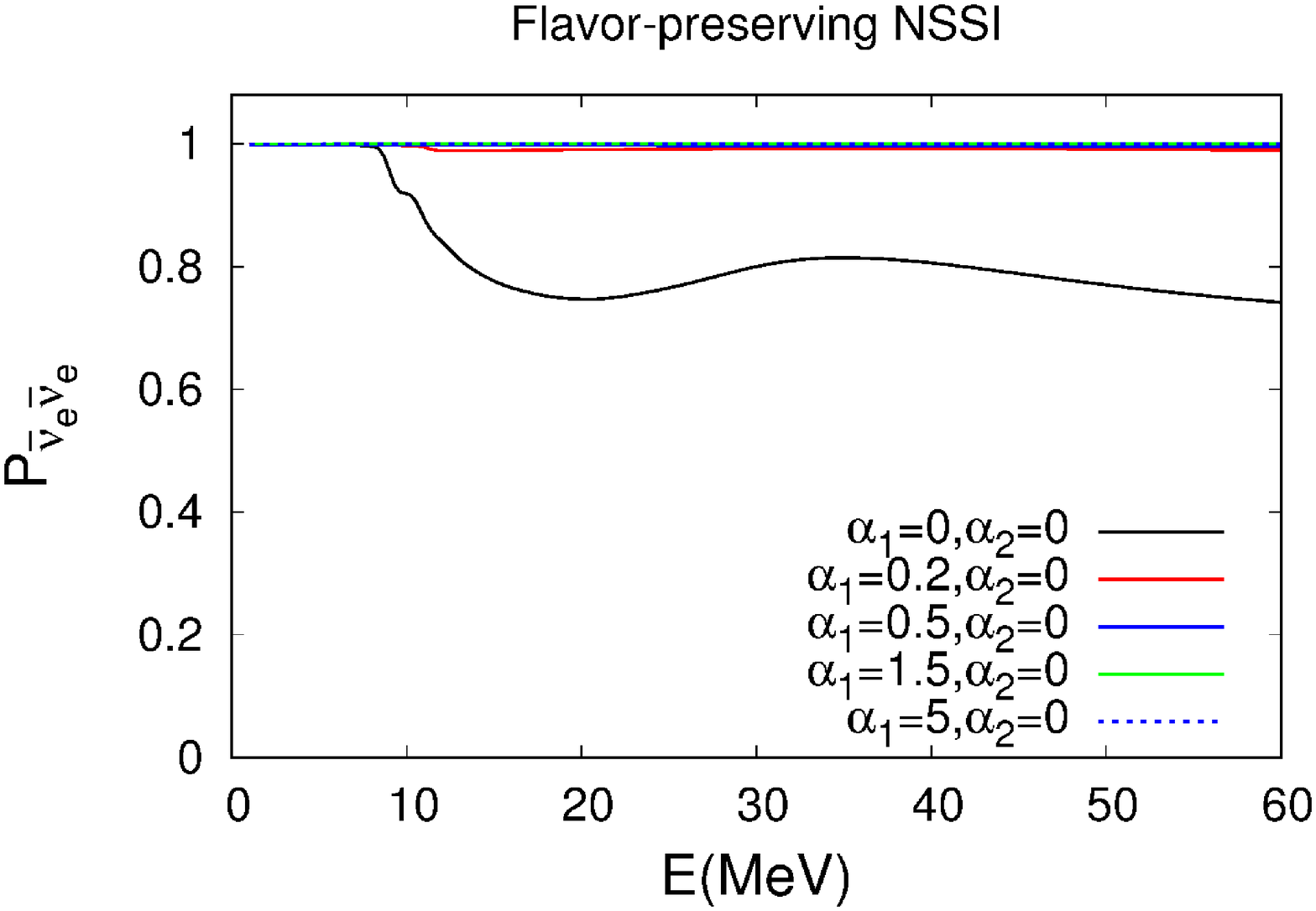}
}
\caption{Survival probability of electron neutrinos (top panels) and antineutrinos (bottom panels) with flavor-preserving NSSI at $t_{pb}=1.0s$. The left panels are the flux averaged probabilities as a function of distance $r$ while the right panels are plotted as  function of energy at $r=400\;{\rm km}$. The combinations of the NSSI parameters are given in the legends.}
\label{fig:520FP_SP}
\end{figure}
The density profiles and neutrino spectra for our calculations comes from the 1-D GR-compatible CCSN simulation for the $10.8\;{\rm M_{\odot}}$ progenitor calculated by Fischer \emph{et al.} \cite{Fischer:2009af}. The matter density profiles are shown in figure (\ref{fig:profiles}). The neutrino emission is assumed to be half-isotropic and the neutrino spectra at $r$ are given by the pinched thermal spectra found by Keil \emph{et al.} \cite{2003ApJ...590..971K}. Therefore we have 
\begin{equation}
d{n_\nu }\left( {r,{\bf{p}}} \right) = \frac{{{L_{\nu ,\infty }}}}{{4{\pi ^2}{\mkern 1mu} R_\nu ^2\langle {E_{\nu ,\infty }}\rangle }}\,{f_\nu }\left(E_{\bf p} \right)d(\cos \theta )d\phi 
\end{equation}
with
\begin{equation}
{f_\nu }\left( E_{\bf p} \right) = \frac{{{{({\gamma _\nu } + 1)}^{{\gamma _\nu } + 1}}}}{{\Gamma ({\gamma _\nu } + 1)}}{\mkern 1mu} \frac{{{{E}_{\bf p}^{{\gamma _\nu }}}}}{{{{\langle {E_{\nu ,\infty }}\rangle }^{{\gamma _\nu } + 1}}}}\exp \left( { - \frac{{({\gamma _\nu } + 1){\mkern 1mu} {E_{\bf p}}}}{{\langle {E_{\nu ,\infty }}\rangle }}} \right),
\end{equation}
where $\theta$ is the angle between the neutrino beams and the radial direction at r, $\phi$ the azimuthal angle of the beam, $L_{\nu,\infty}$ the neutrino luminosity, $\langle E_{\nu,\infty}\rangle$ the mean energy and $\gamma_{\nu}$ the pinch parameter which can be derived from the mean energy $\langle E_{\nu,\infty}\rangle$ and the mean square energy $\langle E^2_{\nu,\infty}\rangle$ via
\begin{equation}
\gamma_{\nu} = \frac{2\langle E_{\nu,\infty} \rangle^2 - \langle E^2_{\nu,\infty} \rangle}{\langle E^2_{\nu,\infty} \rangle-\langle E_{\nu,\infty} \rangle^2}.
\end{equation}
The numerical values for the neutrino luminosities, mean and rms energies for post-bounce times of $t_{pb} = 1.0 \;{\rm s}$ and $t_{pb} = 2.8\;{\rm s}$ are shown in tables (\ref{tab:t=1.0s}) and (\ref{tab:t=2.8s}). These two snapshots are representative of the early to intermediate cooling phase of CCSN explosion and were chosen based on the results from Wu \emph{et al.} \cite{PhysRevD.91.065016} which showed flavor transformations at these two epochs for the $18.0\;{\rm M_{\odot}}$ simulation by Fischer \emph{et al.} \cite{Fischer:2009af} and the similarity of the neutrino spectra in this model with the $10.8\;{\rm M_{\odot}}$ simulation also by Fischer \emph{et al}. The neutrinosphere radius is set to $R_{\nu} = 19\;{\rm km}$ for the $t_{pb} = 1.0\;{\rm s}$ profile and $R_{\nu} = 17\;{\rm km}$ for the $t_{pb}=2.8\;{\rm s}$. For both time slices we compute the evolution starting from $r=100\;{\rm km}$. Our calculation adopts the multi-angle, multi-energy bulb model framework for energies ranging from $1\;{\rm MeV}$ to $60\;{\rm MeV}$ in 200 bins, and the neutrino emission angles ranging from $0^{\circ}$ to $90^{\circ}$ in 200 bins\footnote{Determination of the number of angle bins needed in multi-angle calculations can be difficult. Insufficient angular resolution has been found to cause spurious flavor instabilities\cite{sarikas2012spurious}. However, for the CCSN cooling phase, the matter density is generally not high enough for such artifacts to develop so the required number of angular bins can be reduced. Convergence has been checked to make sure 200 bins are sufficient for both $t_{pb}=1.0\;{\rm s}$ and $t_{pb}=2.8\;{\rm s}$.}. We have also verified our results have converged with the number of energy and angular bins. 

\begin{figure}[h]
\centering
\subfigure{
\includegraphics[width=.3\textwidth]{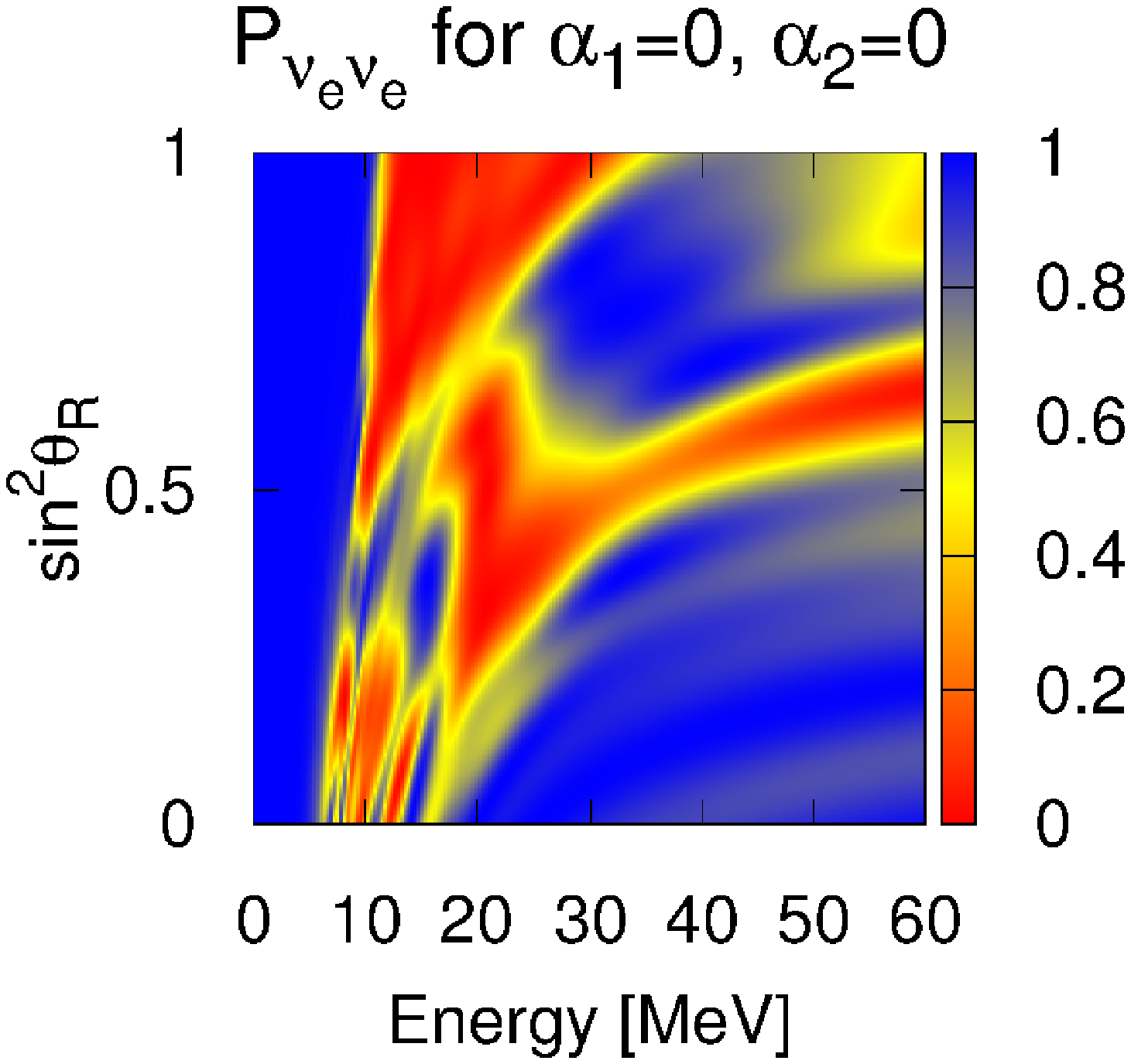}
}
\subfigure{
\includegraphics[width=.3\textwidth]{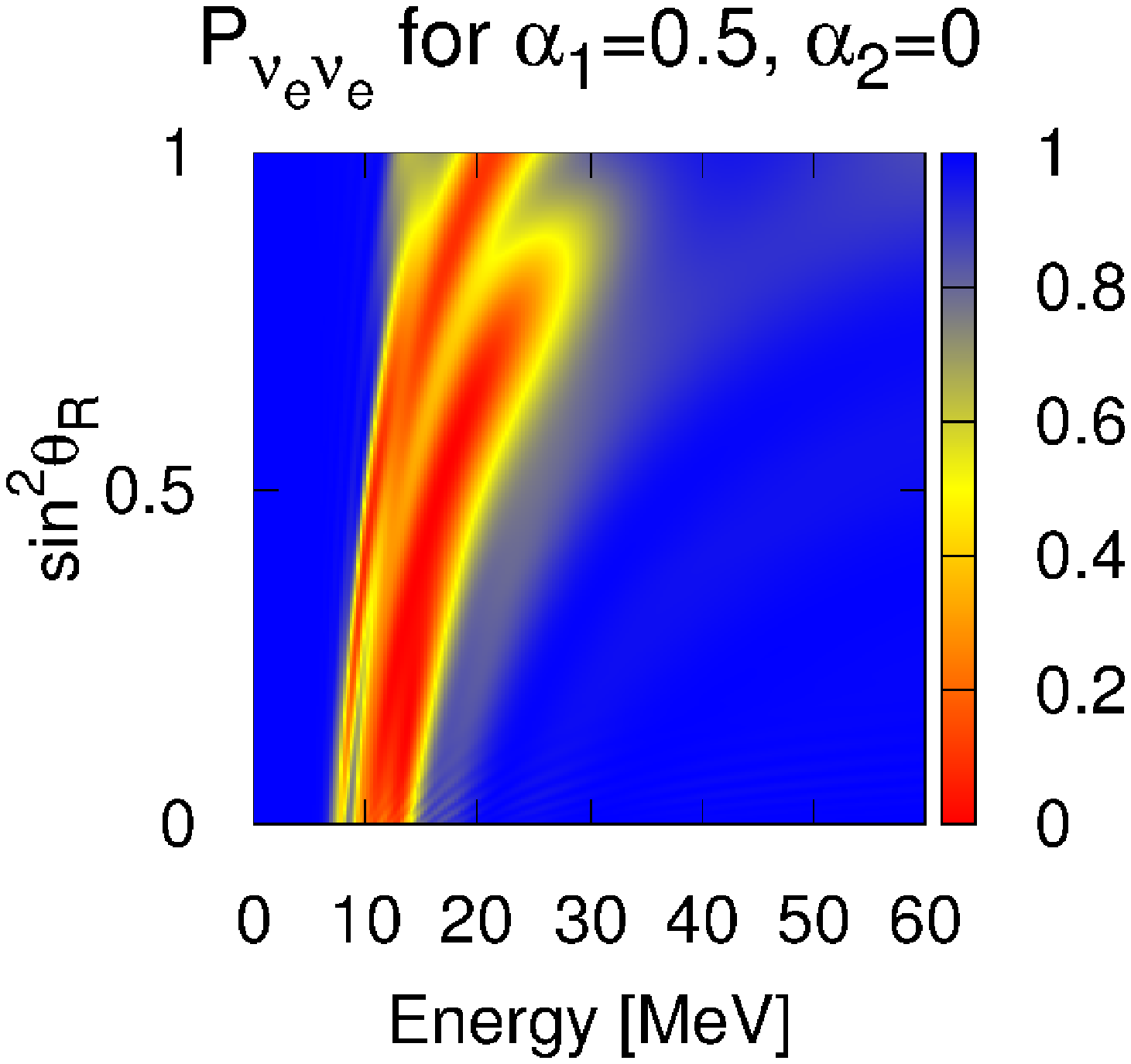}
}
\subfigure{
\includegraphics[width=.3\textwidth]{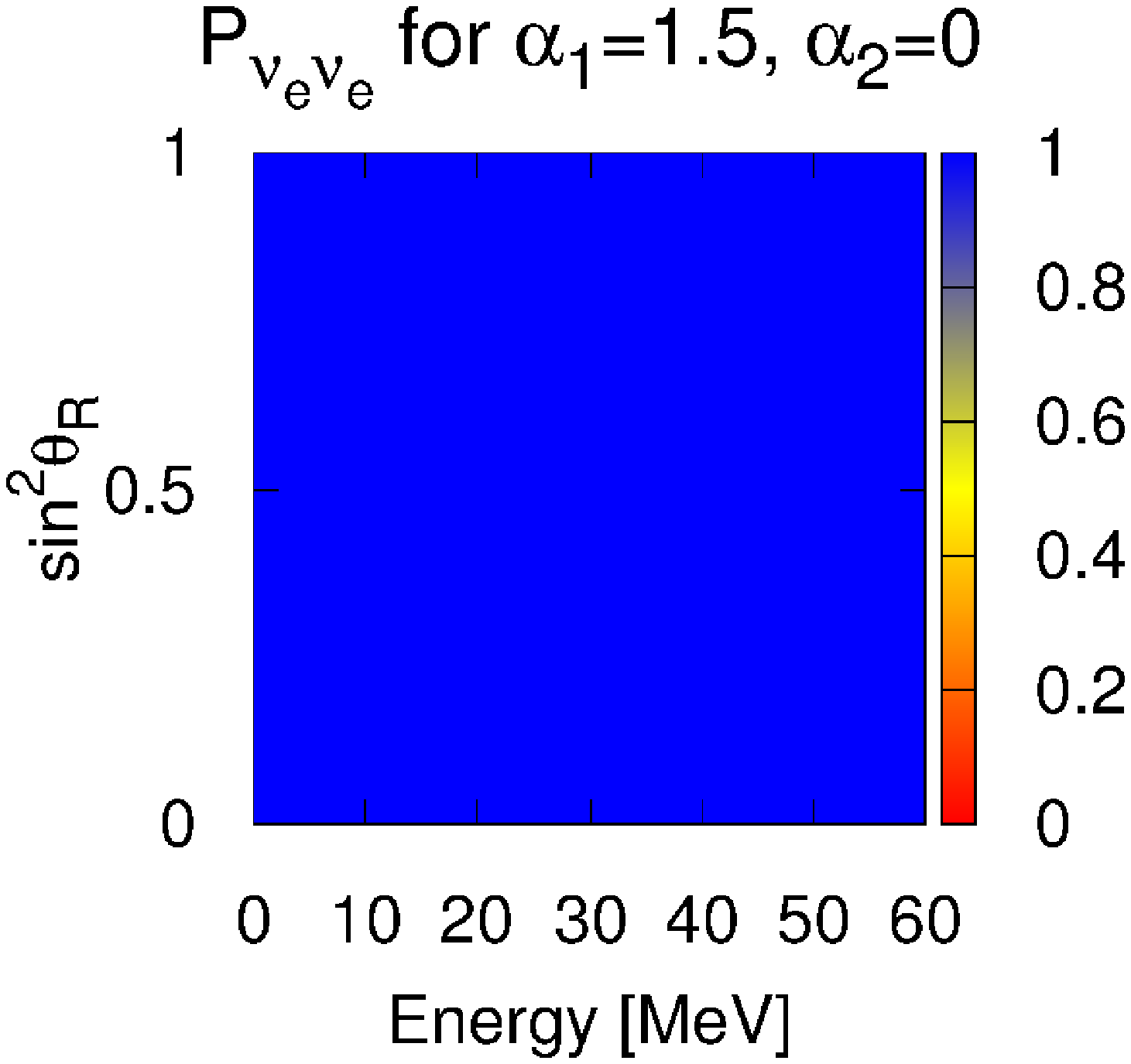}
}
\subfigure{
\includegraphics[width=.3\textwidth]{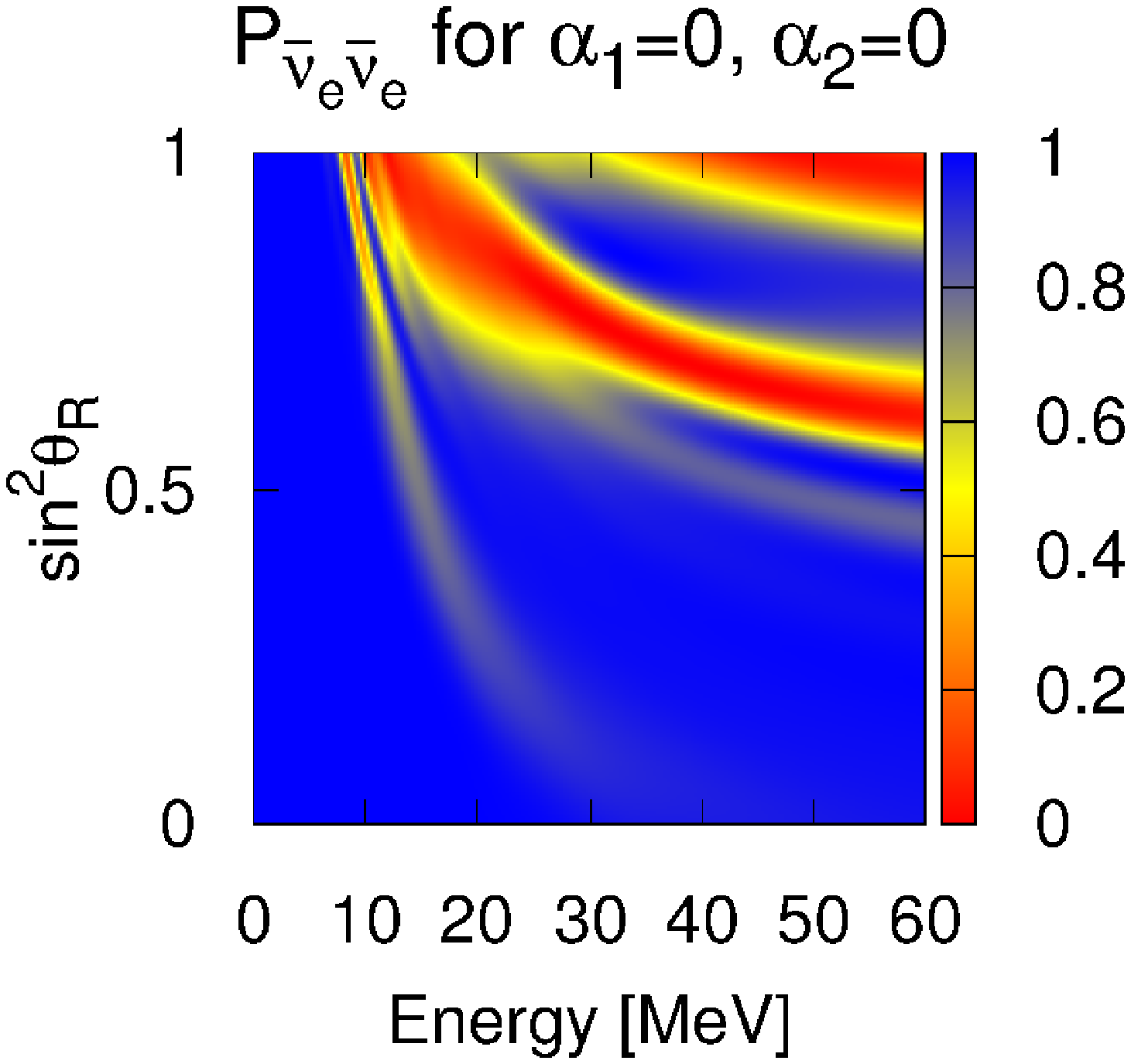}
}
\subfigure{
\includegraphics[width=.3\textwidth]{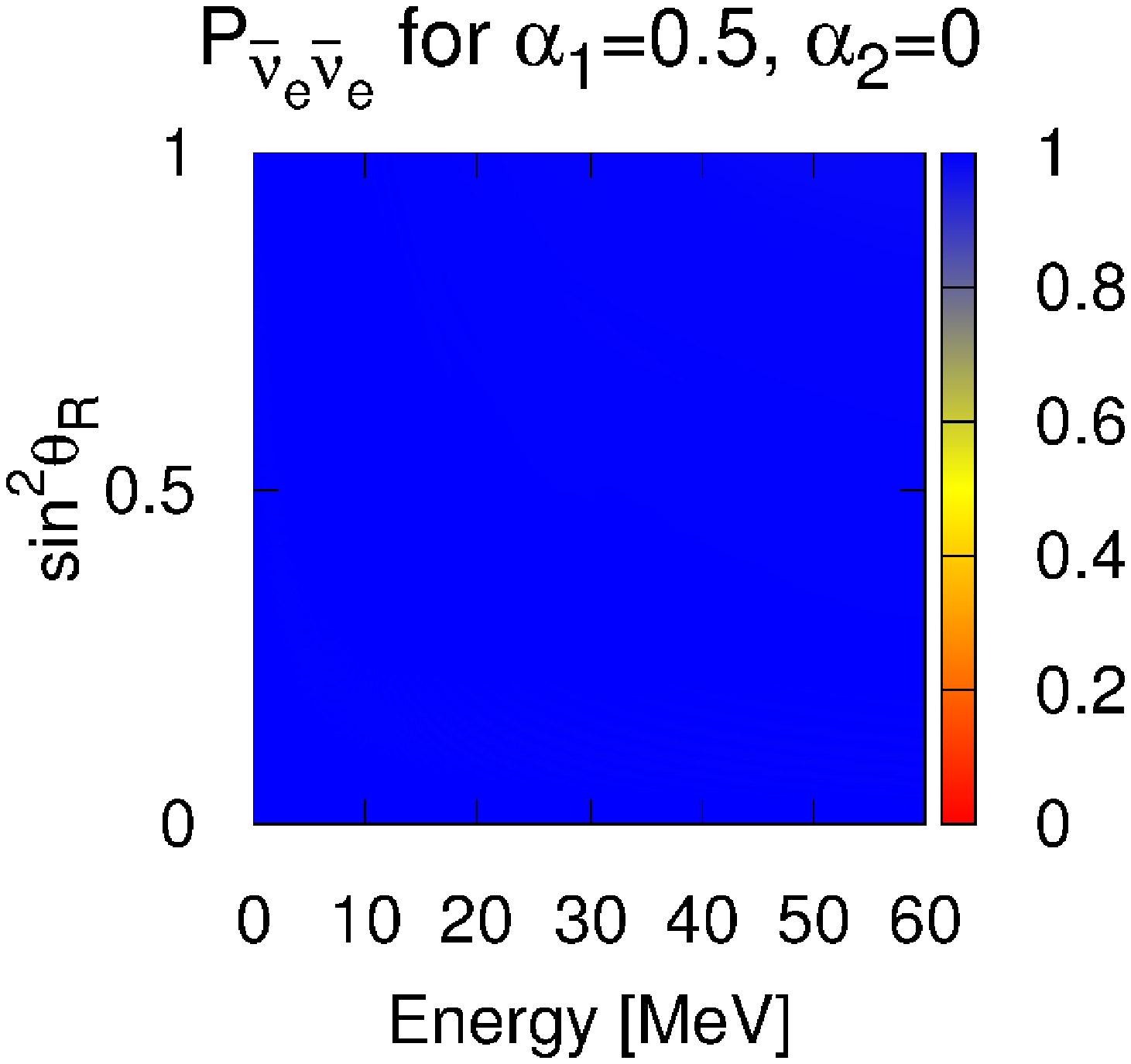}
}
\subfigure{
\includegraphics[width=.3\textwidth]{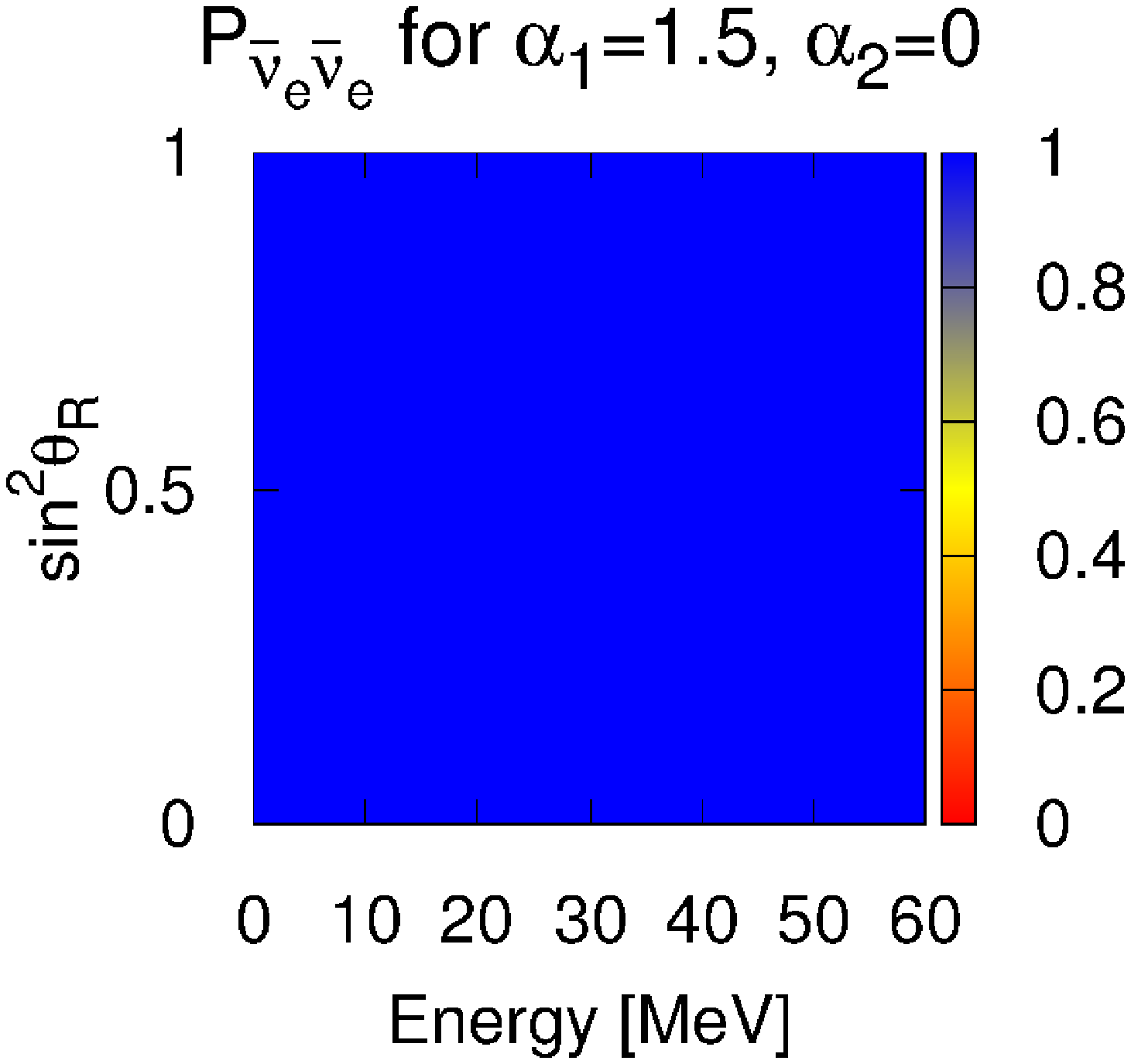}
}
\caption{Top panels: The heatmaps of survival probability of electron neutrinos at $t_{pb}=1.0s$ and $r=400\rm{km}$ as a function of energy and emission angle when there is only flavor-preserving NSSI. Bottom panels: The same but for electron antineutrinos.}
\label{fig:520FP_2D}
\end{figure}


\subsection{Flavor transformation at $t_{pb}=1.0\;\rm{s}$}

\begin{figure}[b]
\centering
\subfigure{
\includegraphics[width=.45\textwidth]{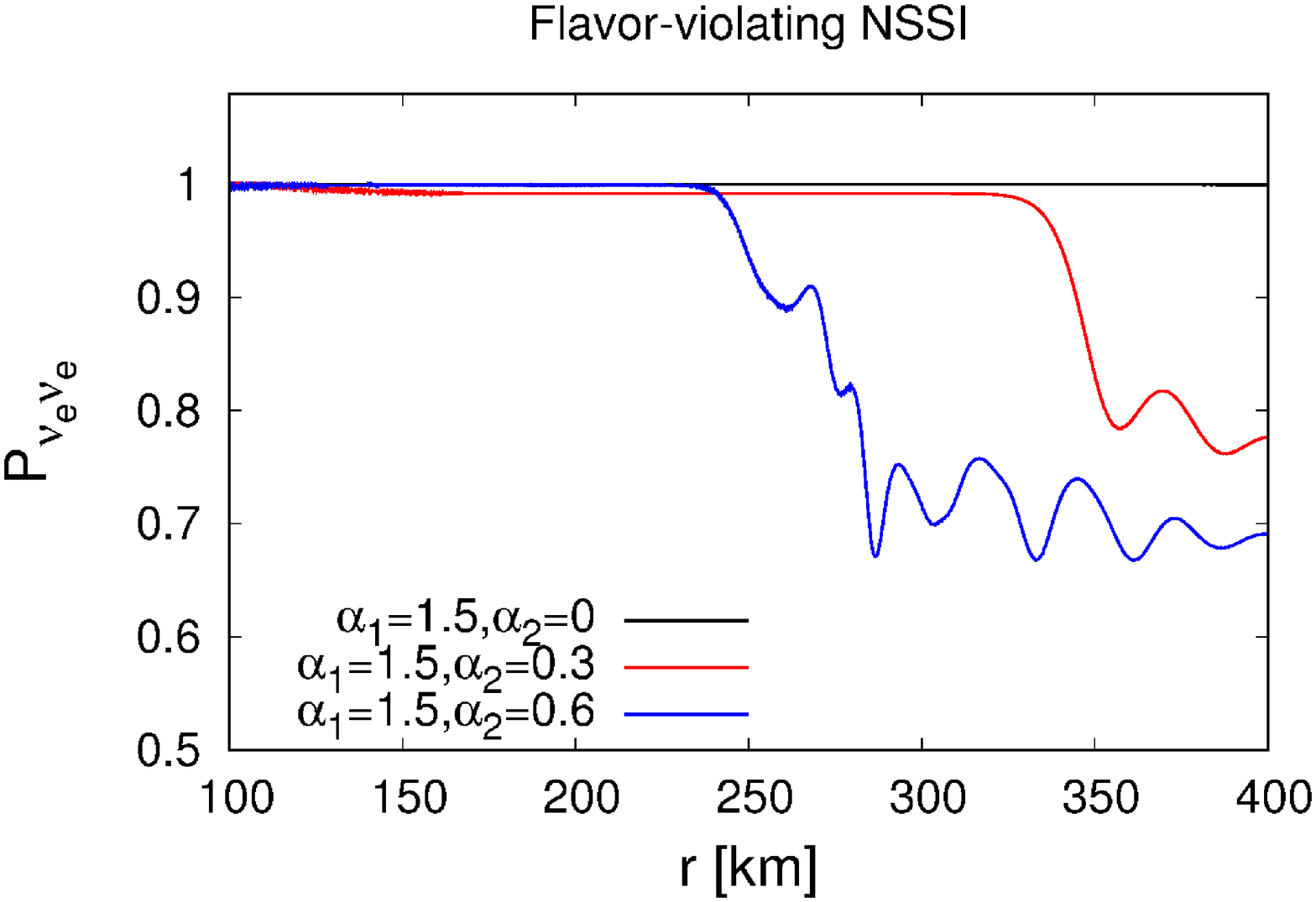}
}
\subfigure{
\includegraphics[width=.45\textwidth]{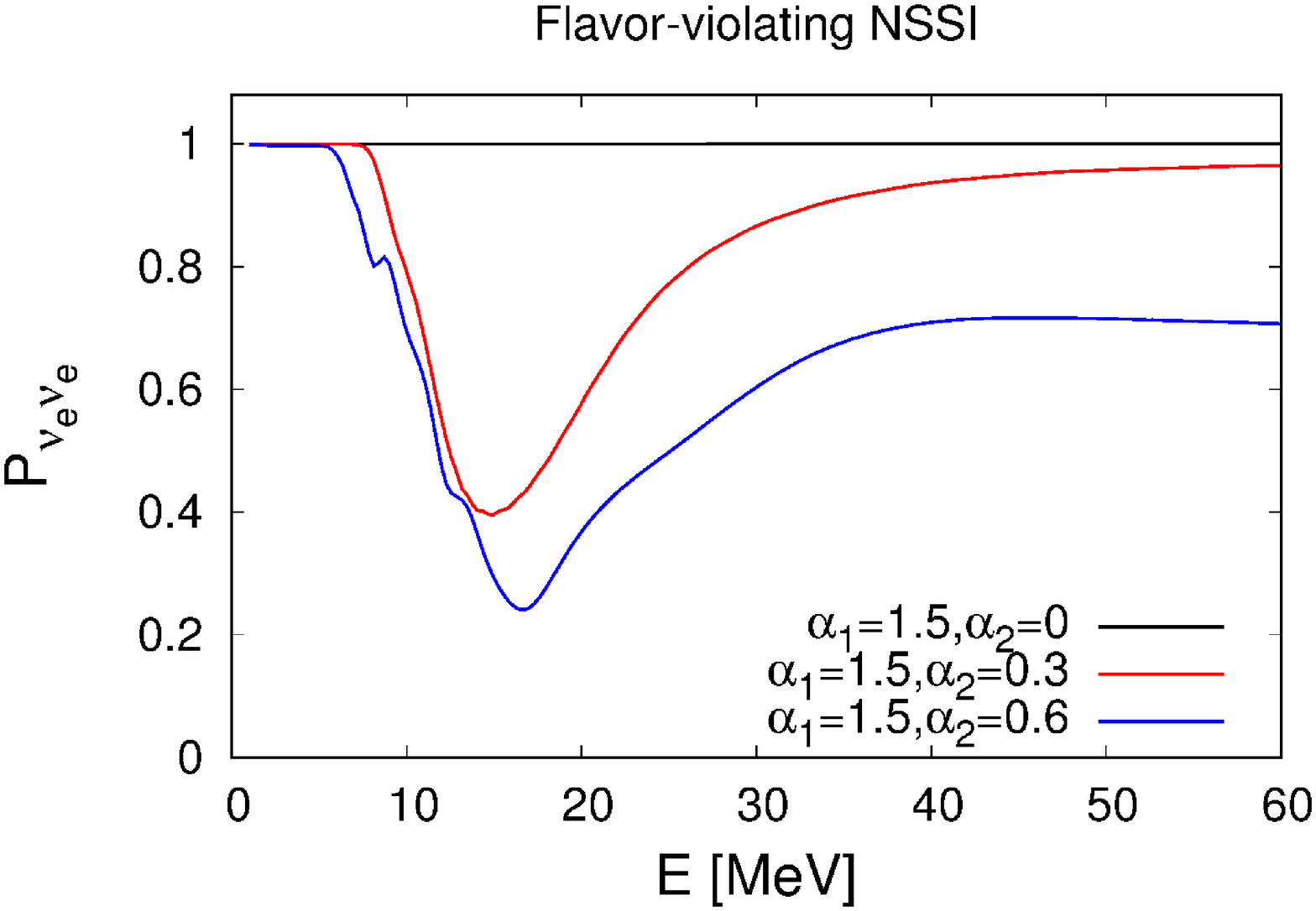}
}
\subfigure{
\includegraphics[width=.45\textwidth]{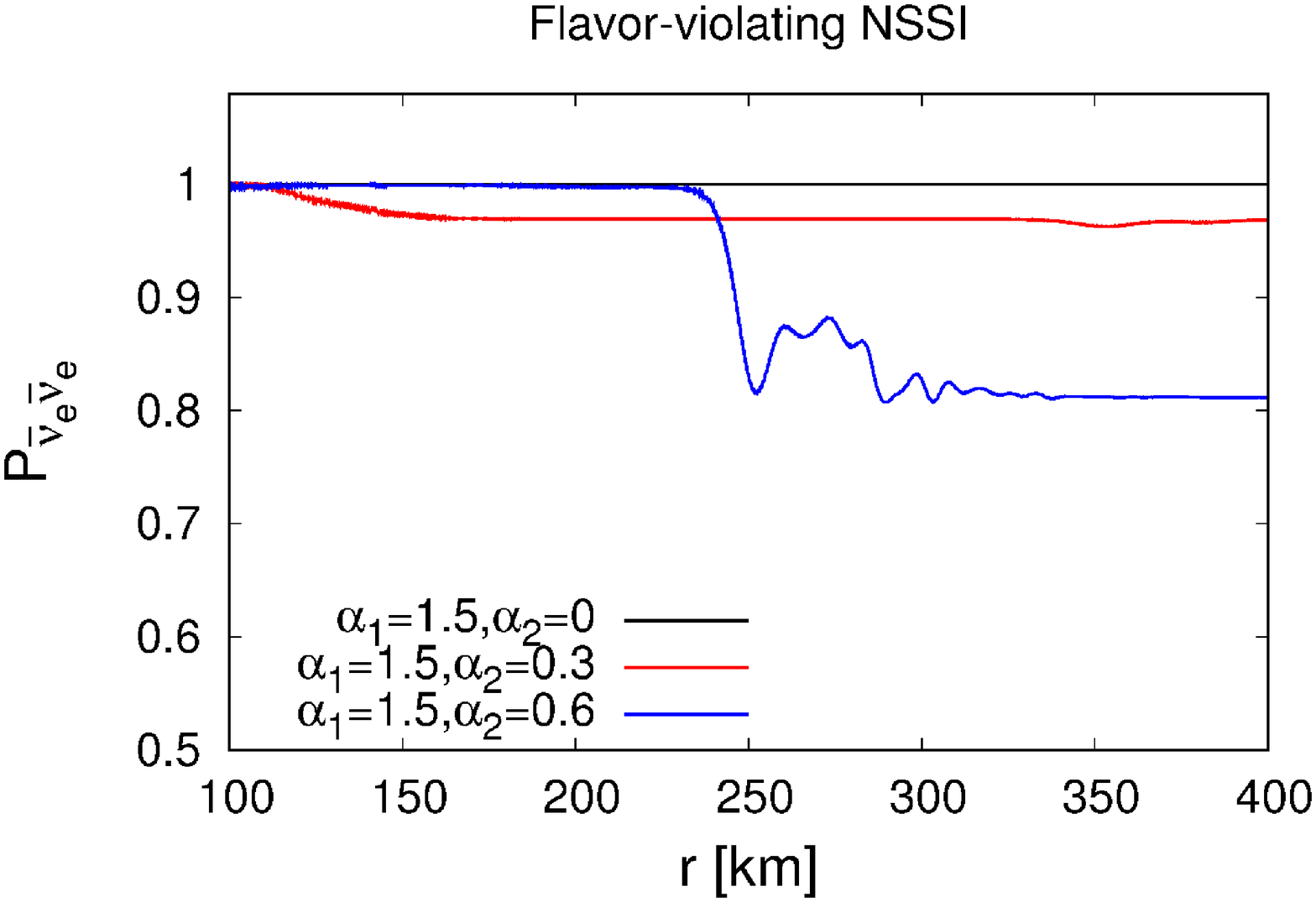}
}
\subfigure{
\includegraphics[width=.45\textwidth]{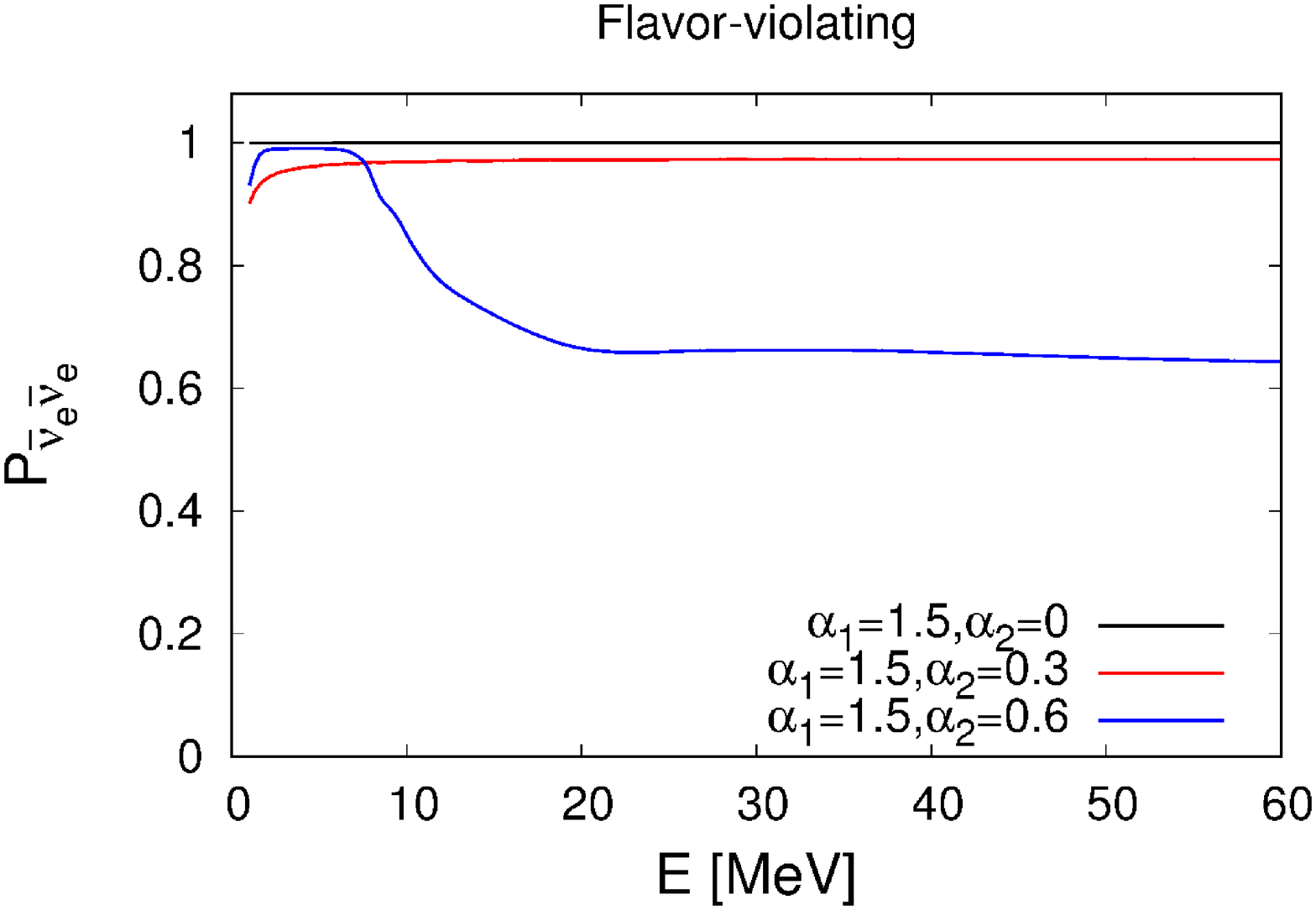}
}
\caption{Top panels: Survival probability of electron neutrinos at $t_{pb}=1.0s$ as a function of distance (left panel) and energy (right panel) at $r=400\;{\rm km}$ with flavor-violating NSSI. The bottom panels are the same but for electron antineutrinos.}
\label{fig:520FV_SP}
\end{figure}
\begin{figure}[h]
\centering
\subfigure{
\includegraphics[width=.3\textwidth]{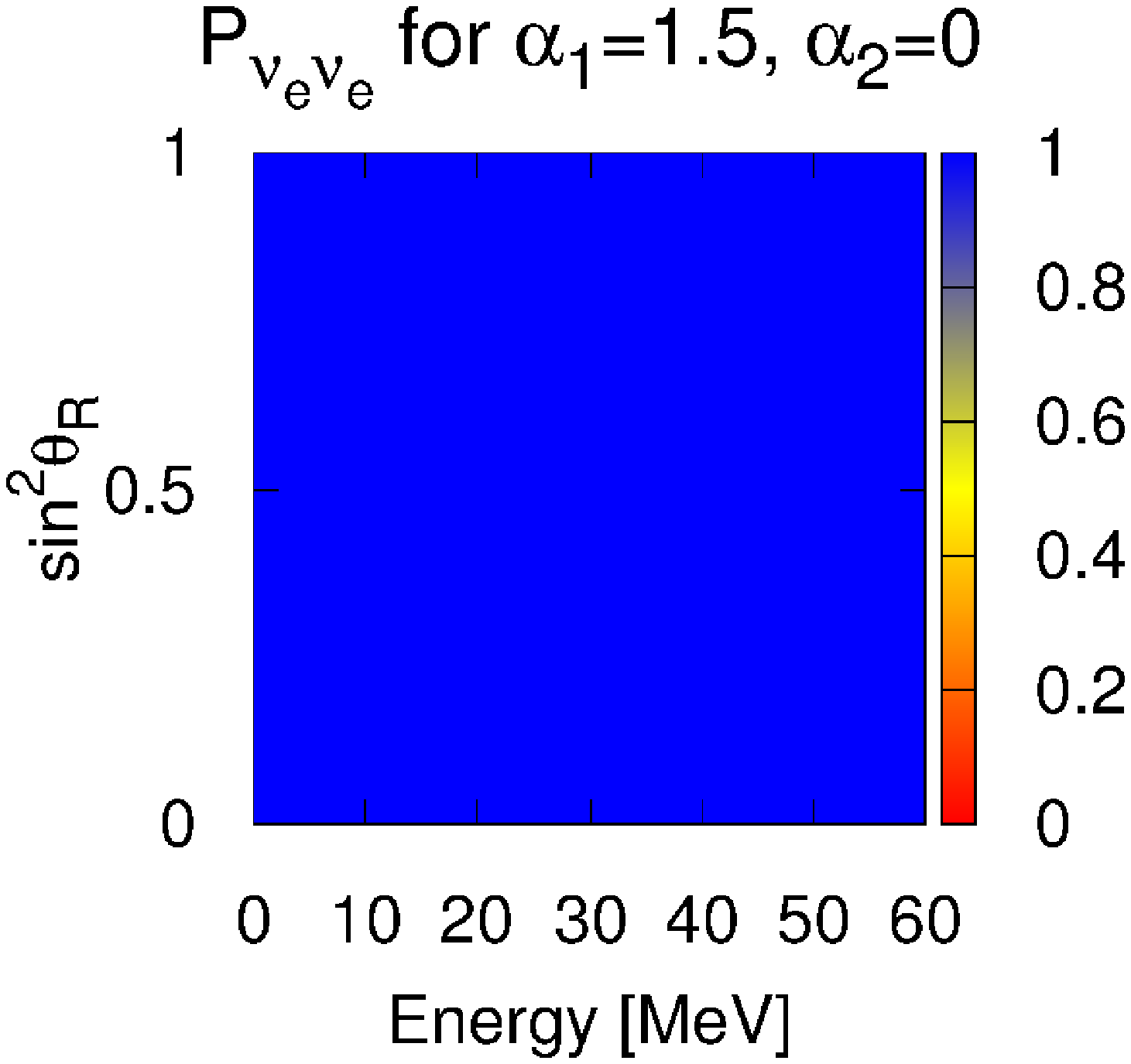}
}
\subfigure{
\includegraphics[width=.3\textwidth]{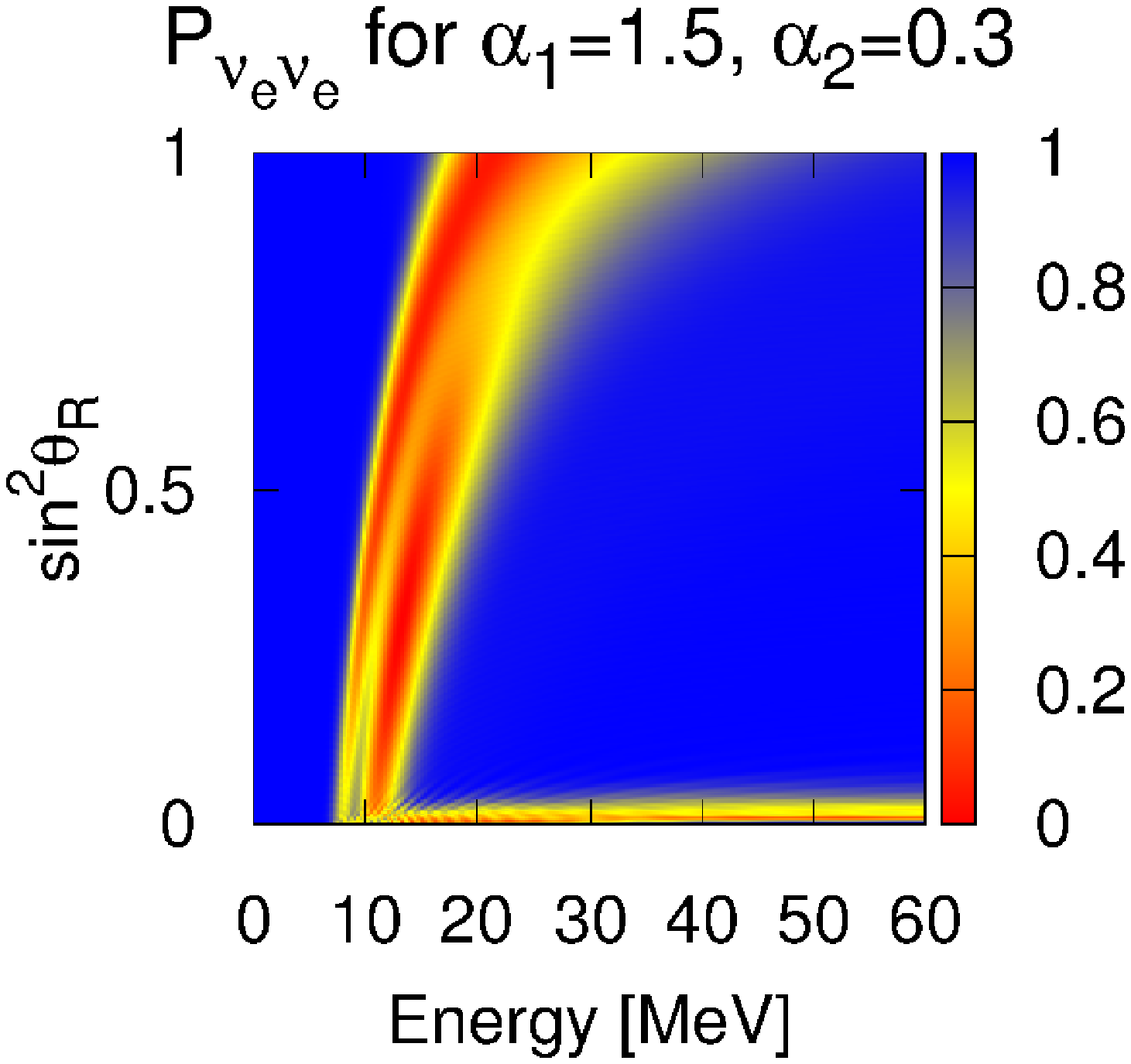}
}
\subfigure{
\includegraphics[width=.3\textwidth]{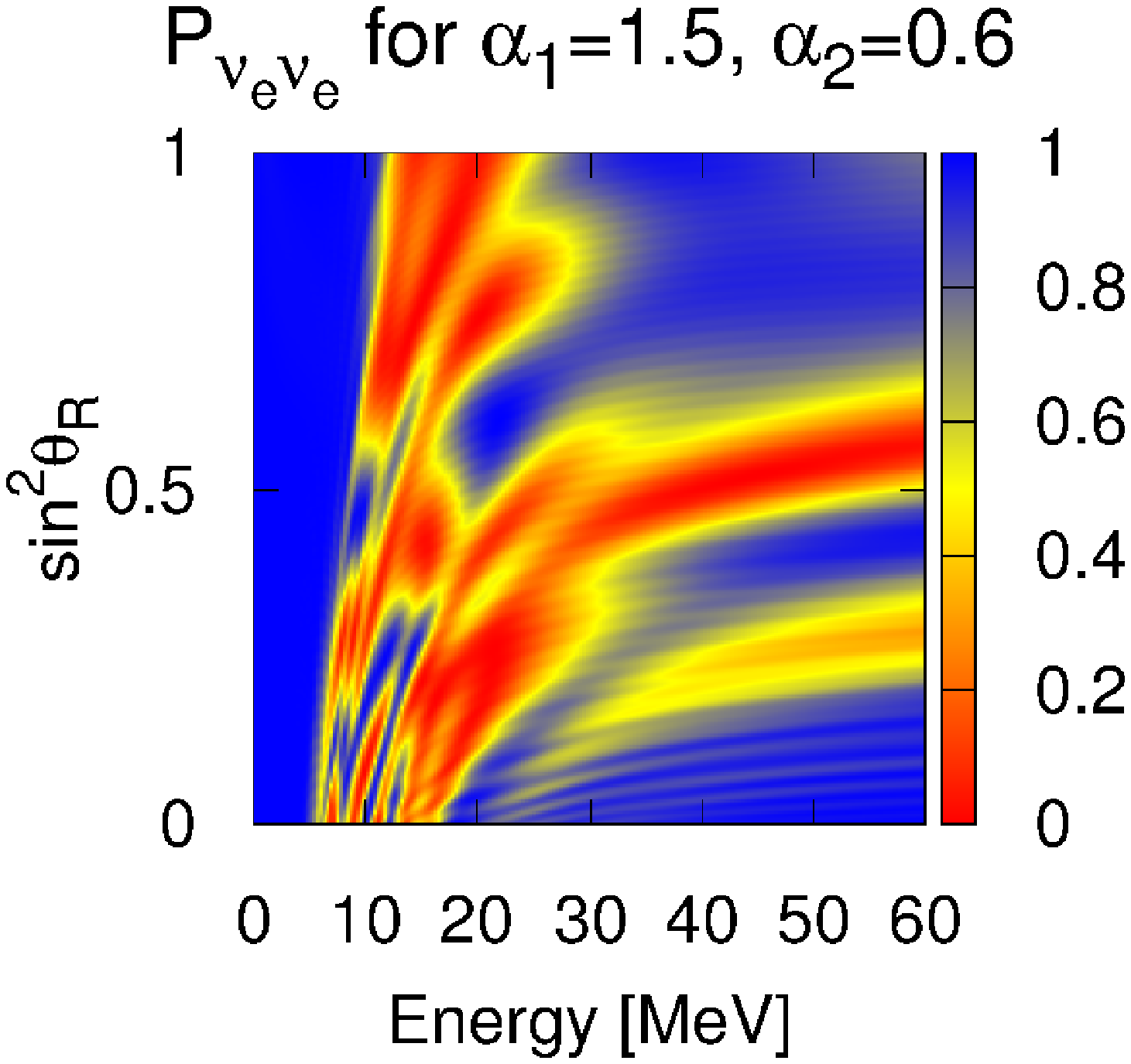}
}
\subfigure{
\includegraphics[width=.3\textwidth]{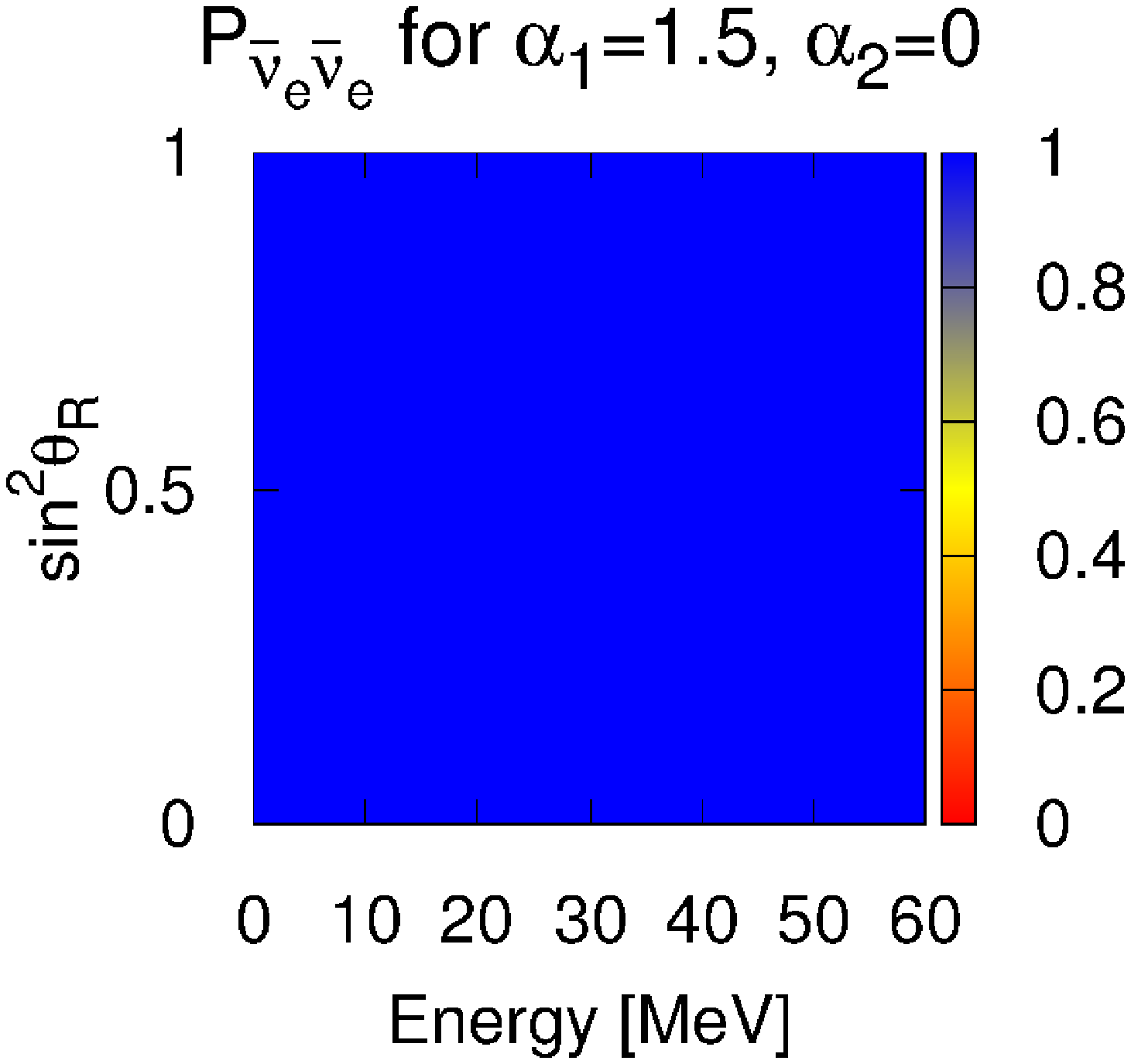}
}
\subfigure{
\includegraphics[width=.3\textwidth]{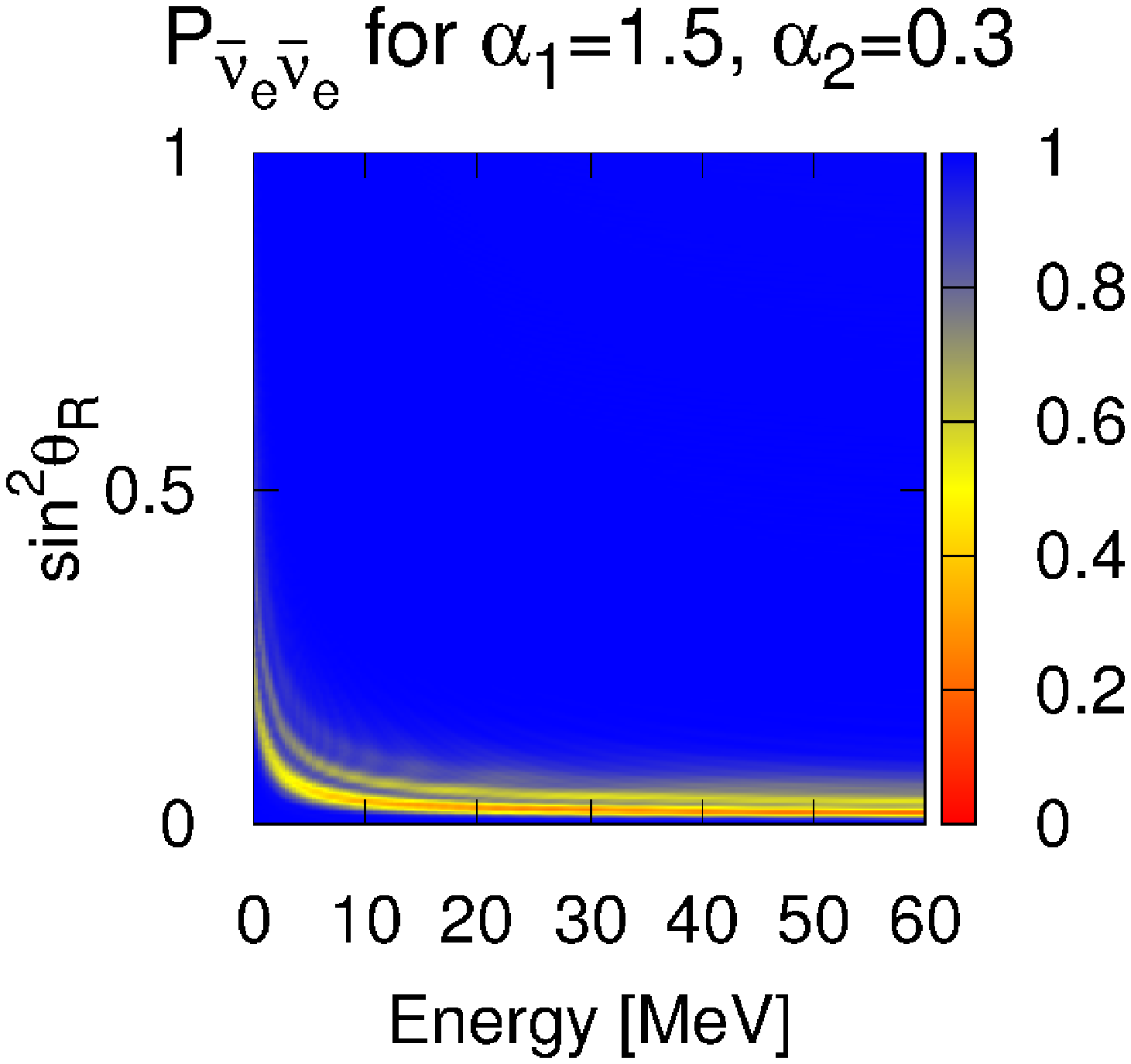}
}
\subfigure{
\includegraphics[width=.3\textwidth]{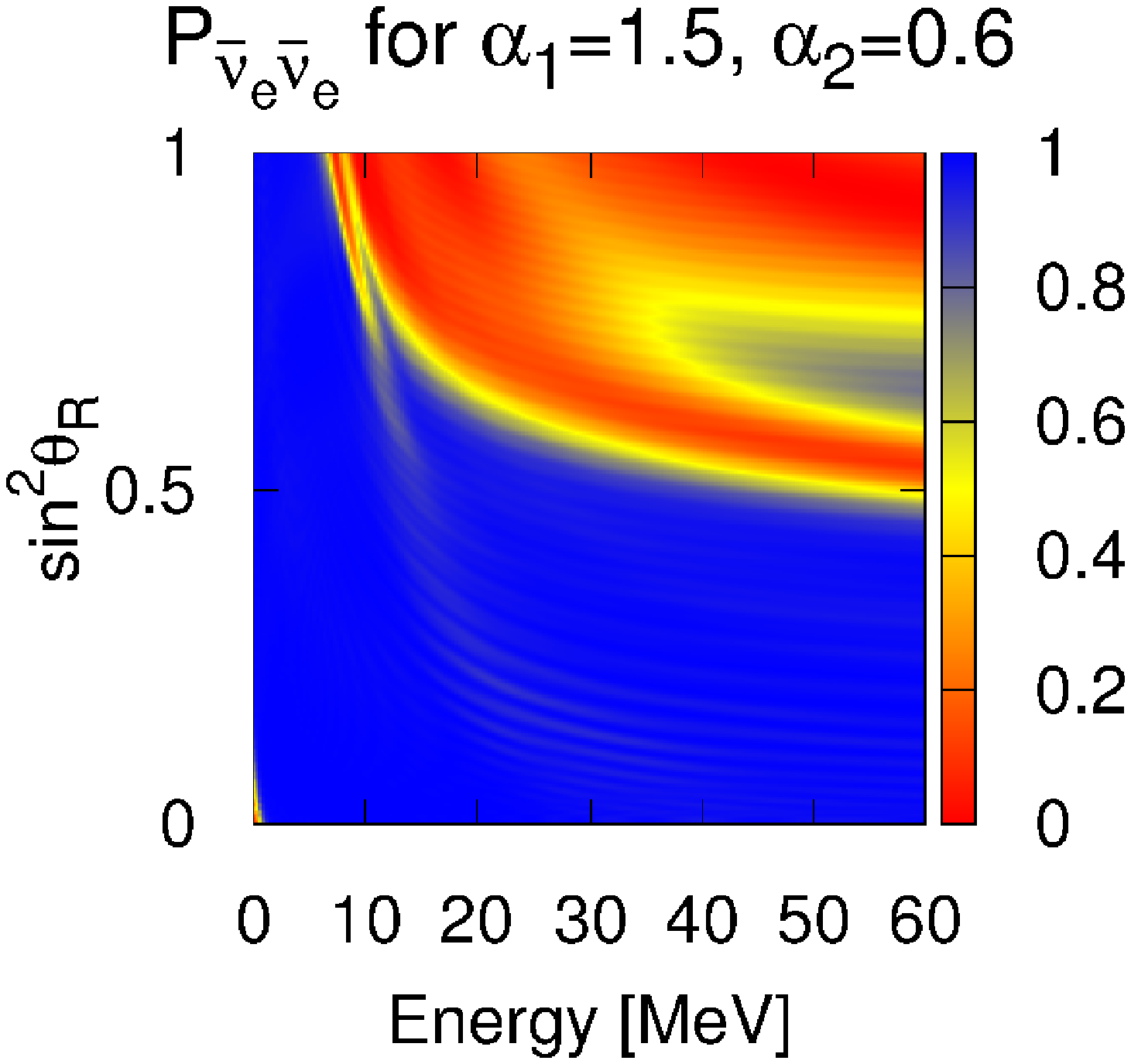}
}
\caption{Top panels: The heatmaps of survival probability of electron neutrinos at $t_{pb}=1.0s$ and $r=400\rm{km}$ as a function of energy and emission angle when there is flavor-violating NSSI. Bottom panels: The same but for electron antineutrinos.}
\label{fig:520FV_2D}
\end{figure}

Figure (\ref{fig:520FP_SP}) shows the numerical results of the survival probabilities of electron neutrino and antineutrino as a function of distance $r$ from the neutrinosphere, for $t_{pb}=1.0\; \rm{s}$ and different values of $\alpha_1$ when $\alpha_2=0$. In the left panels the probabilities are averaged over the energy and angular bins used in the calculation; in the right panels the survival probabilities are shown at $r=400\;{\rm km}$ as a function of neutrino energy averaged over the angular distribution only. We see that when there is no NSSI there is a noticeable amount of electron neutrinos transformation into muon and tau neutrinos, and that there are also flavor transformations in the electron antineutrino sector. This is in agreement with the results from Wu \emph{et al.} \cite{PhysRevD.91.065016}. When we add NSSI we can see the flavor transformation in the neutrino sector is delayed although the average survival probability at $r=400\;{\rm km}$ is essentially unchanged. The spectra of the electron neutrinos at $r=400\;{\rm km}$ also look similar for the three values of $\alpha_1$ shown though larger NSSI seems to suppress the transformation of the higher energy neutrinos. 

The flavor transformation in the antineutrino sector, however, is more affected by NSSI. As the NSSI is turned on, the transformation is immediately suppressed, with the final survival probability going back to $P_{\bar\nu_{e}\bar\nu_{e}}=1$. This suppression effect can be seen more clearly in the sequence of 2-D plots shown figure (\ref{fig:520FP_2D}), where we can see the region of flavor transformation keeps shrinking with an increasing NSSI in both neutrino and antineutrino sectors. 

\begin{table}[t]
\begin{tabular}{l*{3}{c}}
Flavor & \;Luminosity $L_{\nu,\infty}$  & \;Mean Energy $\langle E_{\nu,\infty}\rangle$ &\;rms Energy $\sqrt{ \langle E^2_{\nu,\infty}\rangle }$\\
\hline
$e$ & $2.504\times 10^{51}\;{\rm erg/s}$ & $9.891\;{\rm MeV}$ & $11.12\;{\rm MeV}$ \\
$\mu$,$\tau$ & $2.864\times 10^{51}\;{\rm erg/s}$ & $12.66\;{\rm MeV}$ & $14.99\;{\rm MeV}$ \\
$\bar{e}$ & $2.277\times 10^{51}\;{\rm erg/s}$ & $11.83\;{\rm MeV}$ & $13.65\;{\rm MeV}$ \\
$\bar{\mu}$, $\bar{\tau}$ & $2.875\times 10^{51}\;{\rm erg/s}$ & $12.70\;{\rm MeV}$ & $15.07\;{\rm MeV}$
\end{tabular}
\caption{The luminosities, mean energies, and rms energies used for the $t_{pb}=2.8{\;\rm s}$ calculation.} 
\label{tab:t=2.8s}  
\end{table}
\begin{figure}[t]
\centering
\subfigure{
\includegraphics[width=.45\textwidth]{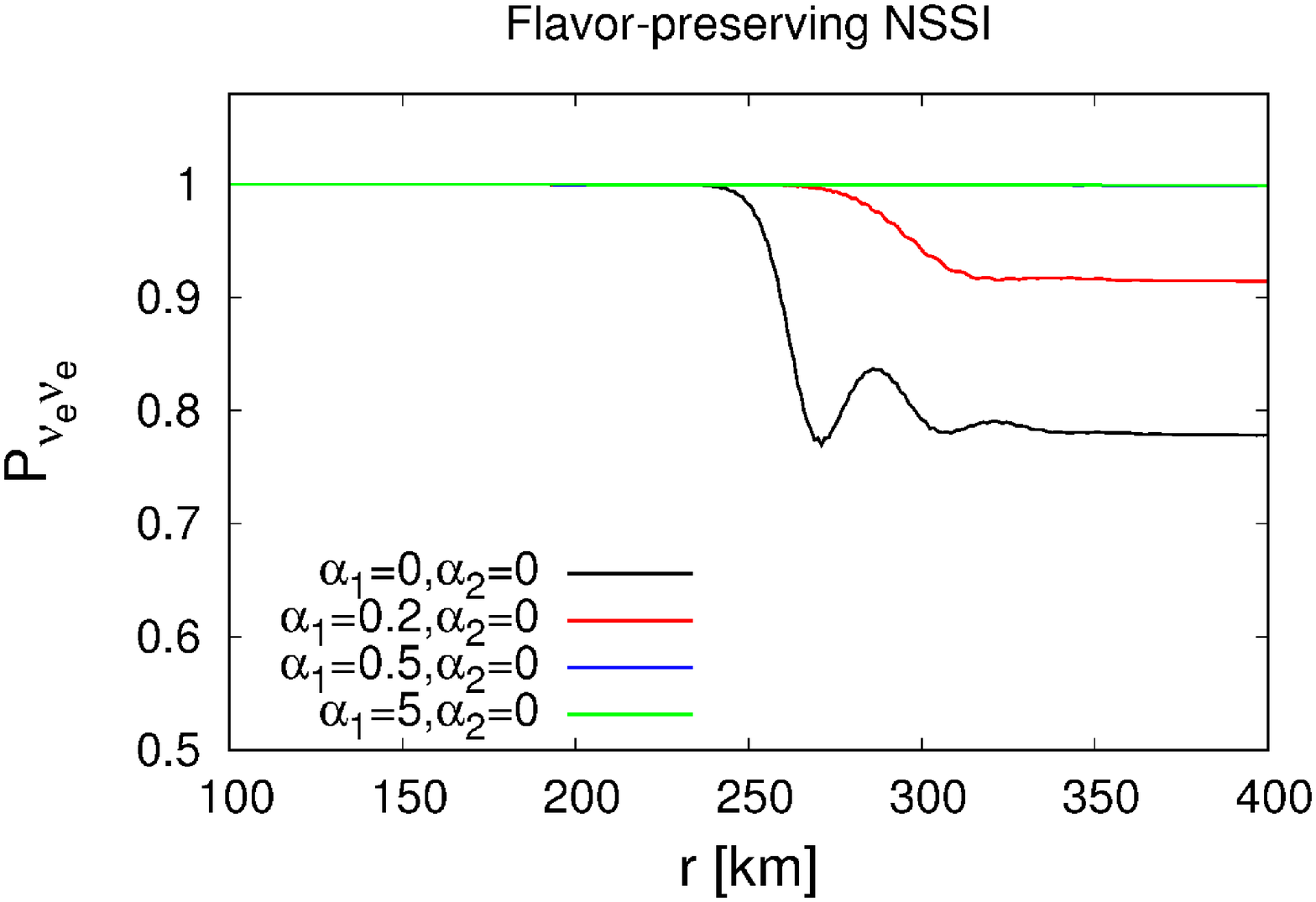}
}
\subfigure{
\includegraphics[width=.45\textwidth]{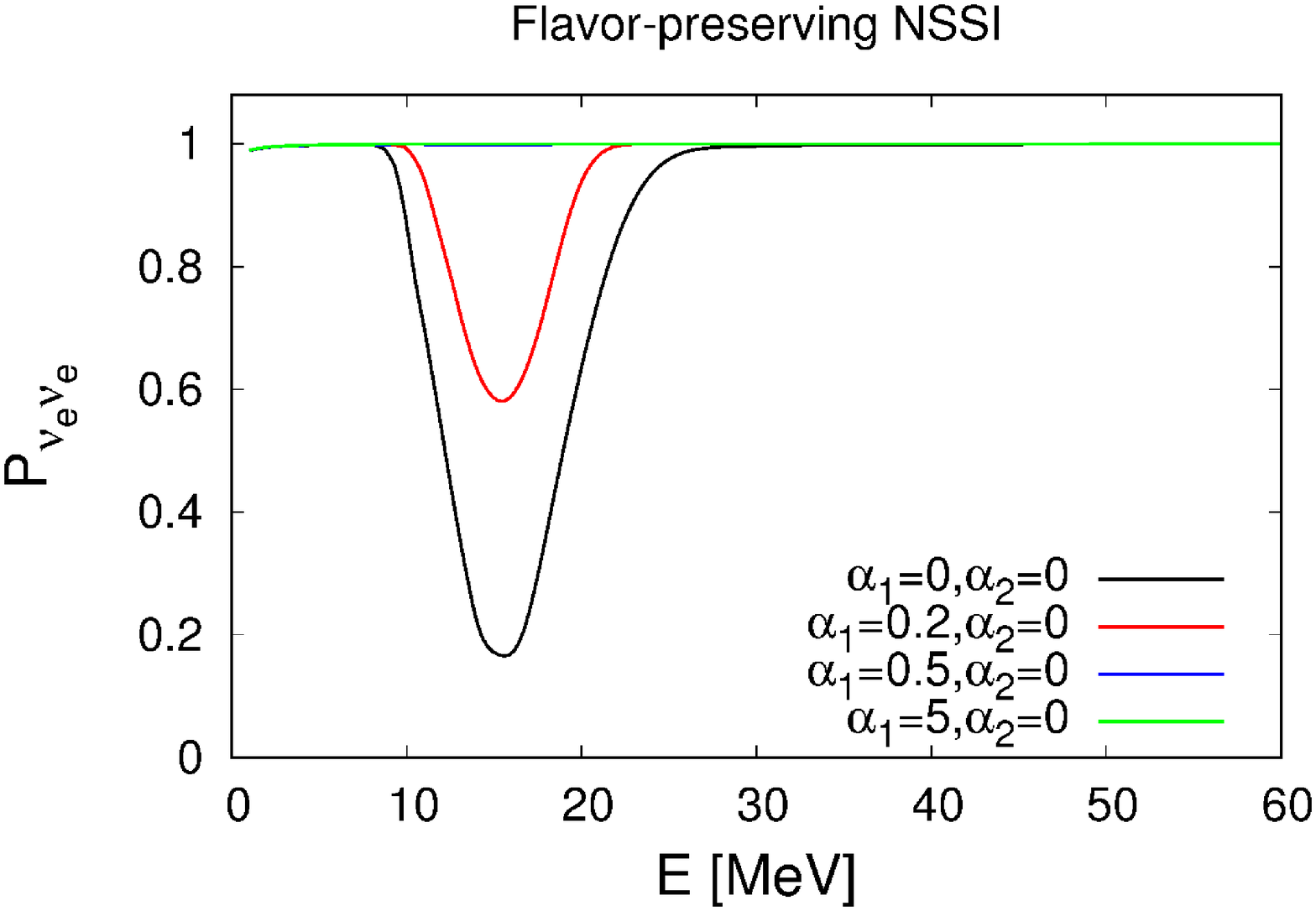}
}
\subfigure{
\includegraphics[width=.45\textwidth]{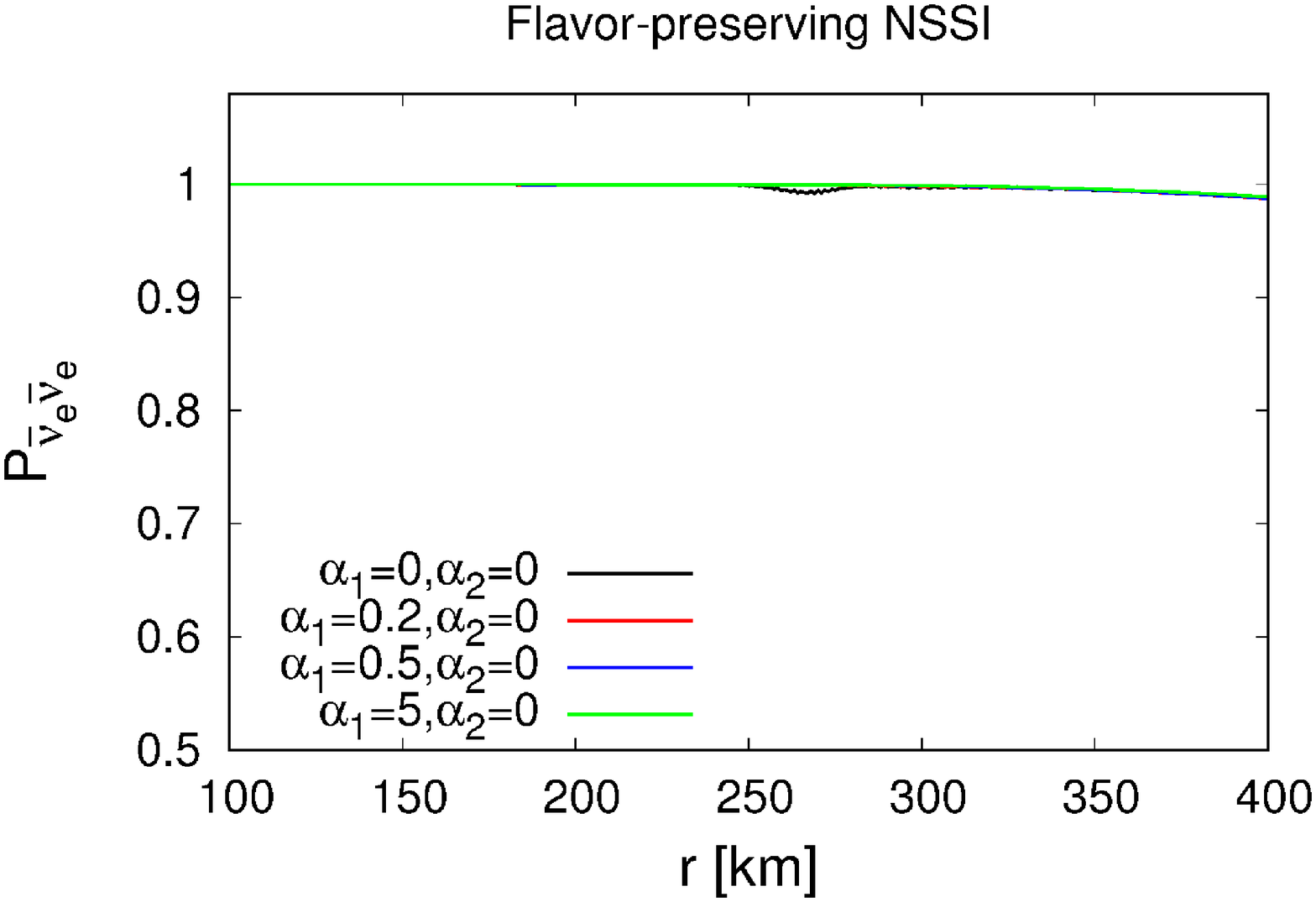}
}
\subfigure{
\includegraphics[width=.45\textwidth]{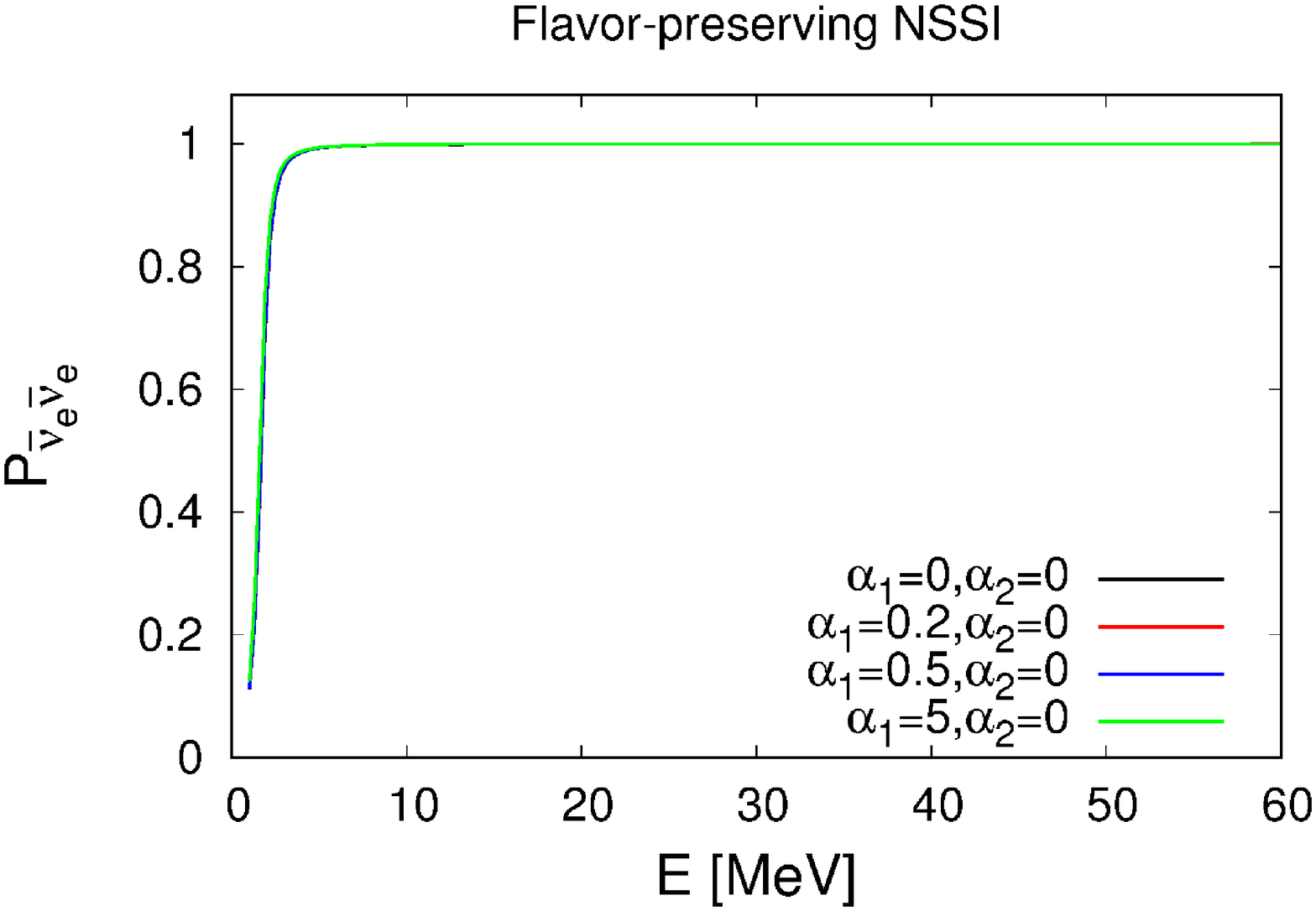}
}
\caption{Top panels: Survival probability of electron neutrinos at $t_{pb}=2.8s$ as a function of distance (left panel) and energy (right panel) at $r=400\;{\rm km}$ with flavor-preserving NSSI. The bottom panels are the same but for electron antineutrinos.}
\label{fig:585FP_SP}
\end{figure}

\begin{figure}[t]
\centering
\subfigure{
\includegraphics[width=.3\textwidth]{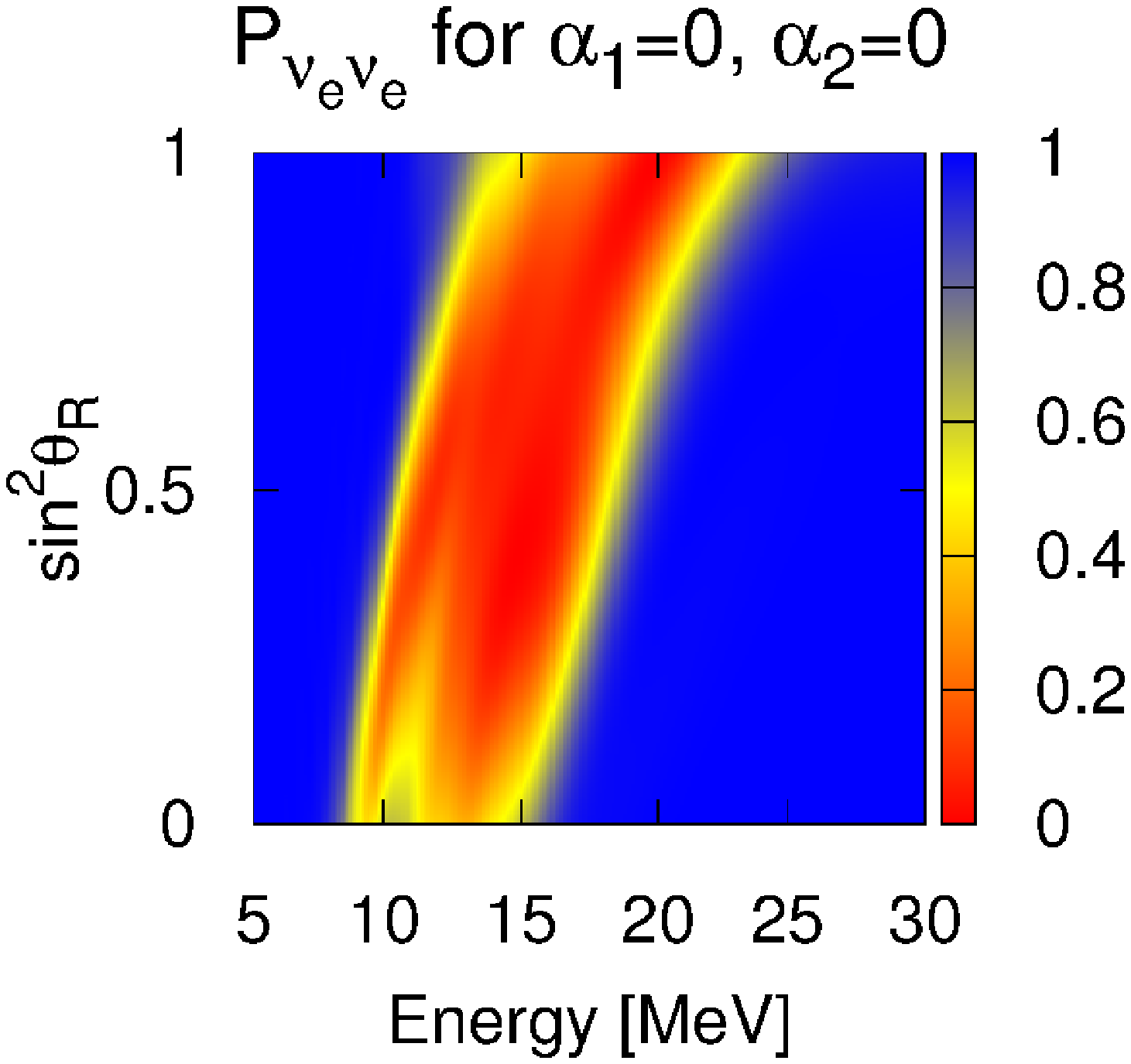}
}
\subfigure{
\includegraphics[width=.3\textwidth]{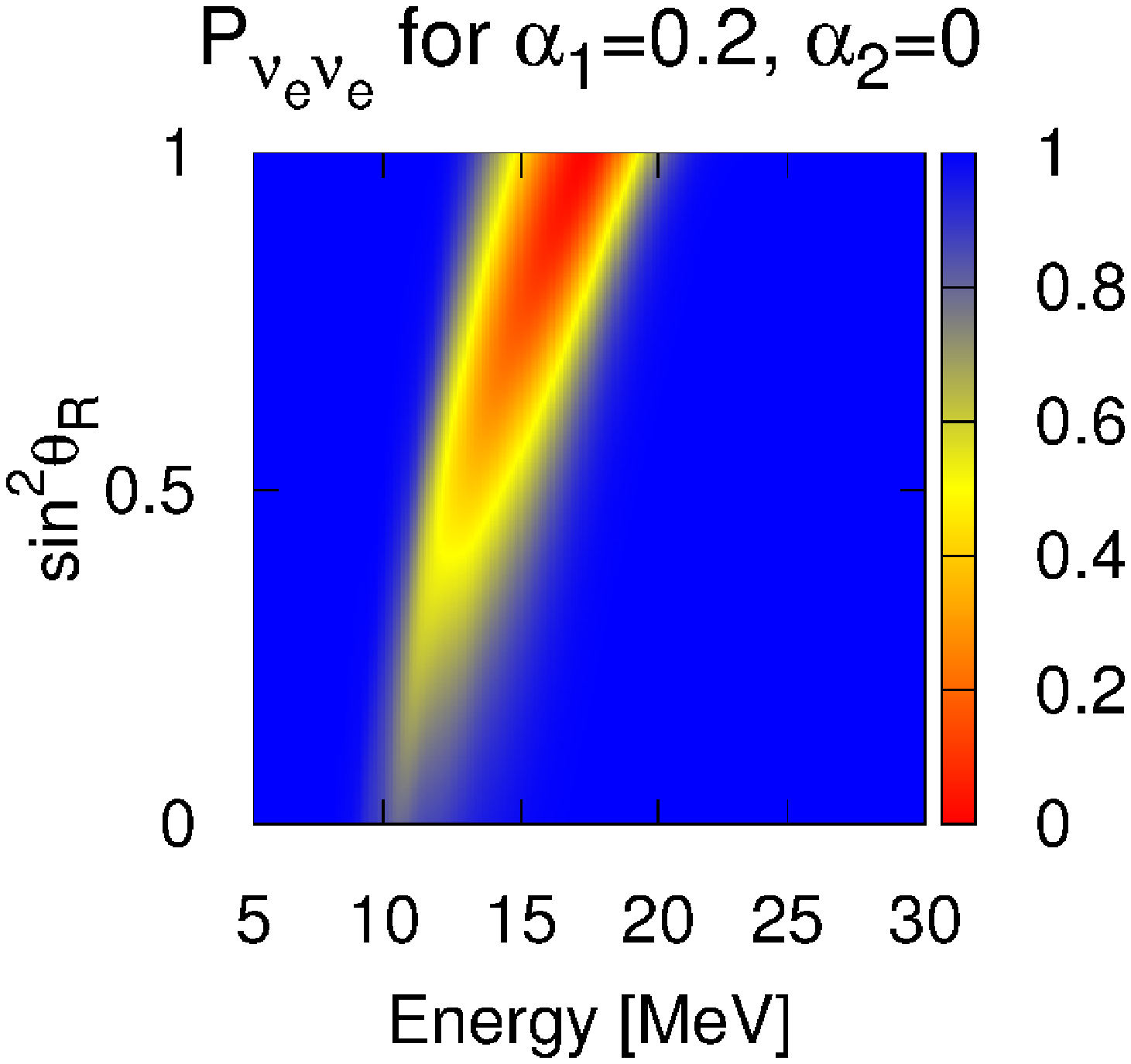}
}
\subfigure{
\includegraphics[width=.3\textwidth]{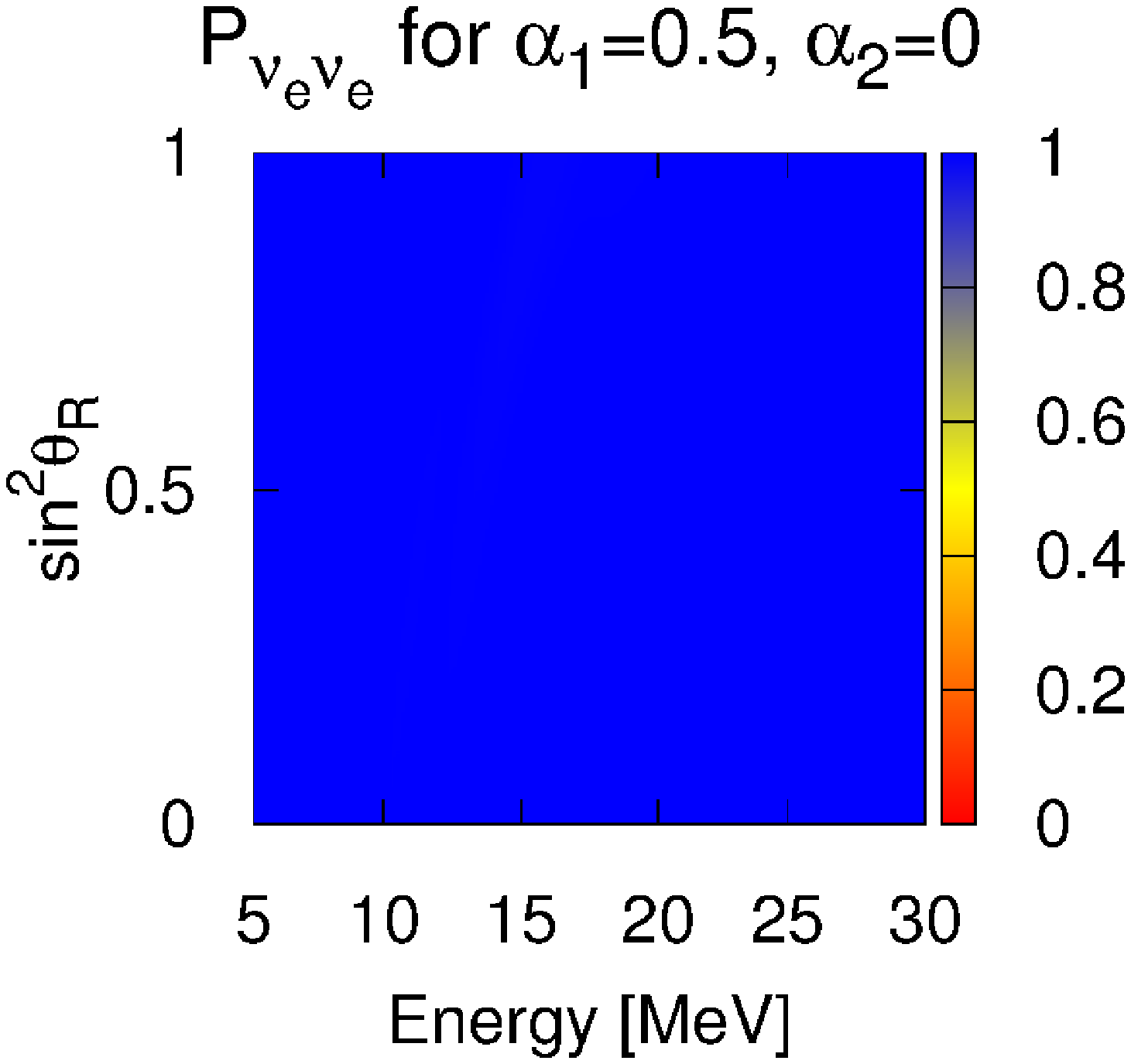}
}
\subfigure{
\includegraphics[width=.3\textwidth]{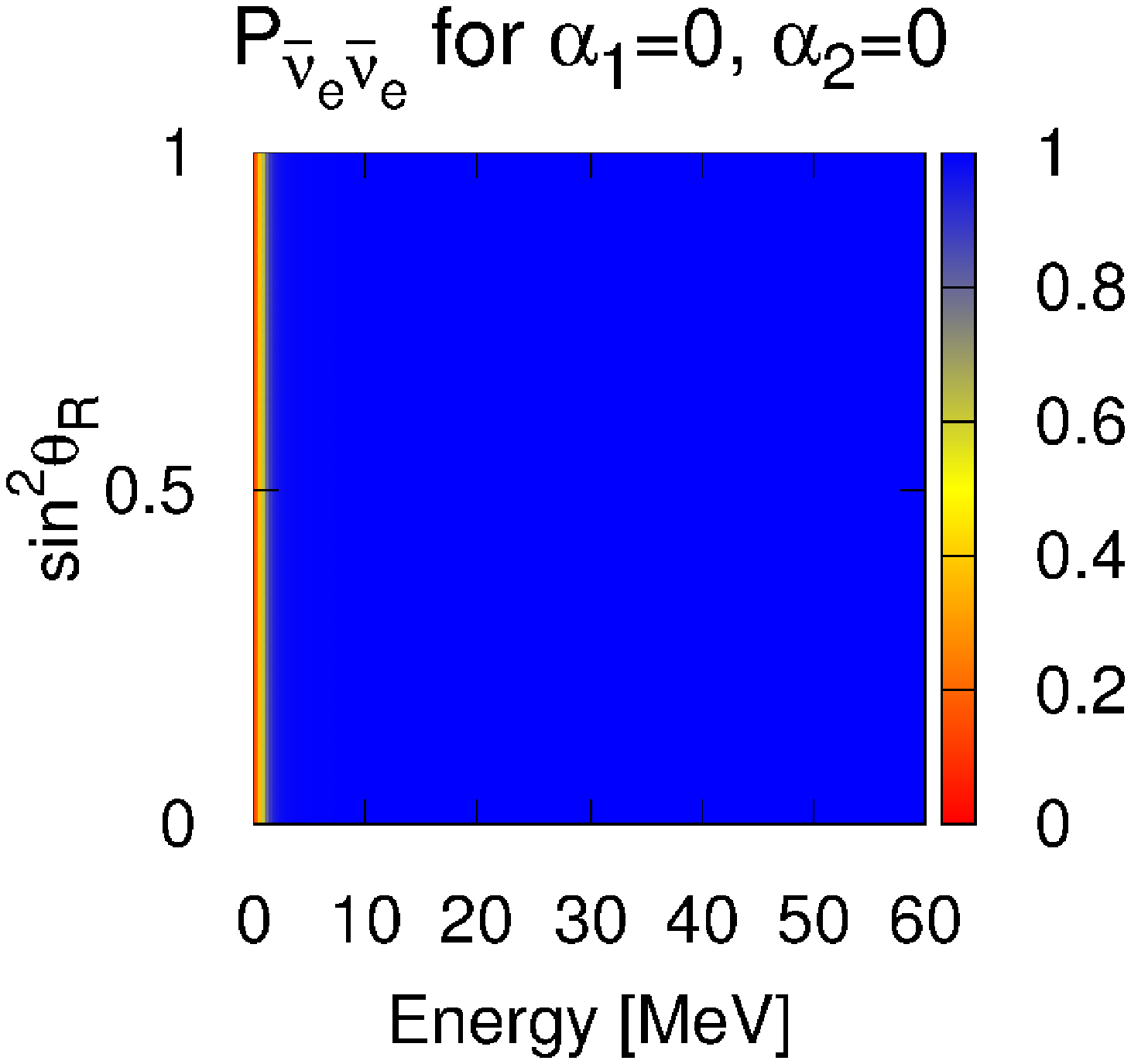}
}
\subfigure{
\includegraphics[width=.3\textwidth]{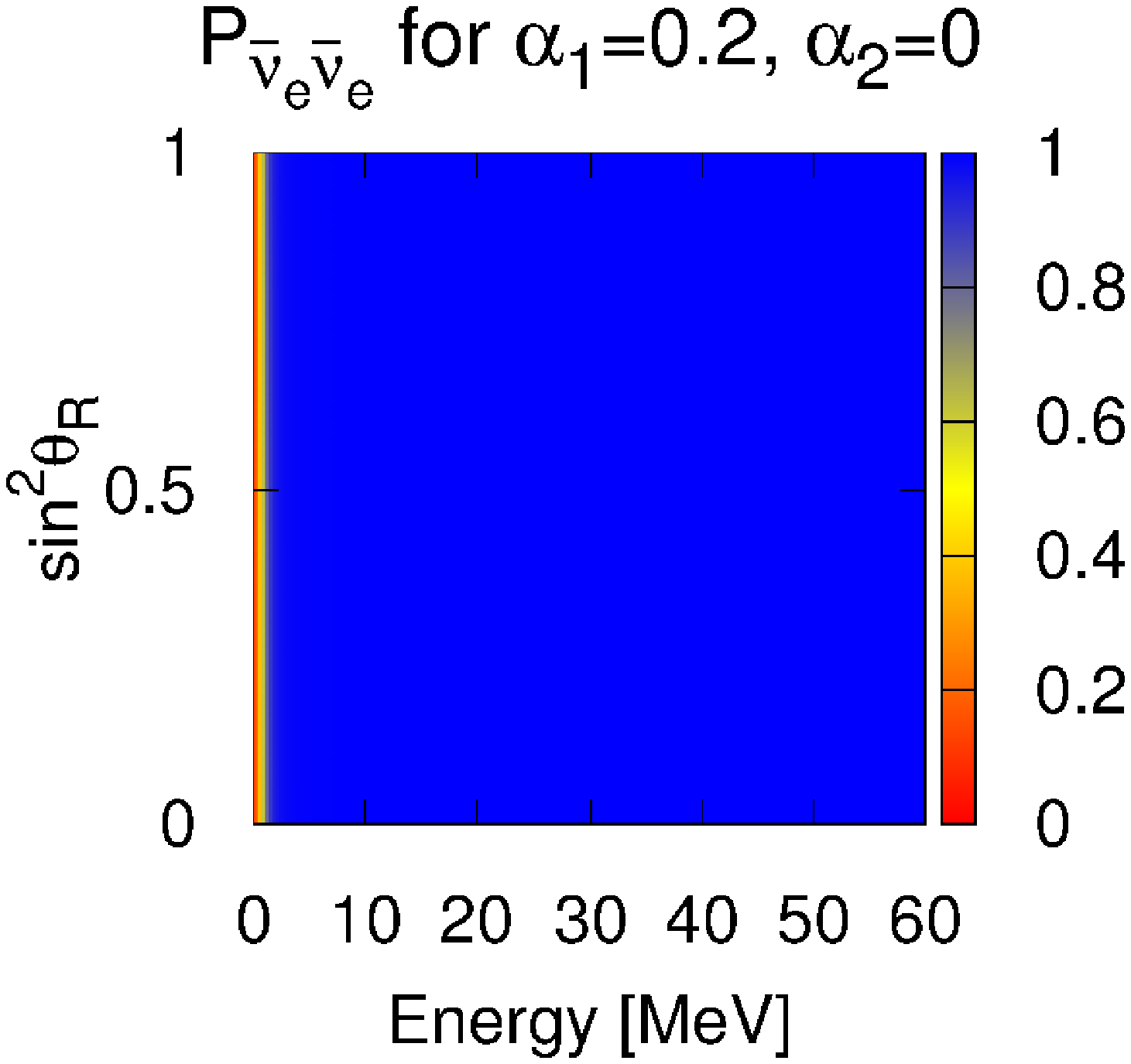}
}
\subfigure{
\includegraphics[width=.3\textwidth]{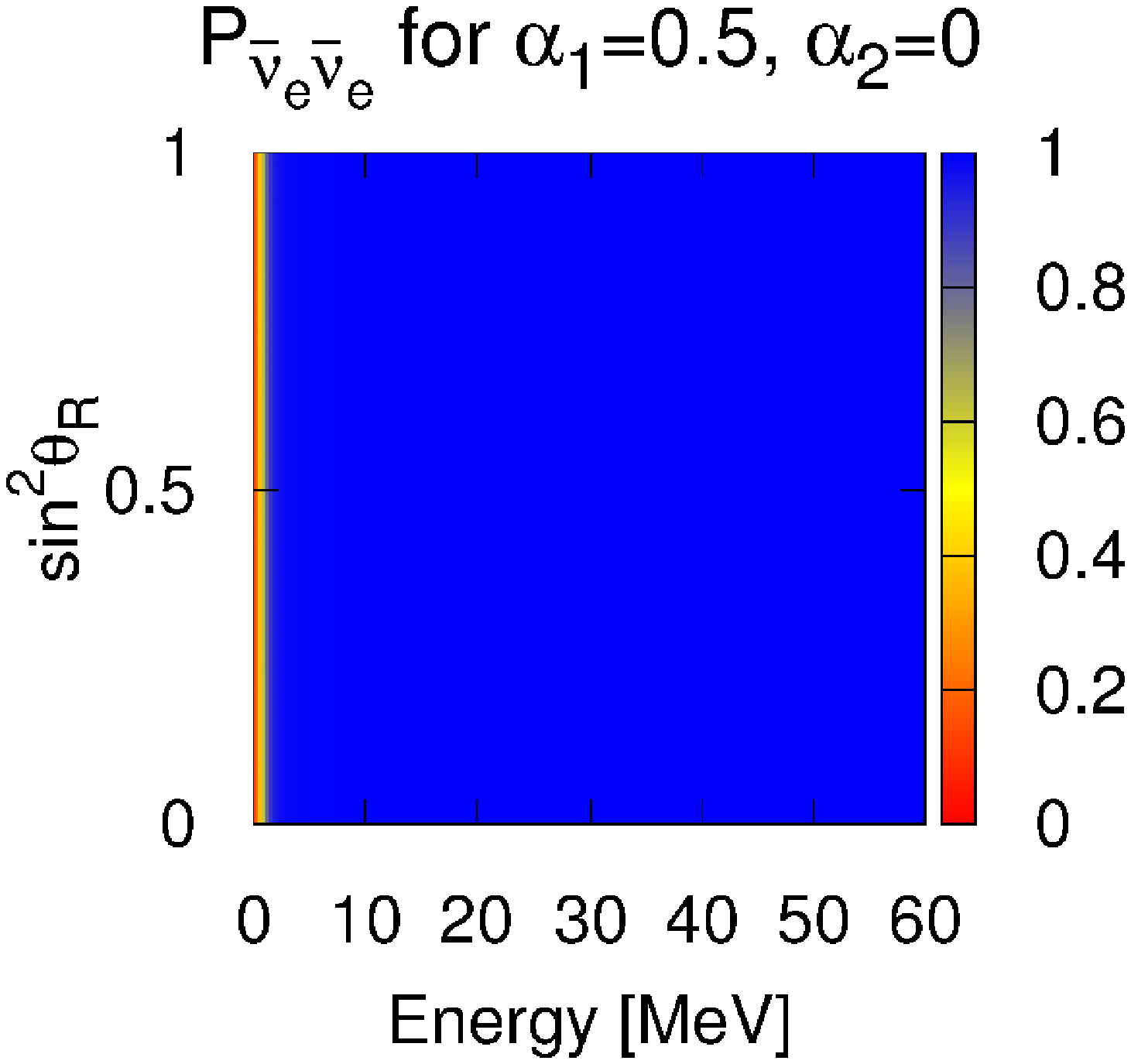}
}
\caption{Top panels: The heatmaps of survival probability of electron neutrinos at $t_{pb}=2.8s$ and $r=400\rm{km}$ as a function of energy and emission angle when there is only flavor-preserving NSSI. Bottom panels: The same but for electron antineutrinos.}
\label{fig:585FP_2D}
\end{figure}

The effect of the NSSI becomes even more interesting when the flavor-violating NSSI parameter $\alpha_2$ is non-zero. Figure \ref{fig:520FV_SP} shows that the flavor-violating NSSI have the effect of undoing the suppression of the flavor-preserving NSSI. As we can see from the blue curve in the figure, the flavor transformation is restored to the original level (i.e. no NSSI) for the combination $\alpha_1=1.5$, $\alpha_2 = 0.6$. At smaller $\alpha_2$, the transformation is only partially restored across the spectrum, as shown by the red curve in the figure. The sequence of 2-D plots shown in figure \ref{fig:520FV_2D} also show the pattern of transformed regions can be largely restored when flavor-violating NSSI is significant.


\subsection{Flavor transformation at $t_{pb}=2.8\;\rm{s}$}

\begin{figure}[b]
\centering
\subfigure{
\includegraphics[width=.45\textwidth]{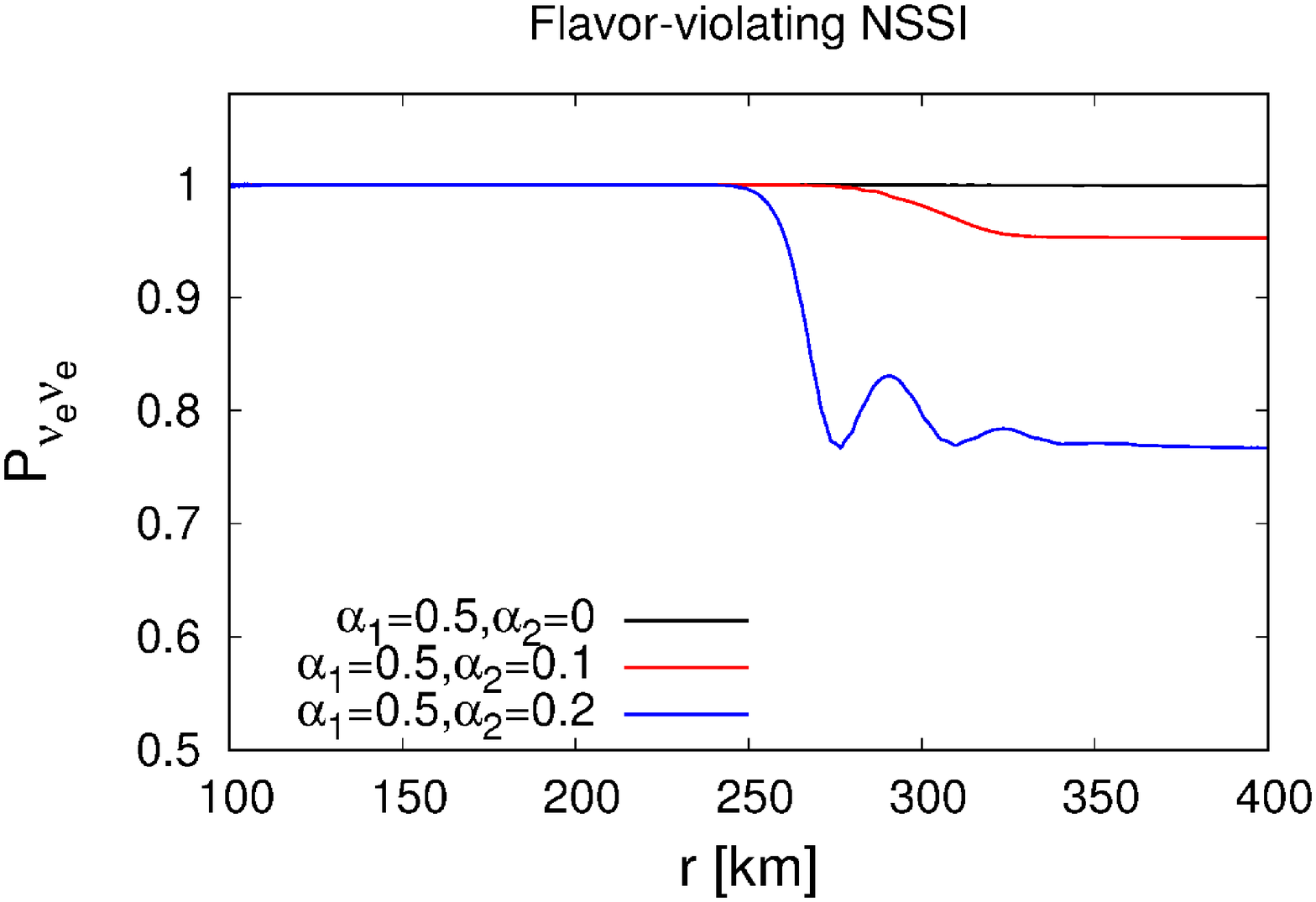}
}
\subfigure{
\includegraphics[width=.45\textwidth]{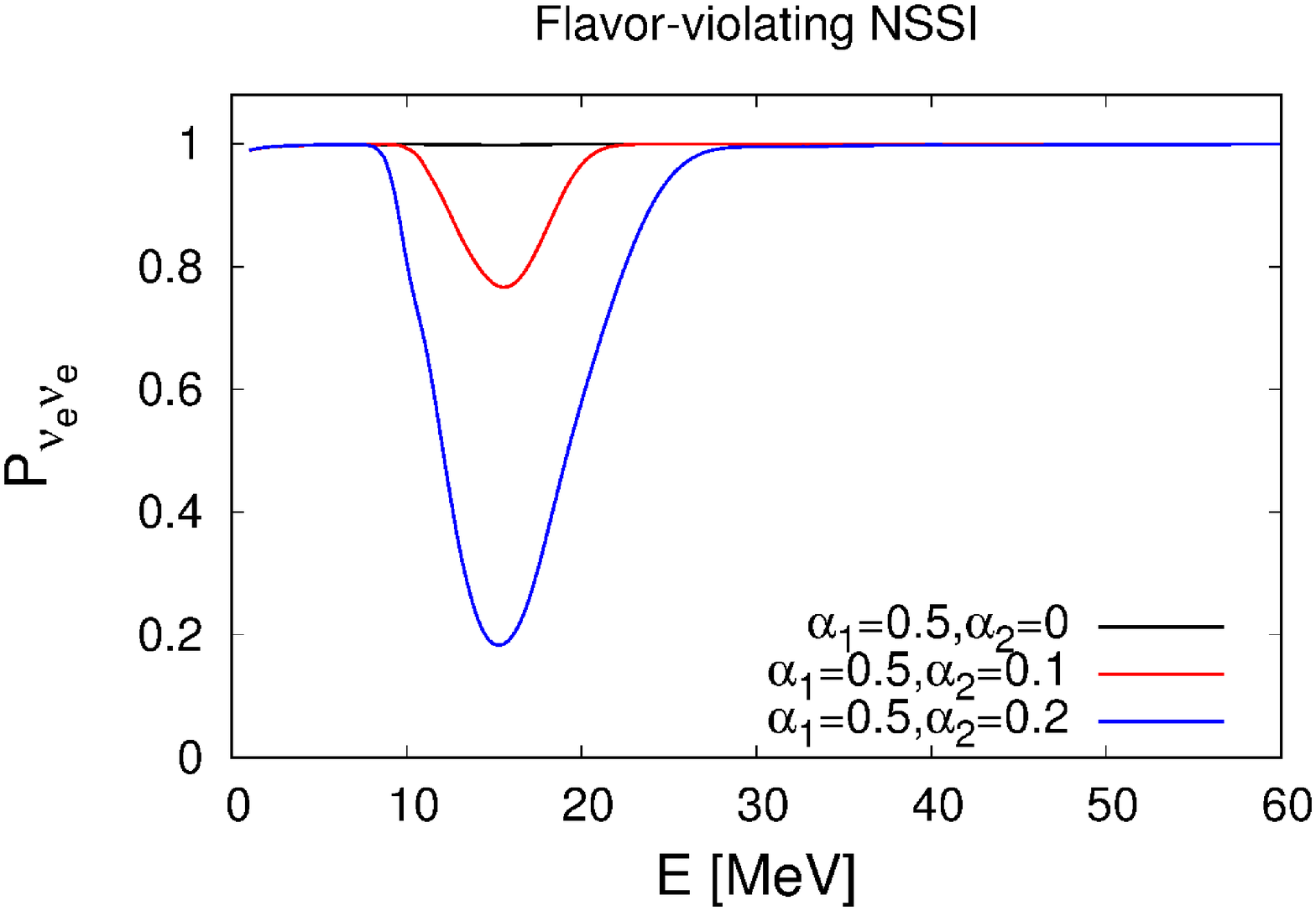}
}
\subfigure{
\includegraphics[width=.45\textwidth]{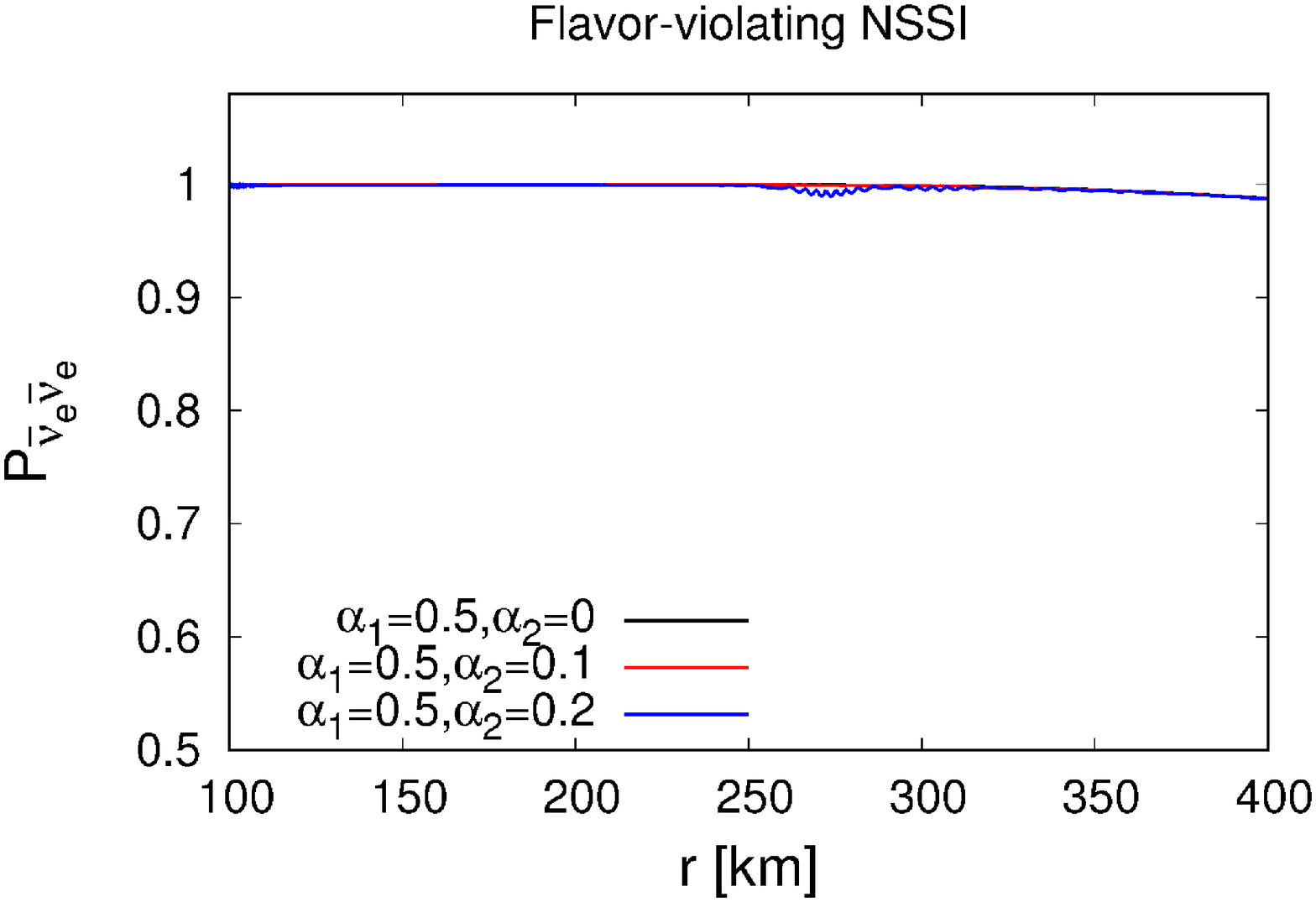}
}
\subfigure{
\includegraphics[width=.45\textwidth]{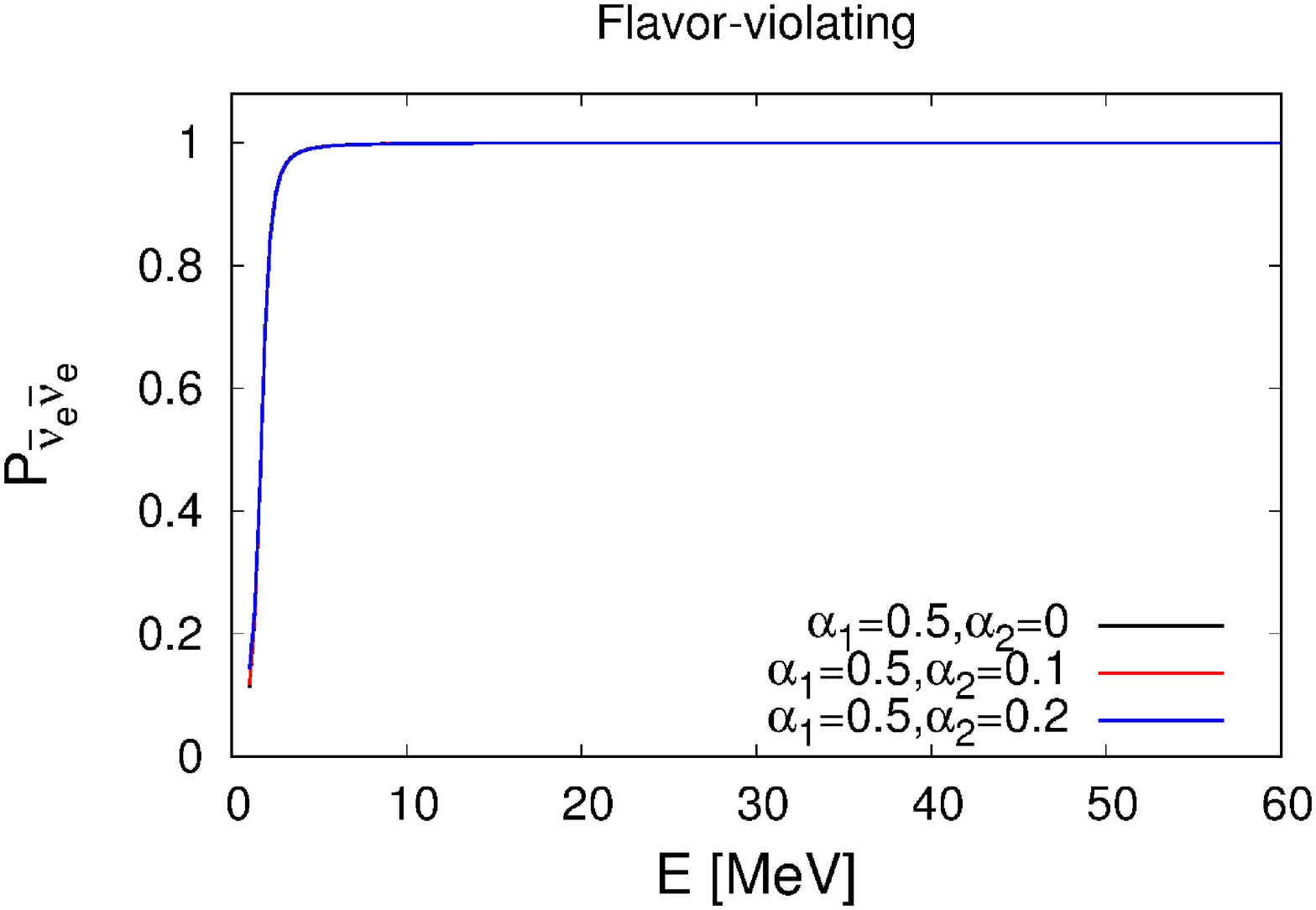}
}
\caption{Top panels: Survival probability of electron neutrinos at $t_{pb}=2.8s$ as a function of distance (left panel) and energy (right panel) at $r=400\;{\rm km}$ with flavor-violating NSSI. Bottom panels: The same but for electron antineutrinos.}
\label{fig:585FV_SP}
\end{figure}
\begin{figure}[t]
\centering
\subfigure{
\includegraphics[width=.3\textwidth]{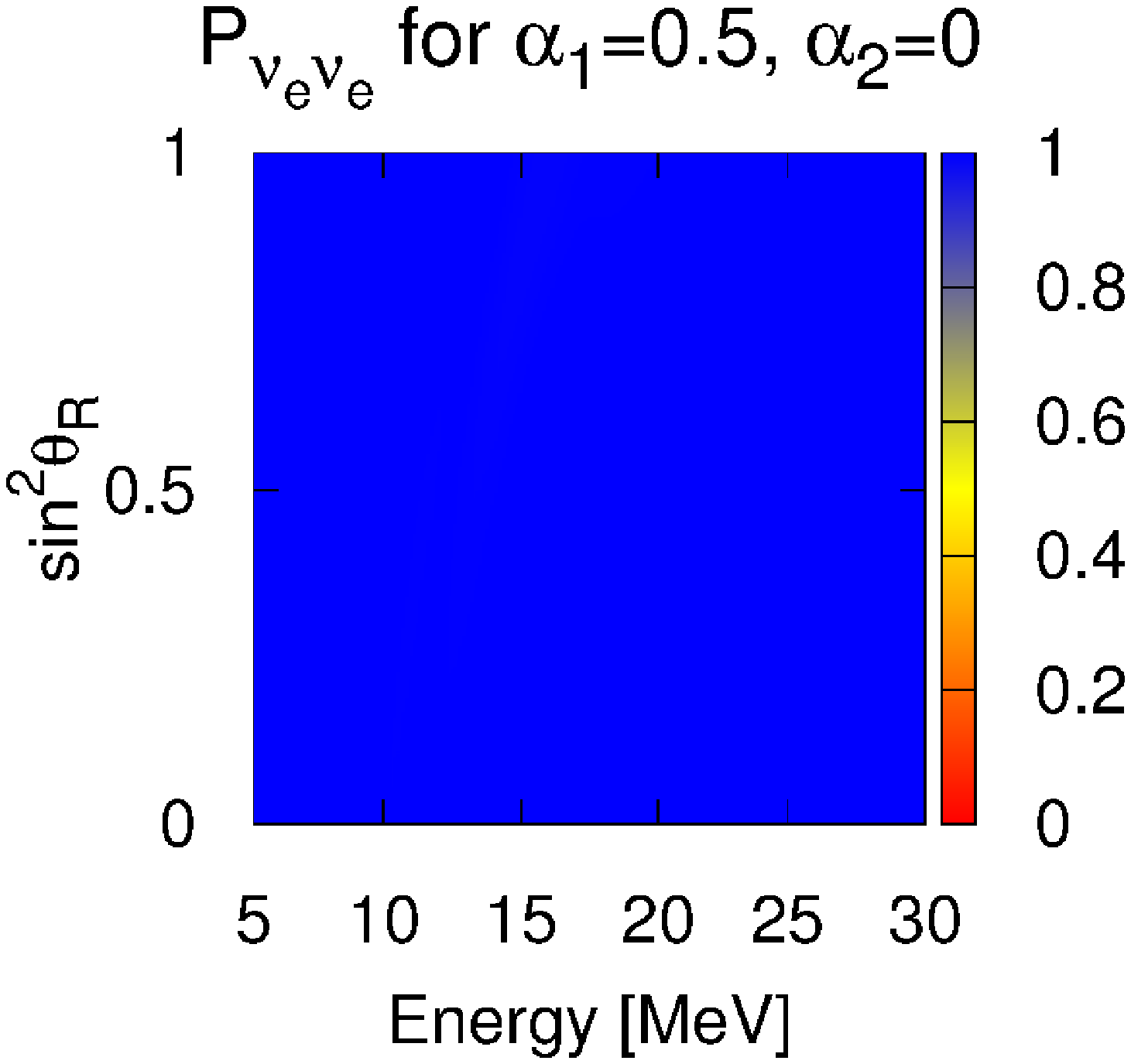}
}
\subfigure{
\includegraphics[width=.3\textwidth]{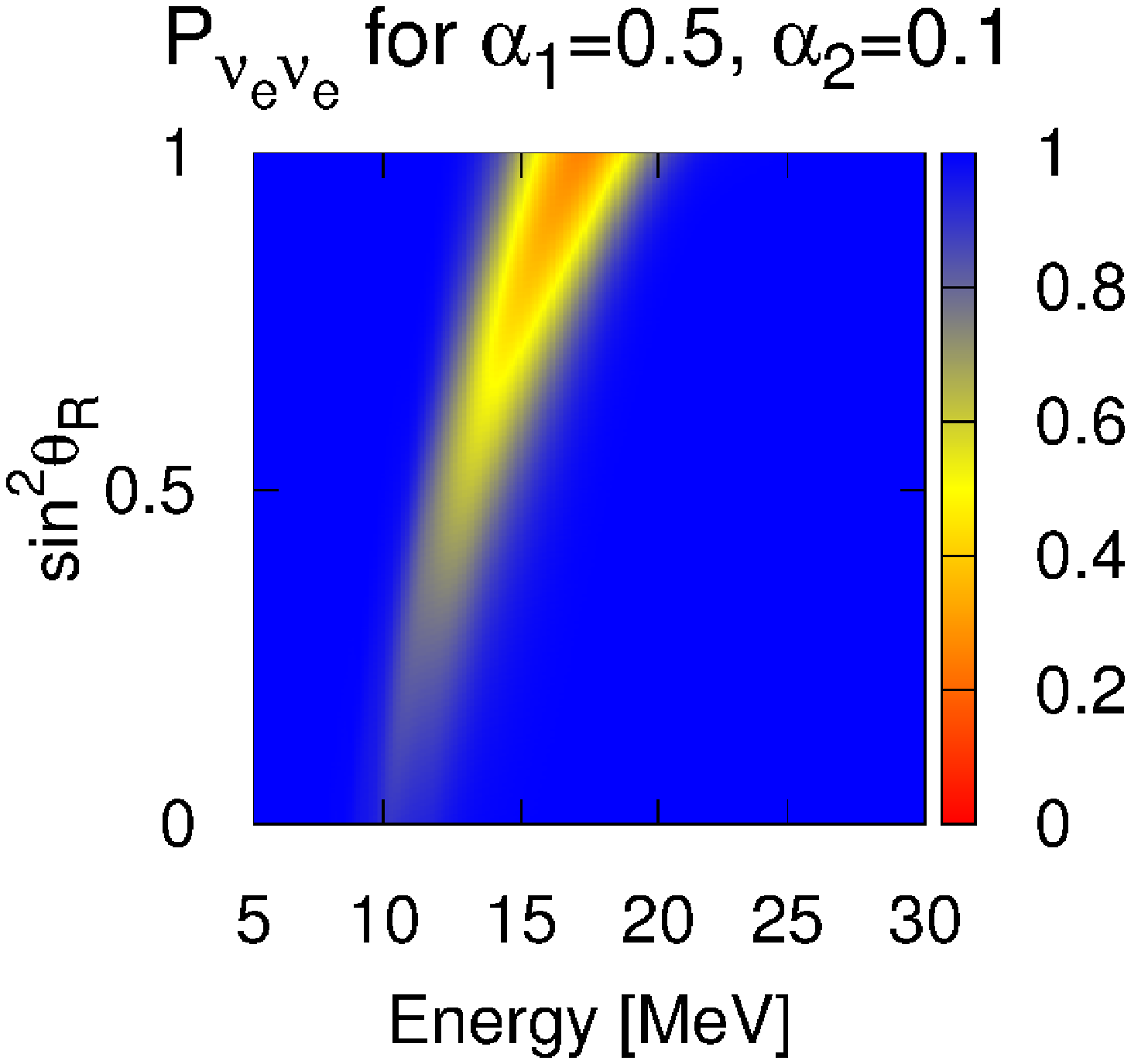}
}
\subfigure{
\includegraphics[width=.3\textwidth]{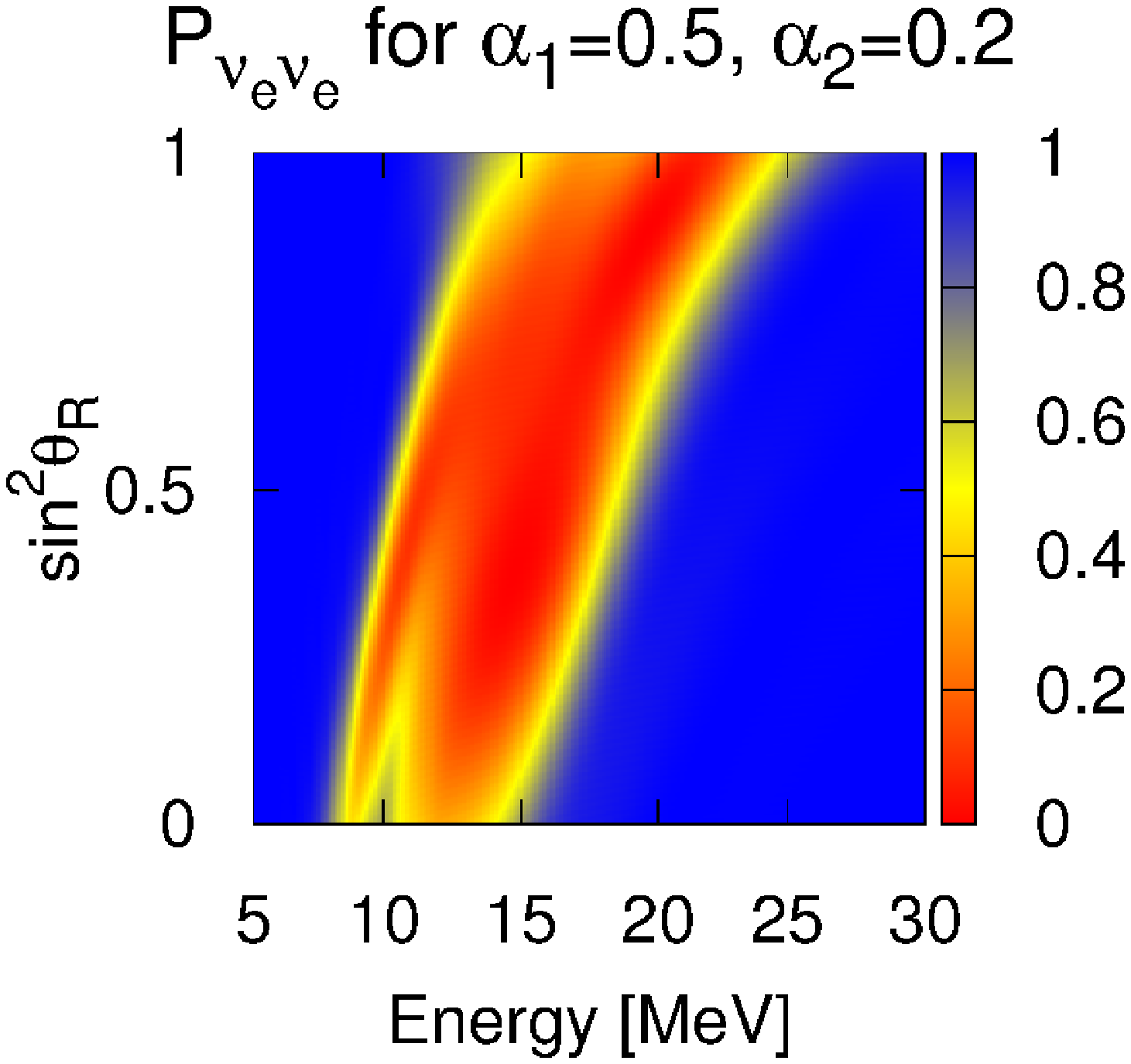}
}
\subfigure{
\includegraphics[width=.3\textwidth]{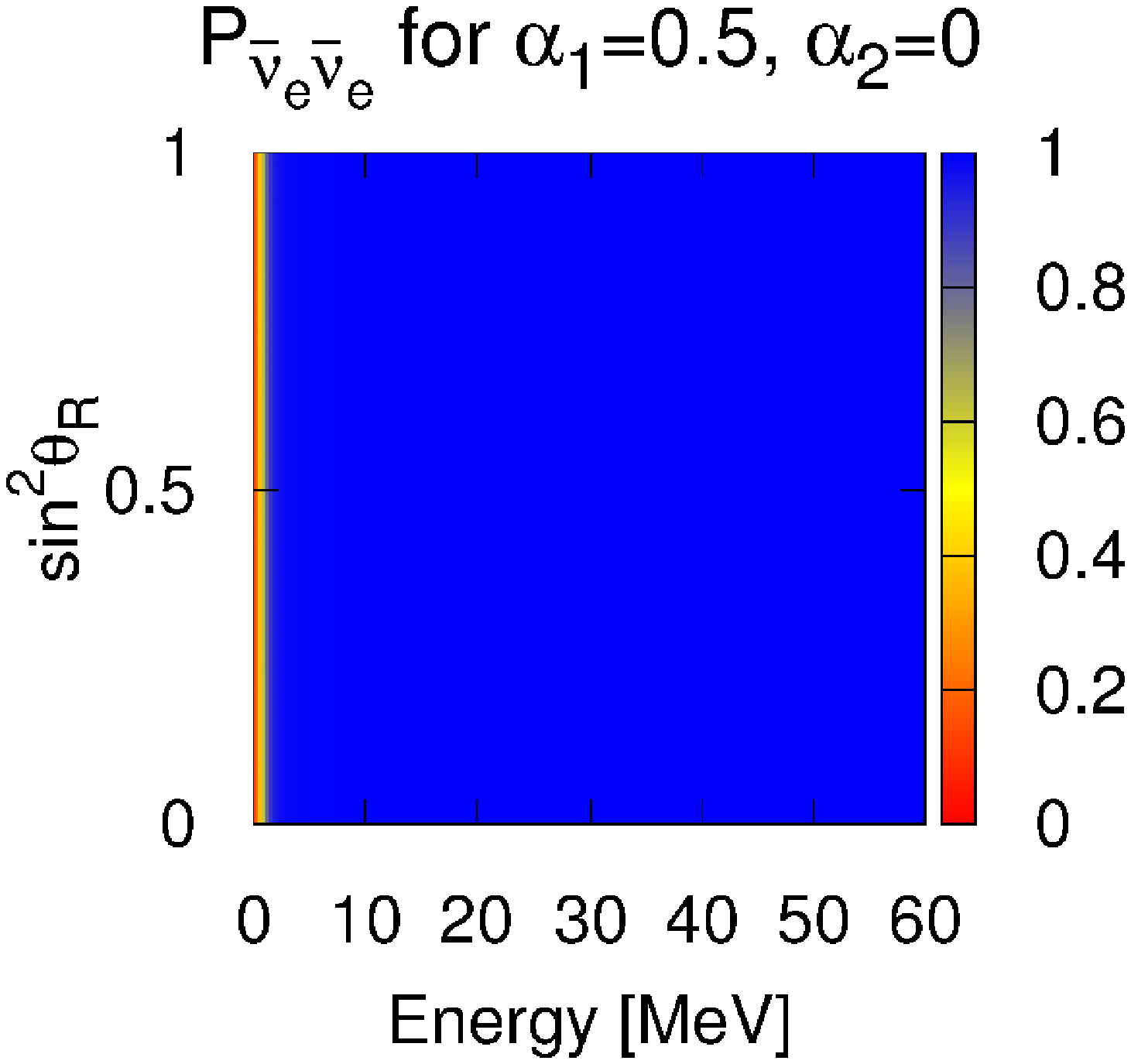}
}
\subfigure{
\includegraphics[width=.3\textwidth]{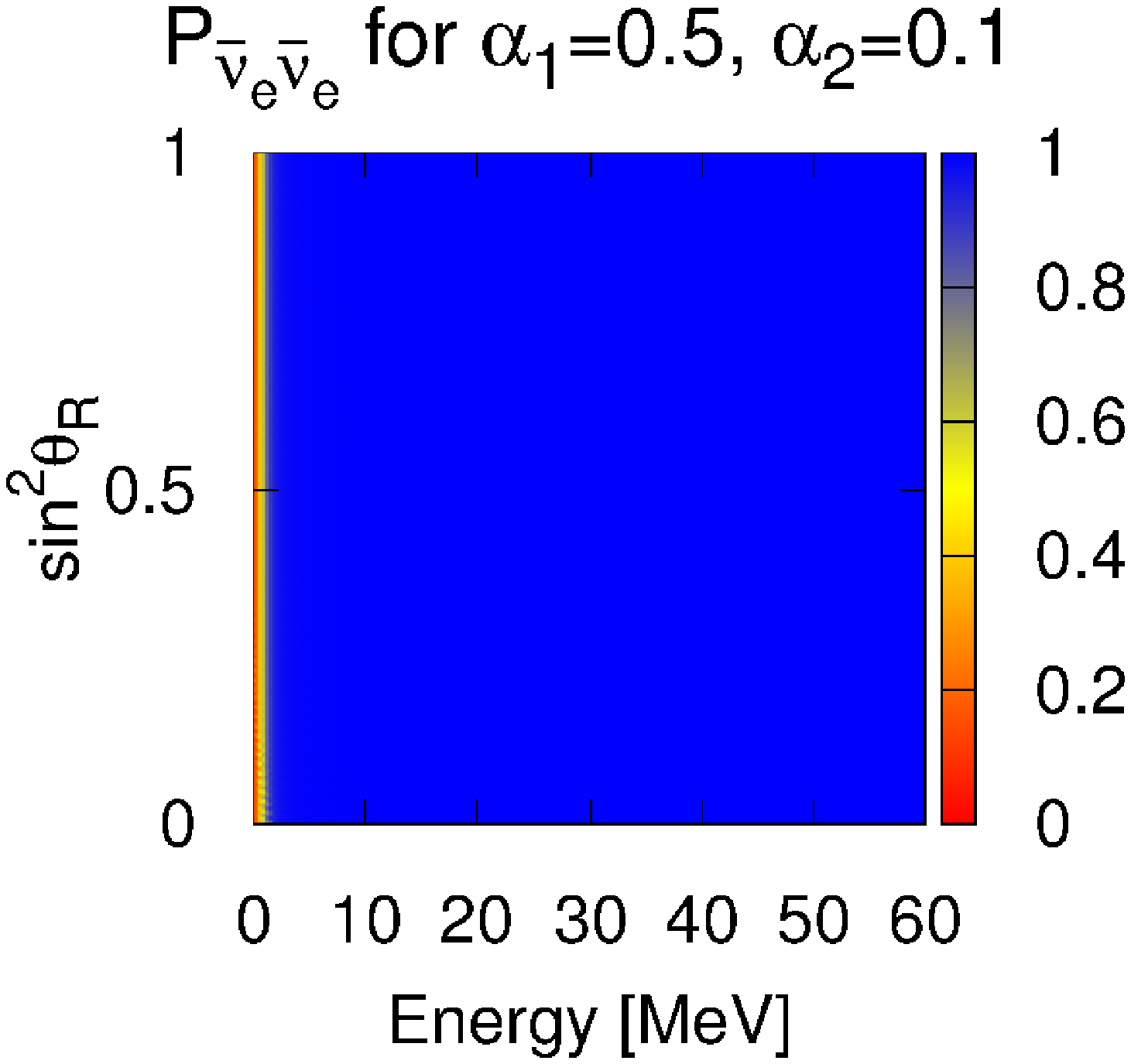}
}
\subfigure{
\includegraphics[width=.3\textwidth]{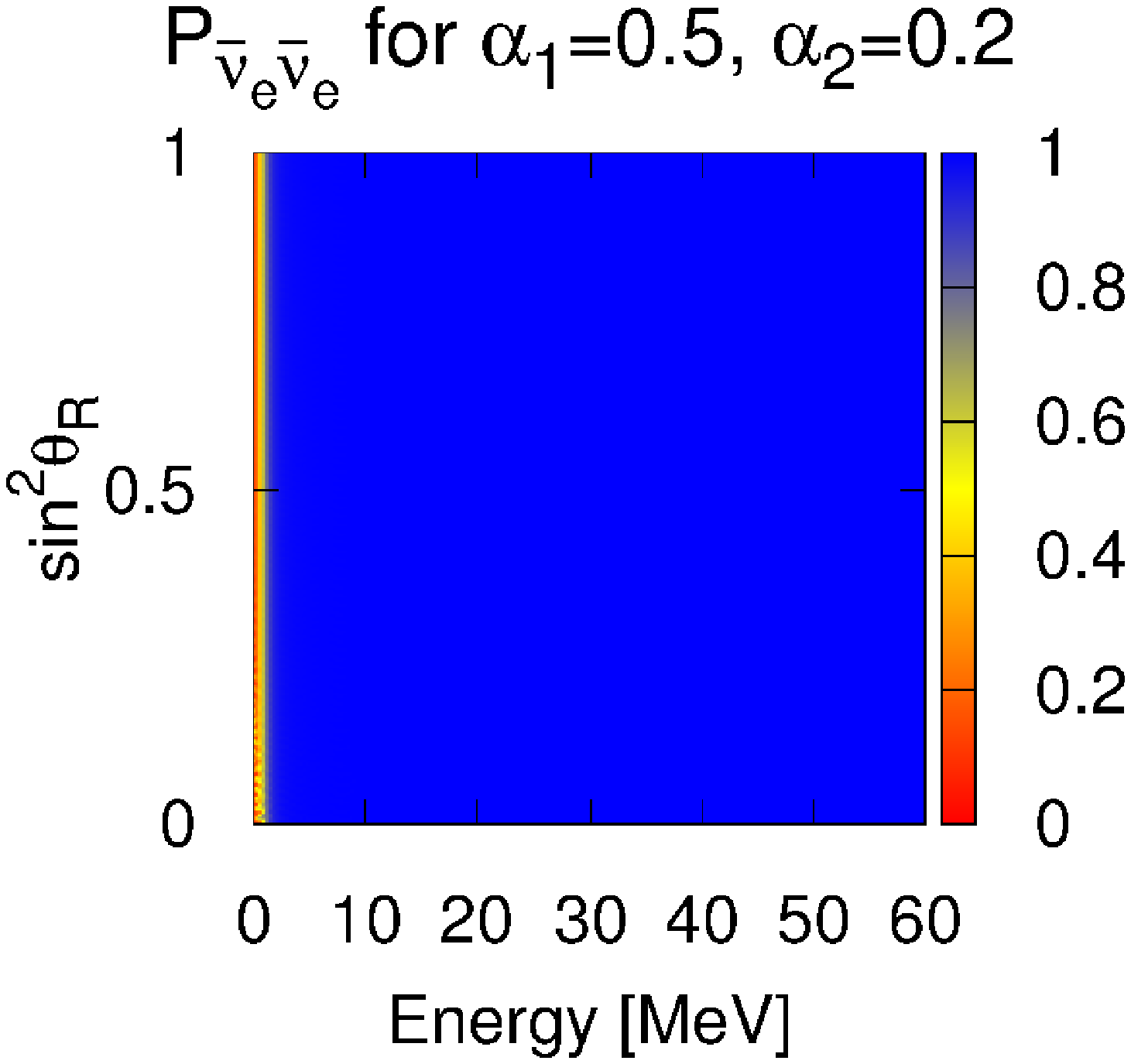}
}
\caption{Top panels: The heatmaps of survival probability of electron neutrinos at $t_{pb}=2.8s$ and $r=400\rm{km}$ as a function of energy and emission angle when there is flavor-violating NSSI. Bottom panels: The same but for electron antineutrinos.}
\label{fig:585FV_2D}
\end{figure}

In order to make sure the ``shut-down'' effect of NSSI is not specific to some certain settings of the supernova environment, we perform the same kind of calculations for the $t_{pb}=2.8\;\rm{s}$ time slice of $10.8\;{\rm M_{\odot}}$ progenitor. In figure (\ref{fig:585FP_SP}) we plot the results with flavor-preserving NSSI only. It shows a similar ``shut-down'' effect in the neutrino sector as at $t_{pb}=1.0\;\rm{s}$. However, flavor transformation does not take place in the antineutrino sector with just the V-A term - this result is consistent with the Wu \emph{et al.} results \cite{PhysRevD.91.065016} - so there is no difference when NSSI is added. From the spectrum at $r=400\;{\rm km}$ we can see the dip in the survival probability becomes shallower as NSSI increased, but the range of flavor transformation remains the same. The sequence of 2-D plots shown in figure (\ref{fig:585FP_2D}) also show a shrinking of the transformed regions due to NSSI, similar to the shrinking seen in the   $t_{pb}=1.0\;\rm{s}$ case. And also as before, the effect of the flavor-violating NSSI is a restoration of flavor transformation to a state as if NSSI does not exist, as shown by figure (\ref{fig:585FV_SP}) and (\ref{fig:585FV_2D}).

\begin{figure}[t]
\centering
\subfigure{
\includegraphics[width=.45\textwidth]{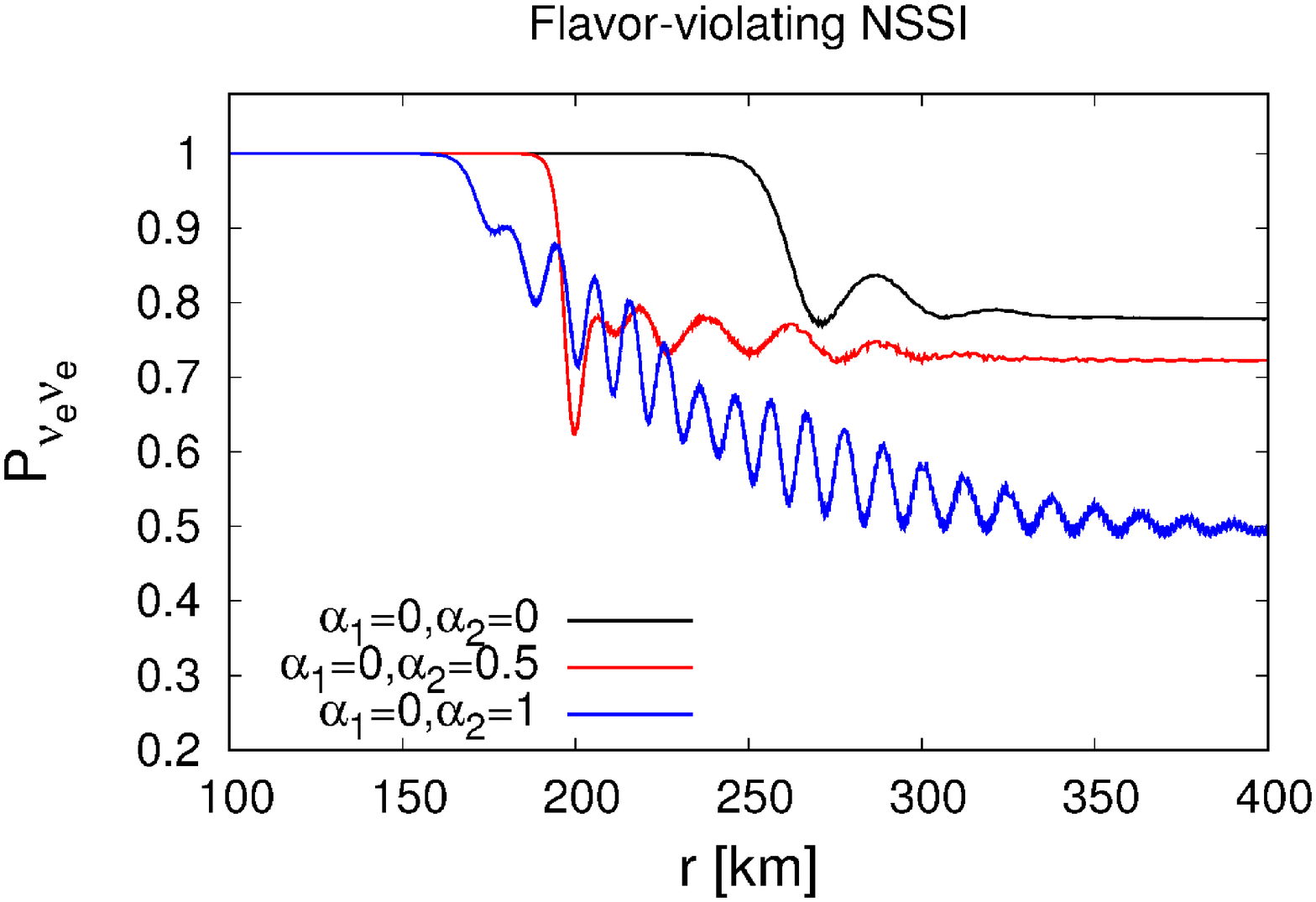}
}
\subfigure{
\includegraphics[width=.45\textwidth]{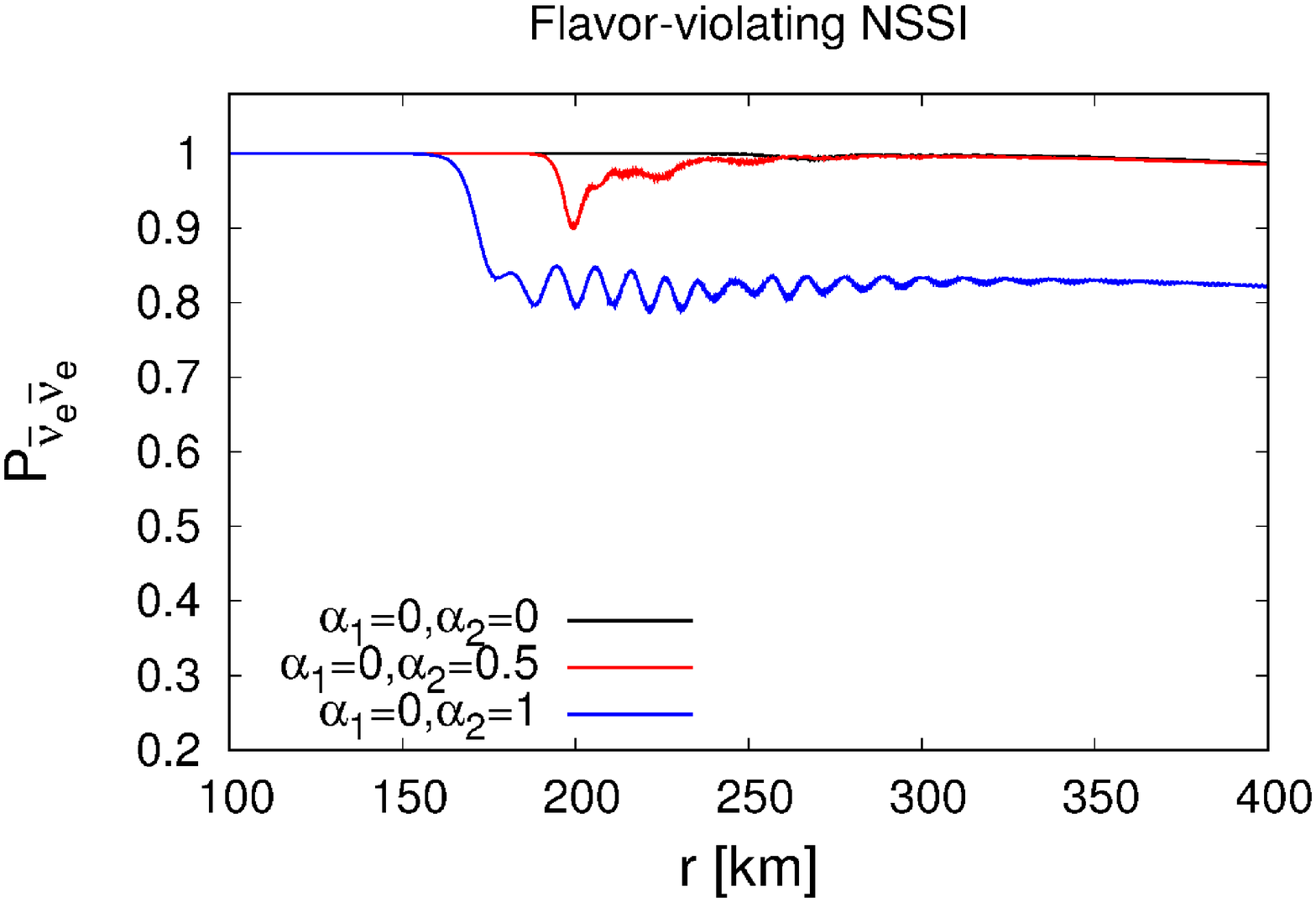}
}
\subfigure{
\includegraphics[width=.45\textwidth]{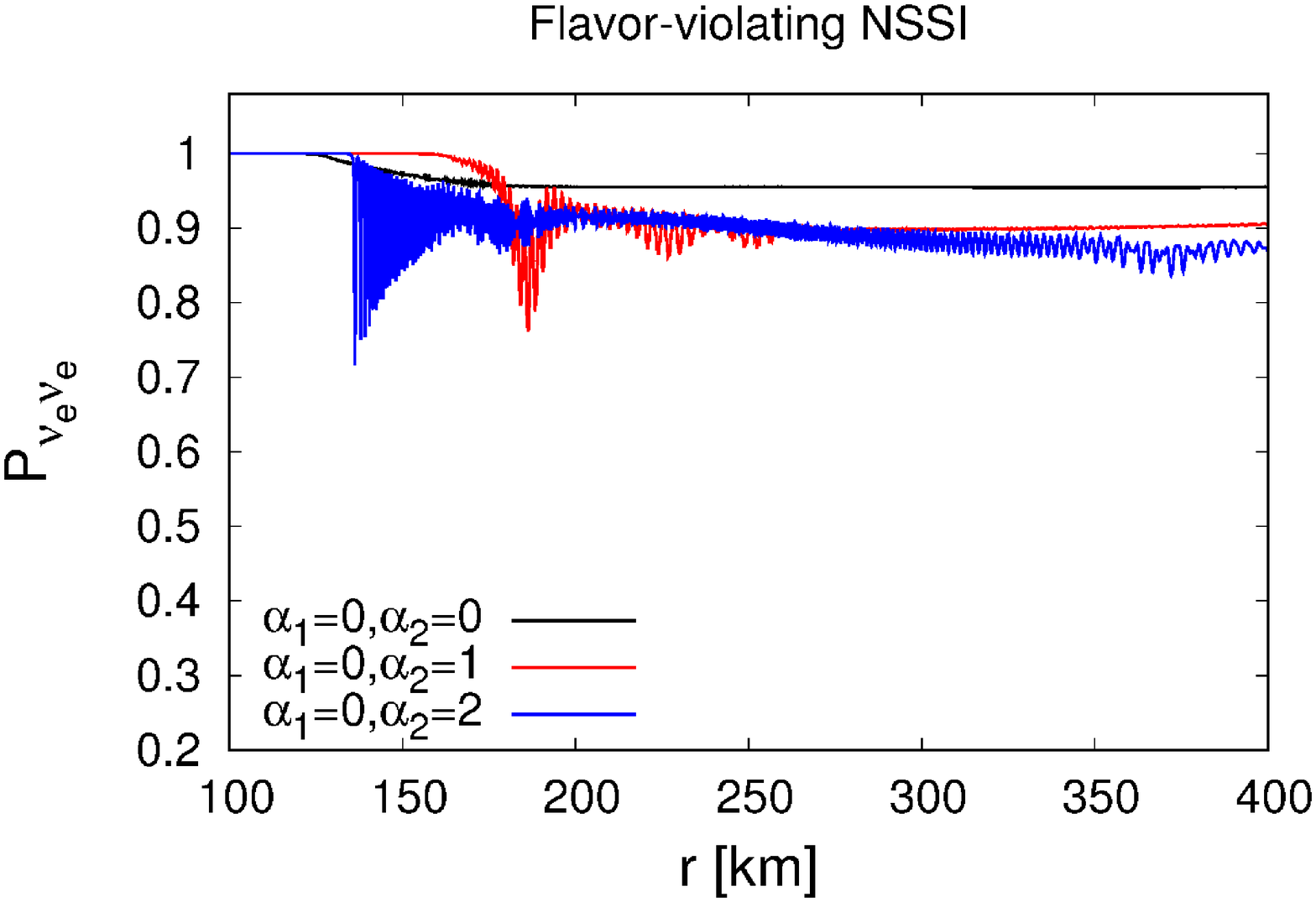}
}
\subfigure{
\includegraphics[width=.45\textwidth]{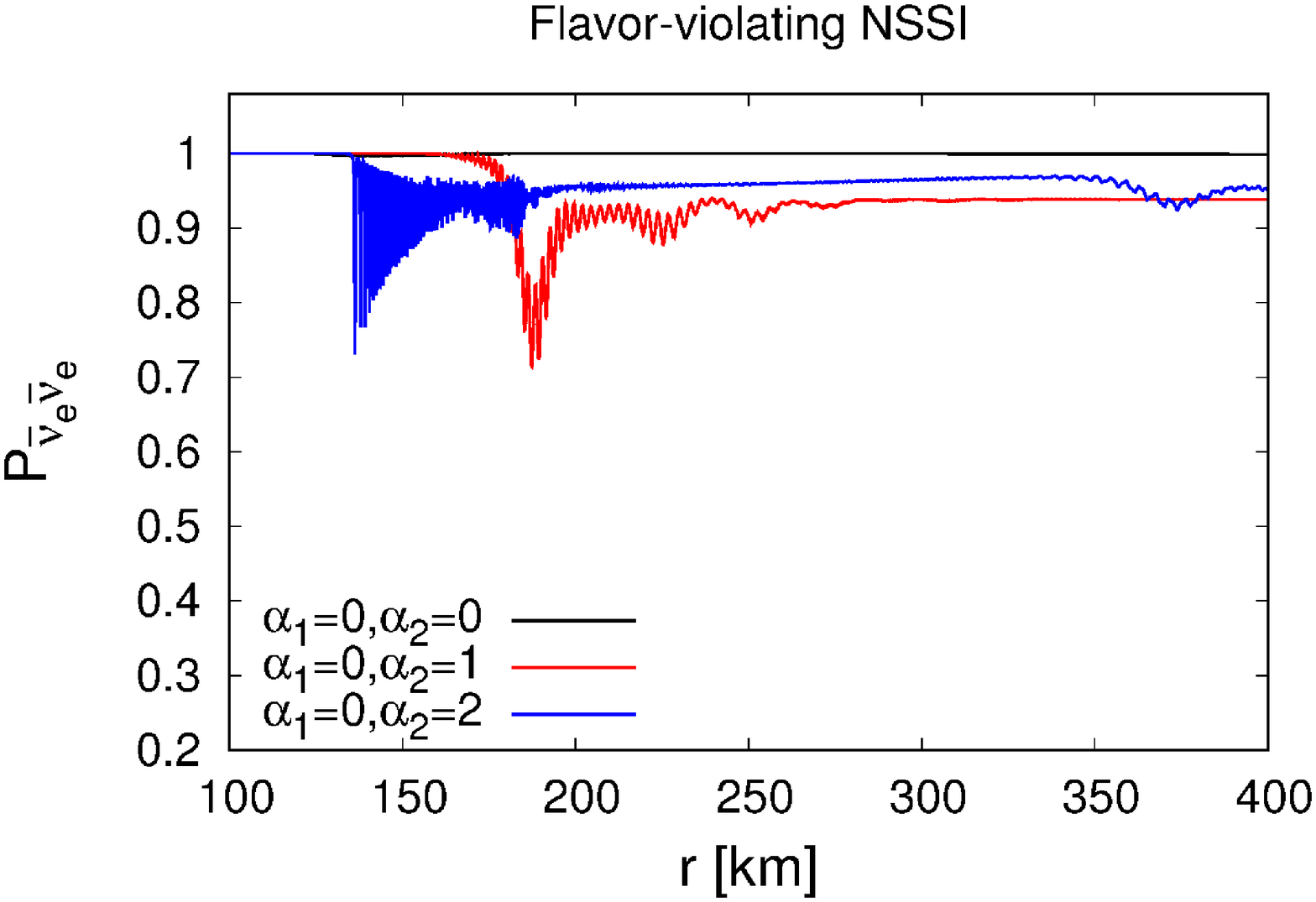}
}
\caption{Top panels: Survival probability of electron neutrinos (left) and antineutrinos (right) at $t_{pb}=2.8s$ as a function of distance with pure flavor violating NSSI for IMO. Bottom panels: The same as top panels but for NMO.}
\label{fig:585pureFV_SP}
\end{figure}
Finally, it is also interesting to look at the effects of a pure flavor-violating NSSI. As seen in figure (\ref{fig:585pureFV_SP}), the pure flavor-violating NSSI is capable of enforcing flavor transformation in the antineutrino sector for the IMO at the post-bounce time of $t_{pb}=2.8\;{\rm s}$, and the flavor transformation in the neutrino sector is also enhanced for this ordering. When the mass ordering is normal the NSSI can also lead to some flavor oscillations for both neutrino and antineutrinos, especially in the region close to the neutrinosphere, although the final survival probabilities are not very different from the result without NSSI even for the case where the flavor-violating parameter $\alpha_2 =2$. These results with non-zero pure flavor-violating NSSI are qualitatively similar to that found by Das, Dighe and Sen with the gauge boson NSSI \cite{Das:2017iuj,dighe2018nonstandard}. This flavor transformation with pure flavor-violating NSSI can be also compared to the results with only the standard V-A interaction found in Wu \emph{et al.} \cite{PhysRevD.91.065016}. 
Using the $18.0\;{\rm M_{\odot}}$ simulation by Fischer \emph{et al.} \cite{Fischer:2009af}, Wu \emph{et al.} observed no transformation in the antineutrinos and only a small amount of transformation in the neutrinos at these late times.  

\begin{figure}[t]
\centering
\subfigure{
\includegraphics[width=.45\textwidth]{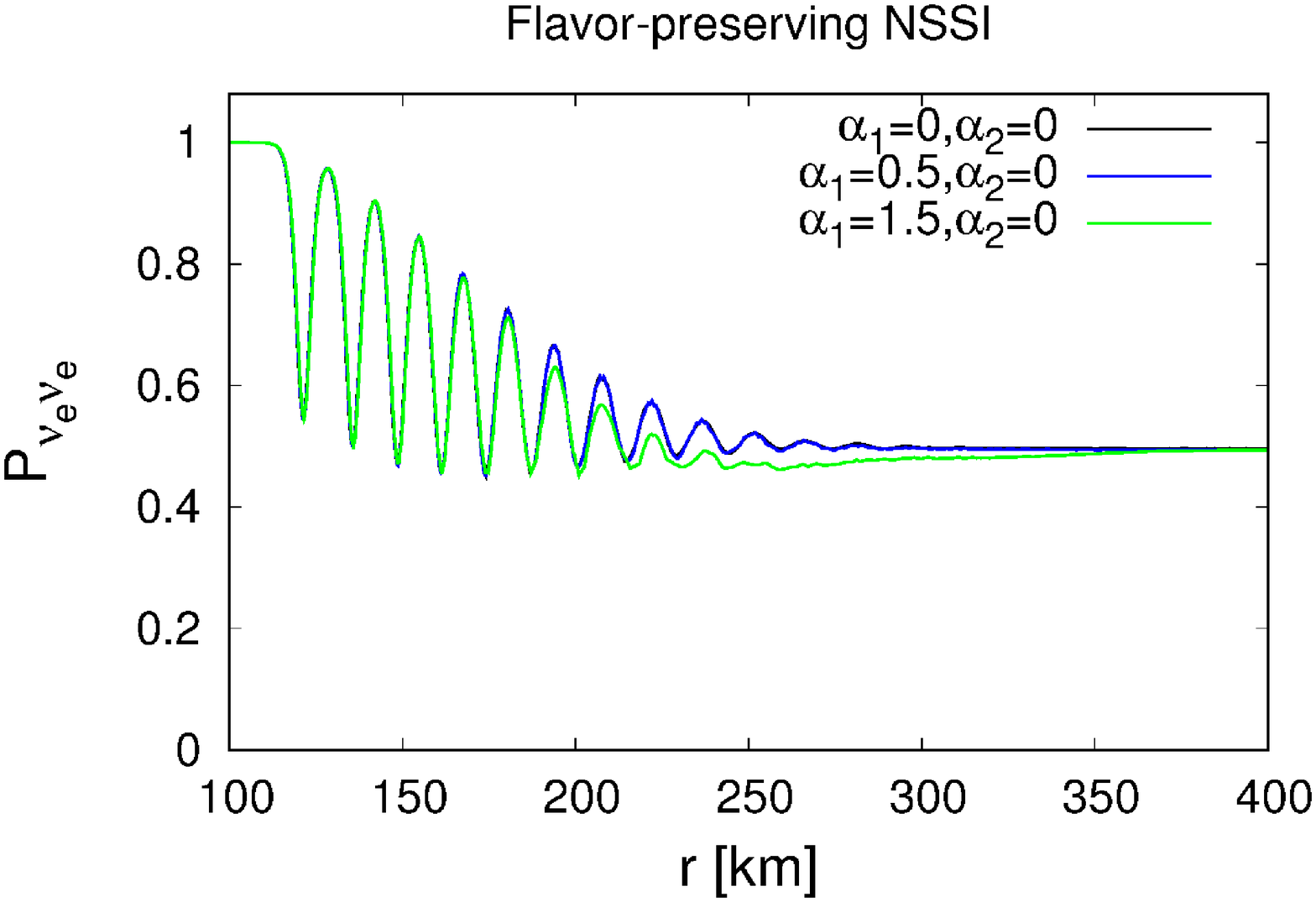}
}
\subfigure{
\includegraphics[width=.45\textwidth]{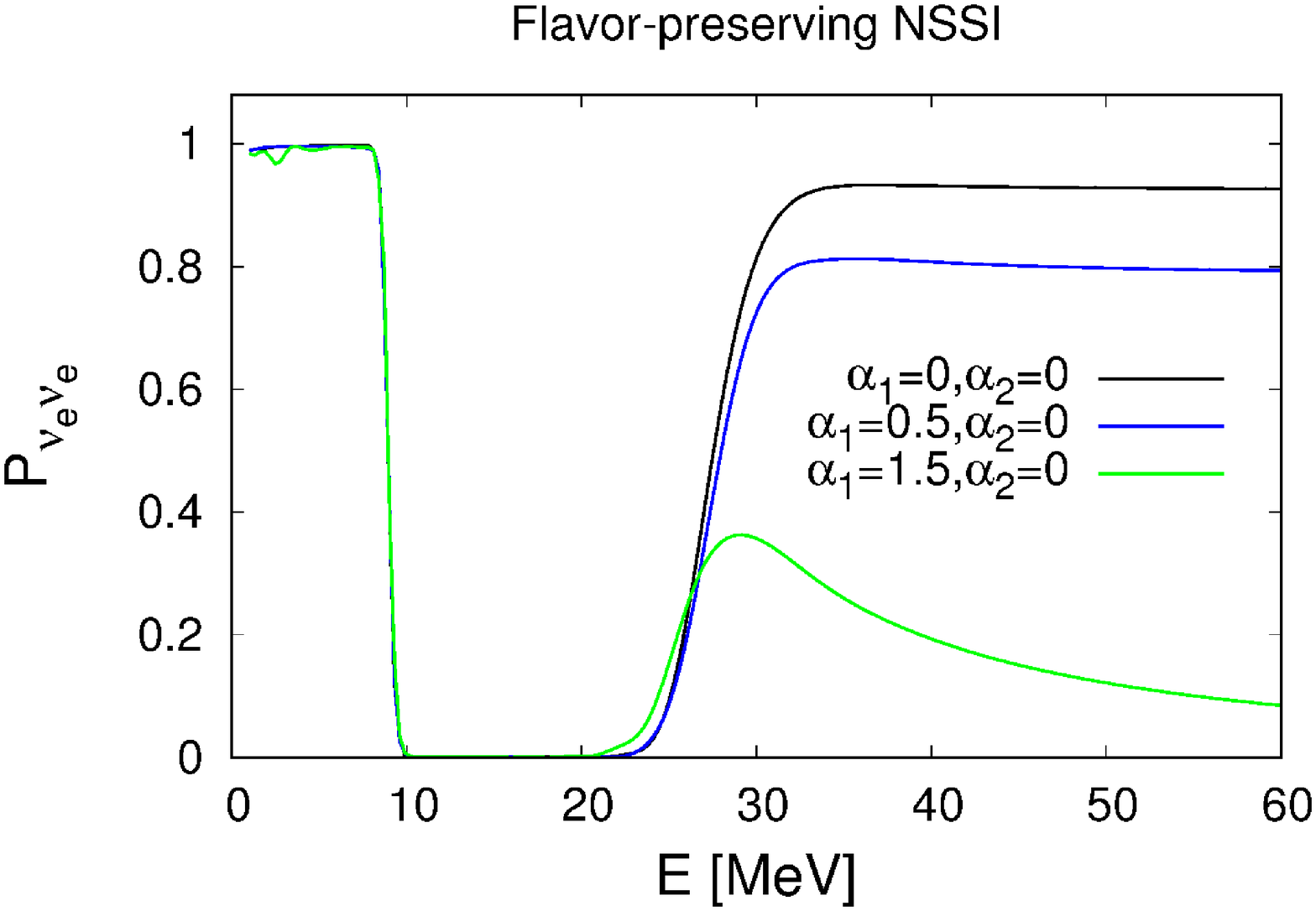}
}
\subfigure{
\includegraphics[width=.45\textwidth]{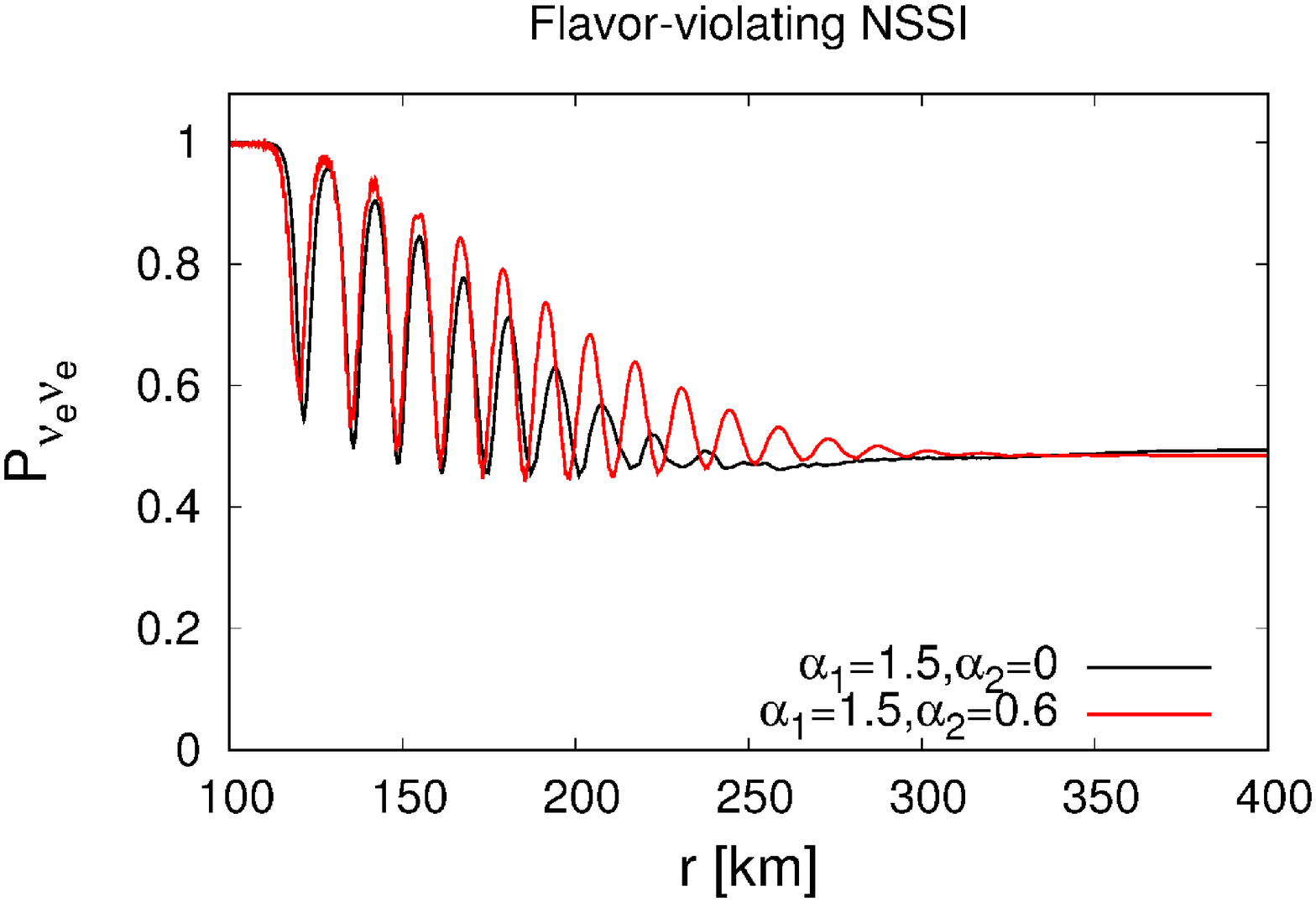}
}
\subfigure{
\includegraphics[width=.45\textwidth]{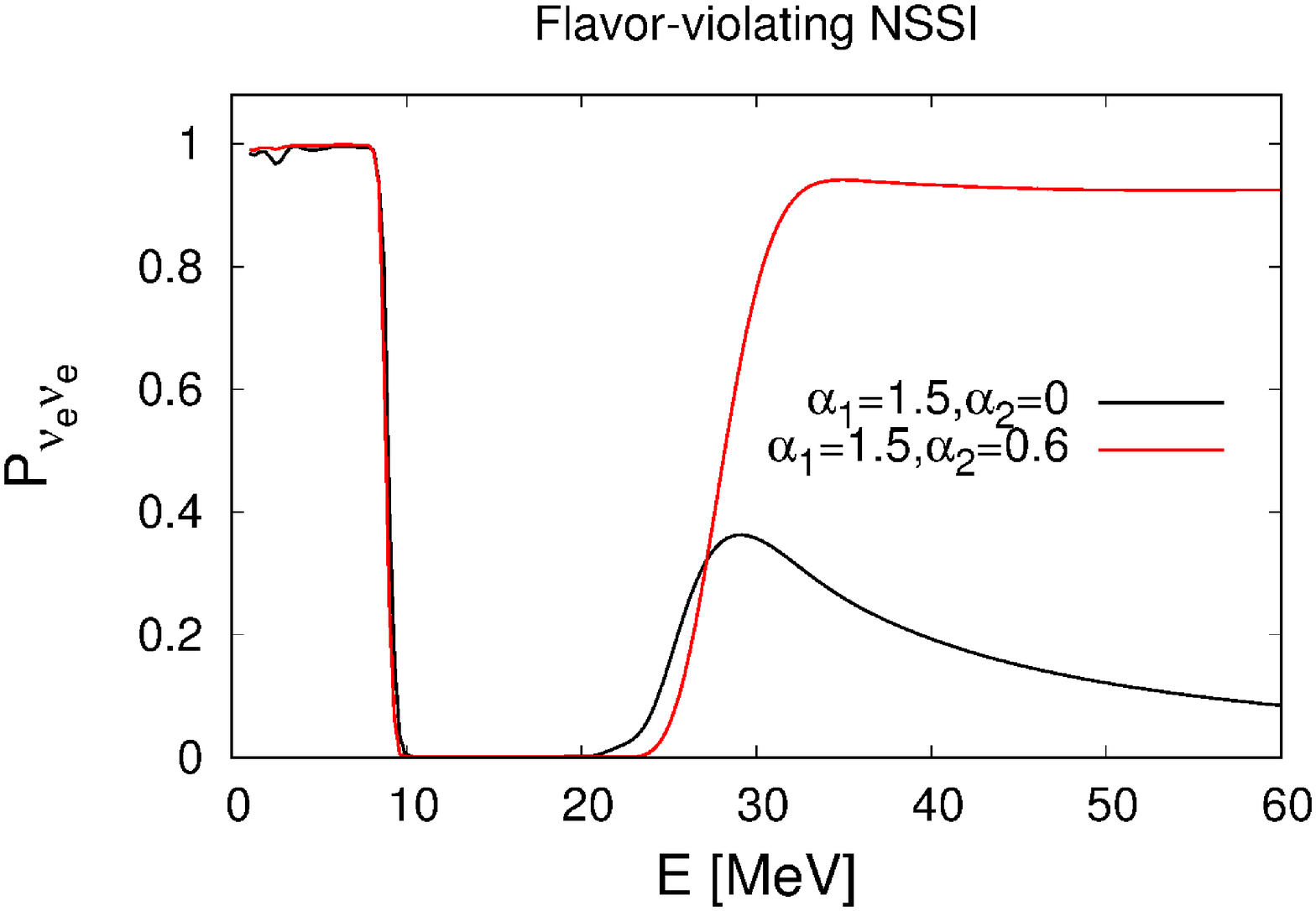}
}
\caption{Top panels: ``Single-angle'' survival probability of electron neutrinos at $t_{pb}=2.8\;\rm{s}$ as a function of distance (left panel) and energy (right panel) at $r=400\;{\rm km}$ with flavor-preserving NSSI. Bottom panels: The same but with flavor-violating terms.}
\label{fig:SA585}
\end{figure}


\subsection{``Single-angle'' vs ``multi-angle'' approach}
In the previous sections we have demonstrated the suppression effect by flavor-preserving NSSI and the effect of undoing the suppression effect by the flavor-violating terms in the NSSI under the ``multi-angle'' framework. One often sees in the literature on supernova neutrinos reference to a ``single-angle'' approximation. This approximation assumes the evolution history of a neutrino is independent of its emission direction and is identical with that of the neutrinos propagating in a chosen direction\footnote{The chosen direction is often set to be either the radial direction or $45^{\circ}$ relative to the radial direction at the neutrinosphere. Here we adopted the radial direction.}. This approximation has been used in previous works about NSSI and supernova neutrinos such as  \cite{Das:2017iuj,Blennow:2008er}. The ``single-angle'' approximation greatly reduces runtimes but its drawback is that it has been known to produce collective flavor transformation which is not seen in ``multi-angle'' calculation due to its artificial synchronization of different angular modes. While in some cases the ``single-angle'' approach gives qualitatively similar results as ``multi-angle'' approach, it also lacks the decoherence effect and can often result in  flavor transformation occurring at much smaller radii than seen in multi-angle calculations \cite{2011PhRvL.106i1101D}. In this section we compare the ``multi-angle'' results with ``single-angle'' counterparts to see whether the effects caused by NSSI can be reproduced more expediently in the single-angle calculations. In the ``single-angle'' approximation all neutrinos with the same energy share the same evolution history regardless of their direction of propagation, so the NSSI Hamiltonian (\ref{scalar}) and (\ref{pseudo}) can be simplified to be \cite{Duan:2006an}
\begin{eqnarray}
{H_{\rm{S}}}\left( r \right) = {\mkern 1mu} \frac{{D\left( {r/{R_\nu }} \right)}}{{2\pi {\mkern 1mu} R_\nu ^2}}\int \left\{ {\bf{\tilde g}} \left[ {\rho}^*(r,E)\frac{{{L_{\nu ,\infty }}}}{{\langle {E_{\nu ,\infty }}\rangle }}\,{f_\nu }\left( E\right) - {\bar \rho }(r,E)\frac{{{L_{\bar \nu ,\infty }}}}{\langle {E_{\bar \nu ,\infty }}\rangle }\,f_{\bar \nu }\left( E\right) \right]{\bf{\tilde g}} \right\}\,dE   \\
{H_P}\left( r \right) = {\mkern 1mu} \frac{{D\left( {r/{R_\nu }} \right)}}{{2\pi {\mkern 1mu} R_\nu ^2}}\int {\left\{ {{\bf{\tilde h}}\left[ {\rho^*(r,E)\frac{{{L_{\nu ,\infty }}}}{{\langle {E_{\nu ,\infty }}\rangle }}\,{f_\nu }\left(E\right) - {{\bar \rho }}(r,E)\frac{{{L_{\bar \nu ,\infty }}}}{{\langle {E_{\bar \nu ,\infty }}\rangle }}\,{f_{\bar \nu }}\left( E \right)} \right]{\bf{\tilde h}}} \right\}\,dE} 
\end{eqnarray}
where 
\begin{equation}
D\left( {r/{R_\nu }} \right) = \frac{1}{2}{\left[ {1 - \sqrt {1 - {{\left( {R_{\nu}/r} \right)}^2}} } \right]^2}
\end{equation}
is the geometric factor obtained after averaging over all the angular modes. $E \equiv E_{\bf p}$ is the energy of the background neutrinos. The expression for the single-angle version of the V-A interactions can be found in Duan \emph{et al.} \cite{Duan:2006an}.

In figure (\ref{fig:SA585}) we plot the survival probabilities for $t_{pb}=2.8\;\rm{s}$ in neutrino sector computed with ``single-angle'' approach. In the upper panels, we only include the flavor-preserving NSSI. Here we can see that unlike in the ``multi-angle'' case, the NSSI do not suppress the flavor transformtion. Instead, in the final spectrum we notice that the flavor-preserving NSSI actually enhances flavor transformation of the neutrinos in the high energy tail. In the lower panels we again add the flavor-violating terms, and just as ``multi-angle'' case the effect of the flavor-preserving NSSI is largely wiped out, since the enhanced transformation in the high energy tail disappears. Thus it appears the presence of flavor-preserving NSSI has different effects in ``single-angle'' and ``multi-angle'' cases but that single-angle does reproduce the correct trend that the  flavor-violating terms always tends to undo any effect caused by flavor-preserving NSSI. The mechanism through which flavor-preserving NSSI shuts down collective oscillations in multi-angle calculation is still a point of interest that needs further investigation.

\section{Summary and Discussion}
\label{sec:summary}
\FloatBarrier
In this paper we have derived the effective neutrino-neutrino Hamiltonian due to a NSSI with a scalar/pseudoscalar field and applied it to the case of neutrino flavor transformations at two epochs of a core-collapse supernova. We find that, as in the case of NSSI due to a new neutrino interaction via a guage boson, there is a suppression effect of the flavor-preserving NSSI which is capable of delaying or shutting down entirely collective flavor oscillation when the strength of the NSSI is comparable to the standard V-A interaction. The presence of flavor-violating terms in the NSSI has the effect of reducing the suppression effect of the flavor-preserving interactions and can even restore the collective flavor oscillations to more-or-less the Standard Model behavior when sufficiently large. When only flavor-violating interactions occur, NSSI can increase the flavor transformation beyond those of V-A alone and even induce oscillations in circumstances where the standard V-A does not. Finally, we find that while the single-angle approximation can give qualitatively similar results to multi-angle calculations as we vary the NSSI parameters, there are large quantitative differences between the two. 

In order to exploit our findings we must successfully identify the signatures of collective flavor oscillation in the signal from a Galactic CCSN. If that can be done, our results indicate that supernova neutrinos can provide several complimentary methods for the determination of neutrino properties should the neutrino be a Majorana fermion and the neutrino-scalar interaction be comparable to the standard V-A interaction (but with small flavor violation). First, the effects of observation of scalar or pseudoscalar NSSI could be used as a complimentary method for identifying the Majorana or Dirac nature of the neutrino. If the NSSI is of the order of the weak interaction, NSSI effects have nothing to do with the neutrino mass so appear even if the mass ordering is normal and the Majorana phases conspire to give an neutrinoless double beta decay effective Majorana mass $m_{\beta\beta}$ which is exactly zero. At the same time, the presence or absence of NSSI signatures in the neutrino signal from a Galactic supernova neutrino burst provides a complimentary tool for measuring, or placing upper limits upon, the coupling strength of NSSI. Current bounds on neutrino-scalar coupling strength are found by a variety of analyses to be $|g|^2<10^{-7}\sim 10^{-6}$ for scalar masses below $100\; \rm{MeV}$, but there are presently no bounds for scalar masses above $300\;\rm{MeV}$ \cite{Pasquini:2015fjv,Heurtier:2016otg}. The effective neutrino-neutrino self-interaction we derived is valid only for scalar fields with large masses so NSSI of supernova neutrinos are able to provide constraints in what is currently a blank area in the neutrino-scalar coupling exclusion plot. Finally, the NSSI we have considered in this paper are flavor symmetric even though they may be flavor-violating. Other than simplicity, there is no reason to expect this property to be true. The interaction strength might be unequal for different neutrino flavors or between different pairs of neutrino flavors. Such flavor asymmetry would introduce new phenomenology, as indicated by the results from Das, Dighe \& Sen and Dighe \& Sen \cite{Das:2017iuj,dighe2018nonstandard} for NSSI due to gauge bosons. 

\section*{Acknowledgements}

The authors are grateful for many useful discussions with Gail McLaughlin and Alexey Vlasenko. This research is supported at NC State by the U.S.~Department of Energy award DE-FG02-10ER41577.

\appendix
\section{The mean field approximation}
\label{sec:meanfield}
\FloatBarrier
In this section we first derive the mean field expressions of the 4-neutrino operators that appear in the NSSI mediated by scalar fields, namely eq. (\ref{4-nu scalar Dirac}) (\ref{4-nu pseudo Dirac}) (\ref{4-nu scalar majorana}) and (\ref{4-nu pseudo majorana}). 
For generality we start by defining a generic 4-fermion operator as follows
\begin{equation}
M_{1234}^{ab} = \left( {{{\bar \psi }_1}{\Gamma ^a}{\psi _2}} \right)\left( {{{\bar \psi }_3}{\Gamma ^b}{\psi _4}} \right),
\end{equation}
here $\Gamma^a$ can be anyone of the 16 $\Gamma$-matrices forming the basis of the vectorial space of all $4\times4$ matrices. Applying the mean field approximation on the 4-fermion operator results in the following expression
\begin{equation}\label{meanfield1}
\begin{array}{l}
M_{1234}^{ab} \approx \left\langle {{{\bar \psi }_1}{\Gamma ^a}{\psi _2}} \right\rangle \left( {{{\bar \psi }_3}{\Gamma ^b}{\psi _4}} \right) + \left\langle {{{\bar \psi }_3}{\Gamma ^b}{\psi _4}} \right\rangle \left( {{{\bar \psi }_1}{\Gamma ^a}{\psi _2}} \right) - \\ \\
\sum\limits_{c,d = S,P,V,A,T} {{C_{ab,cd}}\left[ {\left\langle {{{\bar \psi }_3}{\Gamma ^d}{\psi _2}} \right\rangle \left( {{{\bar \psi }_1}{\Gamma ^c}{\psi _4}} \right) + \left\langle {{{\bar \psi }_1}{\Gamma ^c}{\psi _4}} \right\rangle \left( {{{\bar \psi }_3}{\Gamma ^d}{\psi _2}} \right)} \right]}.
\end{array}
\end{equation}
The first two terms of Eq. (\ref{meanfield1}) represent the regular ``Hartree terms'', while the following terms inside the summation are the ``exchange terms'' arising from the mean field treatment \cite{bruus2004many}. Note that: a \textit{Fierz transformation} has been performed to the ``exchange terms'' since the fermion operators contain spinors, we have dropped the constant term that is present in the mean field expression because it does not have any effect in the evolution equations. In the case of scalar-scalar interaction, we have $a,b=S$. Replacing the generic fermion fields $\psi$ with neutrino fields, we have
\begin{equation}\label{meanfield2}
\begin{array}{l}
\left( {{{\bar \nu }_1}{\nu _2}} \right)\left( {{{\bar \nu }_3}{\nu _4}} \right) \approx \left\langle {{{\bar \nu }_1}{\nu _2}} \right\rangle \left( {{{\bar \nu }_3}{\nu _4}} \right) + \left\langle {{{\bar \nu }_3}{\nu _4}} \right\rangle \left( {{{\bar \nu }_1}{\nu _2}} \right) -\\ \\
\sum\limits_{c,d = S,P,V,A,T} {{C_{SS,cd}}\left[ {\left\langle {{{\bar \nu }_3}{\Gamma ^d}{\nu _2}} \right\rangle \left( {{{\bar \nu }_1}{\Gamma ^c}{\nu _4}} \right) + \left\langle {{{\bar \nu }_1}{\Gamma ^c}{\nu _4}} \right\rangle \left( {{{\bar \nu }_3}{\Gamma ^d}{\nu _2}} \right)} \right]} .
\end{array}
\end{equation}
In the relativistic limit only vector and pseudovector terms can survive the averaging in the single-particle state \cite{Bergmann:1999rz} so we can drop all terms in the right-hand side of equation (\ref{meanfield2}) except for the terms with $V\times V$ or $A\times A$ form. Interestingly, the ``Hartree terms'' are among those who do not survive, which is not the case in the NSSI mediated by gauge bosons where the ``Hartree terms'' are vector-vector type.
Therefore we are left with
\begin{equation}
\begin{array}{l}
\left( {{{\bar \nu }_1}{\nu _2}} \right)\left( {{{\bar \nu }_3}{\nu _4}} \right) \approx  - \frac{1}{4}\left\langle {{{\bar \nu }_1 }{\Gamma ^V}{\nu _4 }} \right\rangle \left( {{{\bar \nu }_3 }{\Gamma ^V}{\nu _2 }} \right) + \frac{1}{4}\left\langle {{{\bar \nu }_1 }{\Gamma ^A}{\nu _4 }} \right\rangle \left( {{{\bar \nu }_3 }{\Gamma ^A}{\nu _2 }} \right) + \left( {1 4  \leftrightarrow 3 2 } \right) \\ \\
 =  - \frac{1}{2}\left\langle {{{\bar \nu }_1 }{\gamma ^\mu }{P_R}{\nu _4 }} \right\rangle \left( {{{\bar \nu }_3 }{\gamma ^\mu }{P_L}{\nu _2 }} \right) - \frac{1}{2}\left\langle {{{\bar \nu }_1 }{\gamma ^\mu }{P_L}{\nu _4 }} \right\rangle \left( {{{\bar \nu }_3 }{\gamma ^\mu }{P_R}{\nu _2 }} \right) + \left( {1 4  \leftrightarrow 3 2 } \right),
\end{array}
\end{equation}
where $\Gamma^{V}\equiv \gamma^{\mu}$, $\Gamma^{A}\equiv \gamma^{\mu}\gamma^{5}$ and $P_{L/R}=\frac{1}{2}(1\mp \gamma^{5})$ are the projection operators. Decomposing the neutrino into $\nu  = {\left( {\begin{array}{*{20}{c}}{{\nu _L}}&{{\nu _R}}\end{array}} \right)^T}$ for Dirac neutrinos, and $\nu  = {\left( {\begin{array}{*{20}{c}}{{\nu _L}}&{{\nu _L ^{C}}}\end{array}} \right)^T}$ for Majorana neutrino, we eventually obtain equations (\ref{4-nu scalar Dirac}) and (\ref{4-nu scalar majorana}). The derivation for the equations (\ref{4-nu pseudo Dirac}) and (\ref{4-nu pseudo majorana}) follows a similar path.

Next we derive the expressions for equations (\ref{normal_current}) and (\ref{conjugate_current}). First we write down the quantized field operator for Majorana neutrino 
\begin{equation}
\nu \left( x \right) = \sum\limits_{h =  \pm 1} {\sum\limits_p {\frac{1}{{2E_{\bf p}V}}\left[ {{a^{\left( h \right)}}\left( p \right){u^{\left( h \right)}}\left( p \right){e^{ - ip \cdot x}} + {a^{\left( h \right)\dag }}\left( p \right){v^{\left( h \right)}}\left( p \right){e^{ip \cdot x}}} \right]} }  \equiv {\nu ^C}\left( x \right),
\end{equation}
where $x\equiv x^{\mu} \equiv (t, {\bf x})$ is the 4-position and $p\equiv p^{\mu}\equiv (E_{\bf p}, {\bf p})$ is the 4-momentum. Then we can decompose the neutrino field into its 2 chirality components $\nu_L(x) = P_L \nu(x)$ and $\nu_L^C(x) = P_R \nu(x)$. If neutrino has mass then both helicity states are present for each of the 2 chirality fields. But in the relativistic limit, for each helicity state, one of the 2 chirality components will be suppressed, resulting in the following equations
\begin{equation}\label{majorana_operator1}
{\nu _L}\left( x \right) = \sum\limits_p {\frac{1}{{2E_{\bf p}V}}\left[ {{a^{\left(  -  \right)}}\left( p \right){u^{\left(  -  \right)}}\left( p \right){e^{ - ip \cdot x}} + {a^{\left(  +  \right)\dag }}\left( p \right){v^{\left(  +  \right)}}\left( p \right){e^{ip \cdot x}}} \right]} ,
\end{equation}
and
\begin{equation}\label{majorana_operator2}
\nu _L^C\left( x \right) = \sum\limits_p {\frac{1}{{2E_{\bf p}V}}\left[ {{a^{\left(  +  \right)}}\left( p \right){u^{\left(  +  \right)}}\left( p \right){e^{ - ip \cdot x}} + {a^{\left(  -  \right)\dag }}\left( p \right){v^{\left(  -  \right)}}\left( p \right){e^{ip \cdot x}}} \right]} ,
\end{equation}
Since Majorana particles are their own antiparticles, we cannot distinguish a Majorana neutrino from an antineutrino by their creation and annihilation operators. Nevertheless it is customary to call Majorana neutrino with negative(positive) helicity \textit{neutrino}(\textit{antineutrino}), therefore we have (flavor subscripts omitted)
\begin{equation}\label{majorana_state}
\left| {\nu \left( \bf p \right)} \right\rangle \equiv \left| {\nu \left( p \right)} \right\rangle  = \frac{1}{{\sqrt {2E_{\bf p}V} }}{a^{\left(  -  \right)\dag }}\left( p \right)\left| 0 \right\rangle ,\;\;
\left| {{\bar \nu} \left( \bf p \right)} \right\rangle \equiv \left| {\bar \nu \left( p \right)} \right\rangle  = \frac{1}{{\sqrt {2E_{\bf p}V} }}{a^{\left(  +  \right)\dag }}\left( p \right)\left| 0 \right\rangle ,
\end{equation}
Note we adopt the finite volume normalization convention from \cite{Giunti:1053706} so that the 4-momentum is summed instead of integrated. The corresponding commutation relations for the creation and annihilation operators are
\begin{equation}\label{commutation}
\left\{ {a_\alpha ^{\left( h \right)}\left( p \right),a_\beta ^{\left( {h'} \right)\dag }\left( {p'} \right)} \right\} = \left( {2E_{\bf p}V} \right){\delta _{\alpha \beta }}{\delta _{hh'}}{\delta _{pp'}},
\end{equation}
with $\alpha,\beta$ denoting the neutrino flavor. Combining equations (\ref{majorana_operator1}), (\ref{majorana_operator2}), (\ref{majorana_state}) and (\ref{commutation}), we can obtain the current equations (\ref{normal_current}) and (\ref{conjugate_current}) with the flavor-superposition states (\ref{flavor_state}).

\section{The effective Hamiltonian}
\label{effective Hamiltonian}
In this section we derive the effective single-particle Hamiltonian for the nonstandard neutrino self-interaction, which is to be used in the flavor evolution equation. For simplicity we consider the case in which there are only neutrinos with momentum ${\bf p}$ in the background, and the momentum of the test neutrino is ${\bf q}$. We start with combining Eqs. (\ref{4-fermion Hamiltonian}) and (\ref{4-nu scalar majorana}). If we only consider the scalar part, the mean field Hamiltonian operator becomes
\begin{equation}
\label{eqn:H_MF}
\mathcal{H}_{\rm S}^{\rm MF} =  - {\tilde g_{\alpha \beta }}\left[ {\left\langle {\bar \nu _{\alpha L}^C{\,\gamma ^\mu }\,\bar \nu _{\eta L}^C} \right\rangle {{\bar \nu }_{\xi L}}{\,\gamma _\mu }\,{\nu _{\beta L}} + \left\langle {{{\bar \nu }_{\xi L}}{\,\gamma _\mu }{\,\nu _{\beta L}}} \right\rangle \,\bar \nu _{\alpha L}^C{\,\gamma ^\mu }\,\bar \nu _{\eta L}^C} + \left( {\alpha \eta  \leftrightarrow \xi \beta } \right) \right]{\tilde g_{\xi \eta }}\, {N_\nu},
\end{equation}
where $N_{\nu}$ is the number of neutrinos in the background, and ${\tilde g}_{\alpha\beta}=\frac{1}{4m_{\phi}} g_{\alpha\beta}$. Here we note that the absence of the ``Hartree terms'' such as $\left\langle \bar \nu _{\alpha L}^C{\,\gamma ^\mu }\,\bar \nu _{\beta L}^C \right\rangle \left({{\bar \nu }_{\xi L}}{\,\gamma _\mu }\,{\nu _{\eta L}}\right)$ in the Eq. (\ref{eqn:H_MF}) is the one of the major differences between a scalar/pseudoscalar NSSI and the NSSI mediated by gauge bosons.  Using Eqs. (\ref{normal_current}) and (\ref{conjugate_current}) we obtain
\begin{equation}
\mathcal{H}_{\rm{S}}^{{\rm{MF}}} =  {{\tilde g}_{\alpha \beta }}\left( {\frac{{{p^\mu }}}{{E_{\bf p}}}} \right)\left[ { c_\eta ^ * {c_\alpha }\,{{\bar \nu }_{\xi L}}{\,\gamma _\mu }{\nu _{\beta L}} - c_\xi ^ * \,{c_\beta }\,\bar \nu _{\alpha L}^C{\,\gamma _\mu }\,\nu _{\eta L}^C +  \left( {\alpha \eta  \leftrightarrow \xi \beta } \right)} \right]{{\tilde g}_{\xi \eta }}\left( {\frac{{{N_\nu }}}{V}} \right)
\end{equation}
The next step is to evaluate the matrix elements by averaging over the single-particle states of the test neutrino with four momentum $q^{\mu}\equiv(E_{\bf q}, \bf{q})$. The $i,j$ element of the Hamiltonian matrix is 
\begin{equation}
{H_{\rm S,ij}} = \int_V {{d^3}x} \left\langle {{\nu _i}\left( {\bf{q}} \right)} \right|\left. {\mathcal{H}_{\rm S}^{\rm MF}} \right|\left. {{\nu _j}\left( {\bf{q}} \right)} \right\rangle  = 2\left( {1 - {\bf{\hat p}} \cdot {\bf{\hat q}}} \right)\,\left( {{{\tilde g}_{\alpha j}}\,{{\tilde g}_{i\eta }}\,{c_\alpha }\,c_\eta ^ *  + {{\tilde g}_{\alpha i}}\,{{\tilde g}_{j\eta }}\,{c_\eta }\,c_\alpha ^ * } \right)\, n_{\nu} ,
\end{equation}
where $i,j$ are the flavor indices and also representing the corresponding element of $H_{\rm S}$. $n_{\nu} = {N_\nu }/{V}$ is the neutrino density. The angular factor $1 - {\bf{\hat p}} \cdot {\bf{\hat q}}$ comes from the inner product of $\left(p^{\mu}/E_{\bf p}\right)$ and $\left(q_{\mu}/E_{\bf q}\right)$. Since in this paper we assume the coupling matrices are real and symmetric, the result can be simplified to be
\begin{equation}
{H_{\rm S}} = 2\left( {1 - {\bf{\hat p}} \cdot {\bf{\hat q}}} \right)\,\left( {{\bf{\tilde g}}\,\rho^ * (\bf{p}) \,{\bf{\tilde g}} + {{\bf{\tilde g}}^T}\,\rho^ * (\bf{p})\,{{\bf{\tilde g}}^T}} \right) {n_\nu } = 4\left( {1 - {\bf{\hat p}} \cdot {\bf{\hat q}}} \right)\,\left( {{\bf{\tilde g}} \,\rho^* \,  (\bf{p})\, {\bf{\tilde g}}} \right)\, {n_\nu },
\end{equation}
where the density matrix $\rho(\bf{p})$ is defined according to Eq. (\ref{density matrices}). Due to the absence of the Hartree terms, we notice there is no term such as $\tilde{\bf g}{\rm Tr}(\rho\tilde{\bf g})$ that appears in the Hamiltonian of the gauge boson case. Finally, the addition of antineutrinos into the background results in an extra term in the Hamiltonian 
\begin{equation}
\label{eqn:H_S}
{H_{\rm S}} = 4\left( {1 - {\bf{\hat p}} \cdot {\bf{\hat q}}} \right)\,{\bf{\tilde g}}\left( {\rho^ * (\bf{p})}\, {n_\nu } - {{{\bar \rho }(\bf{p})}{\,n_{\bar \nu }}} \right){\bf{\tilde g}}.
\end{equation}
In the context of the bulb model we have a collection of neutrino and antineutrino states of different energies and emission angles. To obtain the effective Hamiltonian in the bulb model we need to perform integrations over these distributions which means we must replace $n_{\nu} \to \int {d{n_\nu }\,dE_{\bf p}} $ and $n_{\bar\nu} \to \int {d{n_{\bar\nu }}\,dE_{\bf p}} $ thus leading to Eq. (\ref{scalar}). The derivation of Eq. (\ref{pseudo}) is similar.

\bibliographystyle{apsrev4-1}
\bibliography{nssi}

\begin{thebibliography}{63}%
\makeatletter
\providecommand \@ifxundefined [1]{%
 \@ifx{#1\undefined}
}%
\providecommand \@ifnum [1]{%
 \ifnum #1\expandafter \@firstoftwo
 \else \expandafter \@secondoftwo
 \fi
}%
\providecommand \@ifx [1]{%
 \ifx #1\expandafter \@firstoftwo
 \else \expandafter \@secondoftwo
 \fi
}%
\providecommand \natexlab [1]{#1}%
\providecommand \enquote  [1]{``#1''}%
\providecommand \bibnamefont  [1]{#1}%
\providecommand \bibfnamefont [1]{#1}%
\providecommand \citenamefont [1]{#1}%
\providecommand \href@noop [0]{\@secondoftwo}%
\providecommand \href [0]{\begingroup \@sanitize@url \@href}%
\providecommand \@href[1]{\@@startlink{#1}\@@href}%
\providecommand \@@href[1]{\endgroup#1\@@endlink}%
\providecommand \@sanitize@url [0]{\catcode `\\12\catcode `\$12\catcode
  `\&12\catcode `\#12\catcode `\^12\catcode `\_12\catcode `\%12\relax}%
\providecommand \@@startlink[1]{}%
\providecommand \@@endlink[0]{}%
\providecommand \url  [0]{\begingroup\@sanitize@url \@url }%
\providecommand \@url [1]{\endgroup\@href {#1}{\urlprefix }}%
\providecommand \urlprefix  [0]{URL }%
\providecommand \Eprint [0]{\href }%
\providecommand \doibase [0]{http://dx.doi.org/}%
\providecommand \selectlanguage [0]{\@gobble}%
\providecommand \bibinfo  [0]{\@secondoftwo}%
\providecommand \bibfield  [0]{\@secondoftwo}%
\providecommand \translation [1]{[#1]}%
\providecommand \BibitemOpen [0]{}%
\providecommand \bibitemStop [0]{}%
\providecommand \bibitemNoStop [0]{.\EOS\space}%
\providecommand \EOS [0]{\spacefactor3000\relax}%
\providecommand \BibitemShut  [1]{\csname bibitem#1\endcsname}%
\let\auto@bib@innerbib\@empty
\bibitem [{\citenamefont {{Sigl}}\ and\ \citenamefont
  {{Raffelt}}(1993)}]{1993NuPhB.406..423S}%
  \BibitemOpen
  \bibfield  {author} {\bibinfo {author} {\bibfnamefont {G.}~\bibnamefont
  {{Sigl}}}\ and\ \bibinfo {author} {\bibfnamefont {G.}~\bibnamefont
  {{Raffelt}}},\ }\href {\doibase 10.1016/0550-3213(93)90175-O} {\bibfield
  {journal} {\bibinfo  {journal} {Nuclear Physics B}\ }\textbf {\bibinfo
  {volume} {406}},\ \bibinfo {pages} {423} (\bibinfo {year}
  {1993})}\BibitemShut {NoStop}%
\bibitem [{\citenamefont {{Strack}}\ and\ \citenamefont
  {{Burrows}}(2005)}]{2005PhRvD..71i3004S}%
  \BibitemOpen
  \bibfield  {author} {\bibinfo {author} {\bibfnamefont {P.}~\bibnamefont
  {{Strack}}}\ and\ \bibinfo {author} {\bibfnamefont {A.}~\bibnamefont
  {{Burrows}}},\ }\href {\doibase 10.1103/PhysRevD.71.093004} {\bibfield
  {journal} {\bibinfo  {journal} {{Phys.\ Rev.\ D}}\ }\textbf {\bibinfo
  {volume} {71}},\ \bibinfo {eid} {093004} (\bibinfo {year} {2005})},\ \Eprint
  {http://arxiv.org/abs/hep-ph/0504035} {hep-ph/0504035} \BibitemShut {NoStop}%
\bibitem [{\citenamefont {{Volpe}}\ \emph {et~al.}(2013)\citenamefont
  {{Volpe}}, \citenamefont {{V{\"a}{\"a}n{\"a}nen}},\ and\ \citenamefont
  {{Espinoza}}}]{2013PhRvD..87k3010V}%
  \BibitemOpen
  \bibfield  {author} {\bibinfo {author} {\bibfnamefont {C.}~\bibnamefont
  {{Volpe}}}, \bibinfo {author} {\bibfnamefont {D.}~\bibnamefont
  {{V{\"a}{\"a}n{\"a}nen}}}, \ and\ \bibinfo {author} {\bibfnamefont
  {C.}~\bibnamefont {{Espinoza}}},\ }\href {\doibase
  10.1103/PhysRevD.87.113010} {\bibfield  {journal} {\bibinfo  {journal}
  {Phys.\ Rev.\ D}\ }\textbf {\bibinfo {volume} {87}},\ \bibinfo {eid} {113010}
  (\bibinfo {year} {2013})},\ \Eprint {http://arxiv.org/abs/1302.2374}
  {arXiv:1302.2374 [hep-ph]} \BibitemShut {NoStop}%
\bibitem [{\citenamefont {{Vlasenko}}\ \emph {et~al.}(2014)\citenamefont
  {{Vlasenko}}, \citenamefont {{Fuller}},\ and\ \citenamefont
  {{Cirigliano}}}]{2014PhRvD..89j5004V}%
  \BibitemOpen
  \bibfield  {author} {\bibinfo {author} {\bibfnamefont {A.}~\bibnamefont
  {{Vlasenko}}}, \bibinfo {author} {\bibfnamefont {G.~M.}\ \bibnamefont
  {{Fuller}}}, \ and\ \bibinfo {author} {\bibfnamefont {V.}~\bibnamefont
  {{Cirigliano}}},\ }\href {\doibase 10.1103/PhysRevD.89.105004} {\bibfield
  {journal} {\bibinfo  {journal} {Phys.\ Rev.\ D}\ }\textbf {\bibinfo {volume}
  {89}},\ \bibinfo {eid} {105004} (\bibinfo {year} {2014})},\ \Eprint
  {http://arxiv.org/abs/1309.2628} {arXiv:1309.2628 [hep-ph]} \BibitemShut
  {NoStop}%
\bibitem [{\citenamefont {Duan}\ \emph {et~al.}(2006)\citenamefont {Duan},
  \citenamefont {Fuller}, \citenamefont {Carlson},\ and\ \citenamefont
  {Qian}}]{Duan:2006an}%
  \BibitemOpen
  \bibfield  {author} {\bibinfo {author} {\bibfnamefont {H.}~\bibnamefont
  {Duan}}, \bibinfo {author} {\bibfnamefont {G.~M.}\ \bibnamefont {Fuller}},
  \bibinfo {author} {\bibfnamefont {J.}~\bibnamefont {Carlson}}, \ and\
  \bibinfo {author} {\bibfnamefont {Y.-Z.}\ \bibnamefont {Qian}},\ }\href
  {\doibase 10.1103/PhysRevD.74.105014} {\bibfield  {journal} {\bibinfo
  {journal} {Phys. Rev.}\ }\textbf {\bibinfo {volume} {D 74}},\ \bibinfo
  {pages} {105014} (\bibinfo {year} {2006})},\ \Eprint
  {http://arxiv.org/abs/astro-ph/0606616} {arXiv:astro-ph/0606616 [astro-ph]}
  \BibitemShut {NoStop}%
\bibitem [{\citenamefont {{Horiuchi}}\ and\ \citenamefont
  {{Kneller}}(2017)}]{2017arXiv170901515H}%
  \BibitemOpen
  \bibfield  {author} {\bibinfo {author} {\bibfnamefont {S.}~\bibnamefont
  {{Horiuchi}}}\ and\ \bibinfo {author} {\bibfnamefont {J.~P.}\ \bibnamefont
  {{Kneller}}},\ }\href@noop {} {\bibfield  {journal} {\bibinfo  {journal}
  {ArXiv e-prints}\ } (\bibinfo {year} {2017})},\ \Eprint
  {http://arxiv.org/abs/1709.01515} {arXiv:1709.01515 [astro-ph.HE]}
  \BibitemShut {NoStop}%
\bibitem [{\citenamefont {{Mirizzi}}\ \emph {et~al.}(2016)\citenamefont
  {{Mirizzi}}, \citenamefont {{Tamborra}}, \citenamefont {{Janka}},
  \citenamefont {{Saviano}}, \citenamefont {{Scholberg}}, \citenamefont
  {{Bollig}}, \citenamefont {{H{\"u}depohl}},\ and\ \citenamefont
  {{Chakraborty}}}]{2016NCimR..39....1M}%
  \BibitemOpen
  \bibfield  {author} {\bibinfo {author} {\bibfnamefont {A.}~\bibnamefont
  {{Mirizzi}}}, \bibinfo {author} {\bibfnamefont {I.}~\bibnamefont
  {{Tamborra}}}, \bibinfo {author} {\bibfnamefont {H.-T.}\ \bibnamefont
  {{Janka}}}, \bibinfo {author} {\bibfnamefont {N.}~\bibnamefont {{Saviano}}},
  \bibinfo {author} {\bibfnamefont {K.}~\bibnamefont {{Scholberg}}}, \bibinfo
  {author} {\bibfnamefont {R.}~\bibnamefont {{Bollig}}}, \bibinfo {author}
  {\bibfnamefont {L.}~\bibnamefont {{H{\"u}depohl}}}, \ and\ \bibinfo {author}
  {\bibfnamefont {S.}~\bibnamefont {{Chakraborty}}},\ }\href {\doibase
  10.1393/ncr/i2016-10120-8} {\bibfield  {journal} {\bibinfo  {journal} {Nuovo
  Cimento Rivista Serie}\ }\textbf {\bibinfo {volume} {39}},\ \bibinfo {pages}
  {1} (\bibinfo {year} {2016})},\ \Eprint {http://arxiv.org/abs/1508.00785}
  {arXiv:1508.00785 [astro-ph.HE]} \BibitemShut {NoStop}%
\bibitem [{\citenamefont {{Nunokawa}}\ \emph {et~al.}(1997)\citenamefont
  {{Nunokawa}}, \citenamefont {{Peltoniemi}}, \citenamefont {{Rossi}},\ and\
  \citenamefont {{Valle}}}]{1997PhRvD..56.1704N}%
  \BibitemOpen
  \bibfield  {author} {\bibinfo {author} {\bibfnamefont {H.}~\bibnamefont
  {{Nunokawa}}}, \bibinfo {author} {\bibfnamefont {J.~T.}\ \bibnamefont
  {{Peltoniemi}}}, \bibinfo {author} {\bibfnamefont {A.}~\bibnamefont
  {{Rossi}}}, \ and\ \bibinfo {author} {\bibfnamefont {J.~W.~F.}\ \bibnamefont
  {{Valle}}},\ }\href {\doibase 10.1103/PhysRevD.56.1704} {\bibfield  {journal}
  {\bibinfo  {journal} {Phys.\ Rev.\ D}\ }\textbf {\bibinfo {volume} {56}},\
  \bibinfo {pages} {1704} (\bibinfo {year} {1997})},\ \Eprint
  {http://arxiv.org/abs/hep-ph/9702372} {hep-ph/9702372} \BibitemShut {NoStop}%
\bibitem [{\citenamefont {{McLaughlin}}\ \emph {et~al.}(1999)\citenamefont
  {{McLaughlin}}, \citenamefont {{Fetter}}, \citenamefont {{Balantekin}},\ and\
  \citenamefont {{Fuller}}}]{1999PhRvC..59.2873M}%
  \BibitemOpen
  \bibfield  {author} {\bibinfo {author} {\bibfnamefont {G.~C.}\ \bibnamefont
  {{McLaughlin}}}, \bibinfo {author} {\bibfnamefont {J.~M.}\ \bibnamefont
  {{Fetter}}}, \bibinfo {author} {\bibfnamefont {A.~B.}\ \bibnamefont
  {{Balantekin}}}, \ and\ \bibinfo {author} {\bibfnamefont {G.~M.}\
  \bibnamefont {{Fuller}}},\ }\href {\doibase 10.1103/PhysRevC.59.2873}
  {\bibfield  {journal} {\bibinfo  {journal} {Phys.\ Rev. C}\ }\textbf
  {\bibinfo {volume} {59}},\ \bibinfo {pages} {2873} (\bibinfo {year}
  {1999})},\ \Eprint {http://arxiv.org/abs/astro-ph/9902106} {astro-ph/9902106}
  \BibitemShut {NoStop}%
\bibitem [{\citenamefont {{Peres}}\ and\ \citenamefont
  {{Smirnov}}(2001)}]{2001NuPhB.599....3P}%
  \BibitemOpen
  \bibfield  {author} {\bibinfo {author} {\bibfnamefont {O.~L.~G.}\
  \bibnamefont {{Peres}}}\ and\ \bibinfo {author} {\bibfnamefont {A.~Y.}\
  \bibnamefont {{Smirnov}}},\ }\href {\doibase 10.1016/S0550-3213(01)00012-8}
  {\bibfield  {journal} {\bibinfo  {journal} {Nuclear Physics B}\ }\textbf
  {\bibinfo {volume} {599}},\ \bibinfo {pages} {3} (\bibinfo {year} {2001})},\
  \Eprint {http://arxiv.org/abs/hep-ph/0011054} {hep-ph/0011054} \BibitemShut
  {NoStop}%
\bibitem [{\citenamefont {{Beun}}\ \emph {et~al.}(2006)\citenamefont {{Beun}},
  \citenamefont {{McLaughlin}}, \citenamefont {{Surman}},\ and\ \citenamefont
  {{Hix}}}]{2006PhRvD..73i3007B}%
  \BibitemOpen
  \bibfield  {author} {\bibinfo {author} {\bibfnamefont {J.}~\bibnamefont
  {{Beun}}}, \bibinfo {author} {\bibfnamefont {G.~C.}\ \bibnamefont
  {{McLaughlin}}}, \bibinfo {author} {\bibfnamefont {R.}~\bibnamefont
  {{Surman}}}, \ and\ \bibinfo {author} {\bibfnamefont {W.~R.}\ \bibnamefont
  {{Hix}}},\ }\href {\doibase 10.1103/PhysRevD.73.093007} {\bibfield  {journal}
  {\bibinfo  {journal} {Phys.\ Rev.\ D}\ }\textbf {\bibinfo {volume} {73}},\
  \bibinfo {eid} {093007} (\bibinfo {year} {2006})},\ \Eprint
  {http://arxiv.org/abs/hep-ph/0602012} {hep-ph/0602012} \BibitemShut {NoStop}%
\bibitem [{\citenamefont {{Tamborra}}\ \emph {et~al.}(2012)\citenamefont
  {{Tamborra}}, \citenamefont {{Raffelt}}, \citenamefont {{H{\"u}depohl}},\
  and\ \citenamefont {{Janka}}}]{2012JCAP...01..013T}%
  \BibitemOpen
  \bibfield  {author} {\bibinfo {author} {\bibfnamefont {I.}~\bibnamefont
  {{Tamborra}}}, \bibinfo {author} {\bibfnamefont {G.~G.}\ \bibnamefont
  {{Raffelt}}}, \bibinfo {author} {\bibfnamefont {L.}~\bibnamefont
  {{H{\"u}depohl}}}, \ and\ \bibinfo {author} {\bibfnamefont {H.-T.}\
  \bibnamefont {{Janka}}},\ }\href {\doibase 10.1088/1475-7516/2012/01/013}
  {\bibfield  {journal} {\bibinfo  {journal} {JCAP}\ }\textbf {\bibinfo
  {volume} {1201}},\ \bibinfo {eid} {013} (\bibinfo {year} {2012})},\ \Eprint
  {http://arxiv.org/abs/1110.2104} {arXiv:1110.2104 [astro-ph.SR]} \BibitemShut
  {NoStop}%
\bibitem [{\citenamefont {{Warren}}\ \emph {et~al.}(2014)\citenamefont
  {{Warren}}, \citenamefont {{Meixner}}, \citenamefont {{Mathews}},
  \citenamefont {{Hidaka}},\ and\ \citenamefont
  {{Kajino}}}]{2014PhRvD..90j3007W}%
  \BibitemOpen
  \bibfield  {author} {\bibinfo {author} {\bibfnamefont {M.~L.}\ \bibnamefont
  {{Warren}}}, \bibinfo {author} {\bibfnamefont {M.}~\bibnamefont {{Meixner}}},
  \bibinfo {author} {\bibfnamefont {G.}~\bibnamefont {{Mathews}}}, \bibinfo
  {author} {\bibfnamefont {J.}~\bibnamefont {{Hidaka}}}, \ and\ \bibinfo
  {author} {\bibfnamefont {T.}~\bibnamefont {{Kajino}}},\ }\href {\doibase
  10.1103/PhysRevD.90.103007} {\bibfield  {journal} {\bibinfo  {journal}
  {Phys.\ Rev.\ D}\ }\textbf {\bibinfo {volume} {90}},\ \bibinfo {eid} {103007}
  (\bibinfo {year} {2014})},\ \Eprint {http://arxiv.org/abs/1405.6101}
  {arXiv:1405.6101 [astro-ph.HE]} \BibitemShut {NoStop}%
\bibitem [{\citenamefont {{Wu}}\ \emph {et~al.}(2014)\citenamefont {{Wu}},
  \citenamefont {{Fischer}}, \citenamefont {{Huther}}, \citenamefont
  {{Mart{\'{\i}}nez-Pinedo}},\ and\ \citenamefont
  {{Qian}}}]{2014PhRvD..89f1303W}%
  \BibitemOpen
  \bibfield  {author} {\bibinfo {author} {\bibfnamefont {M.-R.}\ \bibnamefont
  {{Wu}}}, \bibinfo {author} {\bibfnamefont {T.}~\bibnamefont {{Fischer}}},
  \bibinfo {author} {\bibfnamefont {L.}~\bibnamefont {{Huther}}}, \bibinfo
  {author} {\bibfnamefont {G.}~\bibnamefont {{Mart{\'{\i}}nez-Pinedo}}}, \ and\
  \bibinfo {author} {\bibfnamefont {Y.-Z.}\ \bibnamefont {{Qian}}},\ }\href
  {\doibase 10.1103/PhysRevD.89.061303} {\bibfield  {journal} {\bibinfo
  {journal} {Phys.\ Rev.\ D}\ }\textbf {\bibinfo {volume} {89}},\ \bibinfo
  {eid} {061303} (\bibinfo {year} {2014})},\ \Eprint
  {http://arxiv.org/abs/1305.2382} {arXiv:1305.2382 [astro-ph.HE]} \BibitemShut
  {NoStop}%
\bibitem [{\citenamefont {{Esmaili}}\ \emph {et~al.}(2014)\citenamefont
  {{Esmaili}}, \citenamefont {{Peres}},\ and\ \citenamefont
  {{Serpico}}}]{2014PhRvD..90c3013E}%
  \BibitemOpen
  \bibfield  {author} {\bibinfo {author} {\bibfnamefont {A.}~\bibnamefont
  {{Esmaili}}}, \bibinfo {author} {\bibfnamefont {O.~L.~G.}\ \bibnamefont
  {{Peres}}}, \ and\ \bibinfo {author} {\bibfnamefont {P.~D.}\ \bibnamefont
  {{Serpico}}},\ }\href {\doibase 10.1103/PhysRevD.90.033013} {\bibfield
  {journal} {\bibinfo  {journal} {Phys.\ Rev.\ D}\ }\textbf {\bibinfo {volume}
  {90}},\ \bibinfo {eid} {033013} (\bibinfo {year} {2014})},\ \Eprint
  {http://arxiv.org/abs/1402.1453} {arXiv:1402.1453 [hep-ph]} \BibitemShut
  {NoStop}%
\bibitem [{\citenamefont {Mikheyev}\ and\ \citenamefont
  {Smirnov}(1985)}]{Mikheyev:1985aa}%
  \BibitemOpen
  \bibfield  {author} {\bibinfo {author} {\bibfnamefont {S.~P.}\ \bibnamefont
  {Mikheyev}}\ and\ \bibinfo {author} {\bibfnamefont {A.~Y.}\ \bibnamefont
  {Smirnov}},\ }\href@noop {} {\bibfield  {journal} {\bibinfo  {journal} {Yad.\
  Fiz.}\ }\textbf {\bibinfo {volume} {42}},\ \bibinfo {pages} {1441} (\bibinfo
  {year} {1985})},\ \bibinfo {note} {({\it Sov. J. Nucl. Phys.} {\bf 42}
  913)}\BibitemShut {NoStop}%
\bibitem [{\citenamefont {Mikheyev}\ and\ \citenamefont
  {Smirnov}(1986)}]{Mikheyev:1986tj}%
  \BibitemOpen
  \bibfield  {author} {\bibinfo {author} {\bibfnamefont {S.~P.}\ \bibnamefont
  {Mikheyev}}\ and\ \bibinfo {author} {\bibfnamefont {A.~Y.}\ \bibnamefont
  {Smirnov}},\ }in\ \href@noop {} {\emph {\bibinfo {booktitle} {'86 Massive
  Neutrinos in Astrophysics and in Particle Physics}}},\ \bibinfo {editor}
  {edited by\ \bibinfo {editor} {\bibfnamefont {O.}~\bibnamefont {Frackler}}\
  and\ \bibinfo {editor} {\bibfnamefont {J.}~\bibnamefont {Tr\^an
  Thanh~V\^an}}}\ (\bibinfo  {publisher} {Editions Fronti\`eres},\ \bibinfo
  {address} {Gif-sur-Yvette},\ \bibinfo {year} {1986})\ p.\ \bibinfo {pages}
  {355}\BibitemShut {NoStop}%
\bibitem [{\citenamefont {{Wolfenstein}}(1978)}]{1978PhRvD..17.2369W}%
  \BibitemOpen
  \bibfield  {author} {\bibinfo {author} {\bibfnamefont {L.}~\bibnamefont
  {{Wolfenstein}}},\ }\href {\doibase 10.1103/PhysRevD.17.2369} {\bibfield
  {journal} {\bibinfo  {journal} {Phys.\ Rev.\ D}\ }\textbf {\bibinfo {volume}
  {17}},\ \bibinfo {pages} {2369} (\bibinfo {year} {1978})}\BibitemShut
  {NoStop}%
\bibitem [{\citenamefont {Miranda}\ and\ \citenamefont
  {Nunokawa}(2015)}]{1367-2630-17-9-095002}%
  \BibitemOpen
  \bibfield  {author} {\bibinfo {author} {\bibfnamefont {O.~G.}\ \bibnamefont
  {Miranda}}\ and\ \bibinfo {author} {\bibfnamefont {H.}~\bibnamefont
  {Nunokawa}},\ }\href {http://stacks.iop.org/1367-2630/17/i=9/a=095002}
  {\bibfield  {journal} {\bibinfo  {journal} {New Journal of Physics}\ }\textbf
  {\bibinfo {volume} {17}},\ \bibinfo {pages} {095002} (\bibinfo {year}
  {2015})}\BibitemShut {NoStop}%
\bibitem [{\citenamefont {{Ohlsson}}(2013)}]{2013RPPh...76d4201O}%
  \BibitemOpen
  \bibfield  {author} {\bibinfo {author} {\bibfnamefont {T.}~\bibnamefont
  {{Ohlsson}}},\ }\href {\doibase 10.1088/0034-4885/76/4/044201} {\bibfield
  {journal} {\bibinfo  {journal} {Reports on Progress in Physics}\ }\textbf
  {\bibinfo {volume} {76}},\ \bibinfo {eid} {044201} (\bibinfo {year}
  {2013})},\ \Eprint {http://arxiv.org/abs/1209.2710} {arXiv:1209.2710
  [hep-ph]} \BibitemShut {NoStop}%
\bibitem [{\citenamefont {{Valle}}(1987)}]{1987PhLB..199..432V}%
  \BibitemOpen
  \bibfield  {author} {\bibinfo {author} {\bibfnamefont {J.~W.~F.}\
  \bibnamefont {{Valle}}},\ }\href {\doibase 10.1016/0370-2693(87)90947-6}
  {\bibfield  {journal} {\bibinfo  {journal} {Physics Letters B}\ }\textbf
  {\bibinfo {volume} {199}},\ \bibinfo {pages} {432} (\bibinfo {year}
  {1987})}\BibitemShut {NoStop}%
\bibitem [{\citenamefont {{Nunokawa}}\ \emph
  {et~al.}(1996{\natexlab{a}})\citenamefont {{Nunokawa}}, \citenamefont
  {{Qian}}, \citenamefont {{Rossi}},\ and\ \citenamefont
  {{Valle}}}]{1996PhRvD..54.4356N}%
  \BibitemOpen
  \bibfield  {author} {\bibinfo {author} {\bibfnamefont {H.}~\bibnamefont
  {{Nunokawa}}}, \bibinfo {author} {\bibfnamefont {Y.-Z.}\ \bibnamefont
  {{Qian}}}, \bibinfo {author} {\bibfnamefont {A.}~\bibnamefont {{Rossi}}}, \
  and\ \bibinfo {author} {\bibfnamefont {J.~W.~F.}\ \bibnamefont {{Valle}}},\
  }\href {\doibase 10.1103/PhysRevD.54.4356} {\bibfield  {journal} {\bibinfo
  {journal} {Phys.\ Rev.\ D}\ }\textbf {\bibinfo {volume} {54}},\ \bibinfo
  {pages} {4356} (\bibinfo {year} {1996}{\natexlab{a}})},\ \Eprint
  {http://arxiv.org/abs/hep-ph/9605301} {hep-ph/9605301} \BibitemShut {NoStop}%
\bibitem [{\citenamefont {{Nunokawa}}\ \emph
  {et~al.}(1996{\natexlab{b}})\citenamefont {{Nunokawa}}, \citenamefont
  {{Rossi}},\ and\ \citenamefont {{Valle}}}]{1996NuPhB.482..481N}%
  \BibitemOpen
  \bibfield  {author} {\bibinfo {author} {\bibfnamefont {H.}~\bibnamefont
  {{Nunokawa}}}, \bibinfo {author} {\bibfnamefont {A.}~\bibnamefont {{Rossi}}},
  \ and\ \bibinfo {author} {\bibfnamefont {J.~W.~F.}\ \bibnamefont {{Valle}}},\
  }\href {\doibase 10.1016/S0550-3213(96)00541-X} {\bibfield  {journal}
  {\bibinfo  {journal} {Nuclear Physics B}\ }\textbf {\bibinfo {volume}
  {482}},\ \bibinfo {pages} {481} (\bibinfo {year} {1996}{\natexlab{b}})},\
  \Eprint {http://arxiv.org/abs/hep-ph/9606445} {hep-ph/9606445} \BibitemShut
  {NoStop}%
\bibitem [{\citenamefont {{Mansour}}\ and\ \citenamefont
  {{Kuo}}(1998)}]{1998PhRvD..58a3012M}%
  \BibitemOpen
  \bibfield  {author} {\bibinfo {author} {\bibfnamefont {S.~W.}\ \bibnamefont
  {{Mansour}}}\ and\ \bibinfo {author} {\bibfnamefont {T.~K.}\ \bibnamefont
  {{Kuo}}},\ }\href {\doibase 10.1103/PhysRevD.58.013012} {\bibfield  {journal}
  {\bibinfo  {journal} {Phys.\ Rev.\ D}\ }\textbf {\bibinfo {volume} {58}},\
  \bibinfo {eid} {013012} (\bibinfo {year} {1998})},\ \Eprint
  {http://arxiv.org/abs/hep-ph/9711424} {hep-ph/9711424} \BibitemShut {NoStop}%
\bibitem [{\citenamefont {{Fogli}}\ \emph {et~al.}(2002)\citenamefont
  {{Fogli}}, \citenamefont {{Lisi}}, \citenamefont {{Mirizzi}},\ and\
  \citenamefont {{Montanino}}}]{2002PhRvD..66a3009F}%
  \BibitemOpen
  \bibfield  {author} {\bibinfo {author} {\bibfnamefont {G.~L.}\ \bibnamefont
  {{Fogli}}}, \bibinfo {author} {\bibfnamefont {E.}~\bibnamefont {{Lisi}}},
  \bibinfo {author} {\bibfnamefont {A.}~\bibnamefont {{Mirizzi}}}, \ and\
  \bibinfo {author} {\bibfnamefont {D.}~\bibnamefont {{Montanino}}},\ }\href
  {\doibase 10.1103/PhysRevD.66.013009} {\bibfield  {journal} {\bibinfo
  {journal} {Phys.\ Rev.\ D}\ }\textbf {\bibinfo {volume} {66}},\ \bibinfo
  {eid} {013009} (\bibinfo {year} {2002})},\ \Eprint
  {http://arxiv.org/abs/hep-ph/0202269} {hep-ph/0202269} \BibitemShut {NoStop}%
\bibitem [{\citenamefont {Esteban-Pretel}\ \emph {et~al.}(2007)\citenamefont
  {Esteban-Pretel}, \citenamefont {Tom\`as},\ and\ \citenamefont
  {Valle}}]{PhysRevD.76.053001}%
  \BibitemOpen
  \bibfield  {author} {\bibinfo {author} {\bibfnamefont {A.}~\bibnamefont
  {Esteban-Pretel}}, \bibinfo {author} {\bibfnamefont {R.}~\bibnamefont
  {Tom\`as}}, \ and\ \bibinfo {author} {\bibfnamefont {J.~W.~F.}\ \bibnamefont
  {Valle}},\ }\href {\doibase 10.1103/PhysRevD.76.053001} {\bibfield  {journal}
  {\bibinfo  {journal} {Phys. Rev. D}\ }\textbf {\bibinfo {volume} {76}},\
  \bibinfo {pages} {053001} (\bibinfo {year} {2007})}\BibitemShut {NoStop}%
\bibitem [{\citenamefont {Blennow}\ \emph {et~al.}(2008)\citenamefont
  {Blennow}, \citenamefont {Mirizzi},\ and\ \citenamefont
  {Serpico}}]{Blennow:2008er}%
  \BibitemOpen
  \bibfield  {author} {\bibinfo {author} {\bibfnamefont {M.}~\bibnamefont
  {Blennow}}, \bibinfo {author} {\bibfnamefont {A.}~\bibnamefont {Mirizzi}}, \
  and\ \bibinfo {author} {\bibfnamefont {P.~D.}\ \bibnamefont {Serpico}},\
  }\href {\doibase 10.1103/PhysRevD.78.113004} {\bibfield  {journal} {\bibinfo
  {journal} {Phys. Rev.}\ }\textbf {\bibinfo {volume} {D 78}},\ \bibinfo
  {pages} {113004} (\bibinfo {year} {2008})},\ \Eprint
  {http://arxiv.org/abs/0810.2297} {arXiv:0810.2297 [hep-ph]} \BibitemShut
  {NoStop}%
\bibitem [{\citenamefont {{Esteban-Pretel}}\ \emph {et~al.}(2010)\citenamefont
  {{Esteban-Pretel}}, \citenamefont {{Tom{\`a}s}},\ and\ \citenamefont
  {{Valle}}}]{2010PhRvD..81f3003E}%
  \BibitemOpen
  \bibfield  {author} {\bibinfo {author} {\bibfnamefont {A.}~\bibnamefont
  {{Esteban-Pretel}}}, \bibinfo {author} {\bibfnamefont {R.}~\bibnamefont
  {{Tom{\`a}s}}}, \ and\ \bibinfo {author} {\bibfnamefont {J.~W.~F.}\
  \bibnamefont {{Valle}}},\ }\href {\doibase 10.1103/PhysRevD.81.063003}
  {\bibfield  {journal} {\bibinfo  {journal} {Phys.\ Rev.\ D}\ }\textbf
  {\bibinfo {volume} {81}},\ \bibinfo {eid} {063003} (\bibinfo {year}
  {2010})},\ \Eprint {http://arxiv.org/abs/0909.2196} {arXiv:0909.2196
  [hep-ph]} \BibitemShut {NoStop}%
\bibitem [{\citenamefont {Stapleford}\ \emph {et~al.}(2016)\citenamefont
  {Stapleford}, \citenamefont {{V{\"a}{\"a}n{\"a}nen}}, \citenamefont
  {Kneller}, \citenamefont {McLaughlin},\ and\ \citenamefont
  {Shapiro}}]{Stapleford:2016jgz}%
  \BibitemOpen
  \bibfield  {author} {\bibinfo {author} {\bibfnamefont {C.~J.}\ \bibnamefont
  {Stapleford}}, \bibinfo {author} {\bibfnamefont {D.~J.}\ \bibnamefont
  {{V{\"a}{\"a}n{\"a}nen}}}, \bibinfo {author} {\bibfnamefont {J.~P.}\
  \bibnamefont {Kneller}}, \bibinfo {author} {\bibfnamefont {G.~C.}\
  \bibnamefont {McLaughlin}}, \ and\ \bibinfo {author} {\bibfnamefont {B.~T.}\
  \bibnamefont {Shapiro}},\ }\href {\doibase 10.1103/PhysRevD.94.093007}
  {\bibfield  {journal} {\bibinfo  {journal} {Phys. Rev.}\ }\textbf {\bibinfo
  {volume} {D 94}},\ \bibinfo {pages} {093007} (\bibinfo {year} {2016})},\
  \Eprint {http://arxiv.org/abs/1605.04903} {arXiv:1605.04903 [hep-ph]}
  \BibitemShut {NoStop}%
\bibitem [{\citenamefont {Bilenky}\ \emph {et~al.}(1993)\citenamefont
  {Bilenky}, \citenamefont {Bilenky},\ and\ \citenamefont
  {Santamaria}}]{Bilenky:1992xn}%
  \BibitemOpen
  \bibfield  {author} {\bibinfo {author} {\bibfnamefont {M.~S.}\ \bibnamefont
  {Bilenky}}, \bibinfo {author} {\bibfnamefont {S.~M.}\ \bibnamefont
  {Bilenky}}, \ and\ \bibinfo {author} {\bibfnamefont {A.}~\bibnamefont
  {Santamaria}},\ }\href {\doibase 10.1016/0370-2693(93)90703-K} {\bibfield
  {journal} {\bibinfo  {journal} {Phys. Lett.}\ }\textbf {\bibinfo {volume} {B
  301}},\ \bibinfo {pages} {287} (\bibinfo {year} {1993})}\BibitemShut
  {NoStop}%
\bibitem [{\citenamefont {Bilenky}\ and\ \citenamefont
  {Santamaria}(1994)}]{Bilenky:1994ma}%
  \BibitemOpen
  \bibfield  {author} {\bibinfo {author} {\bibfnamefont {M.~S.}\ \bibnamefont
  {Bilenky}}\ and\ \bibinfo {author} {\bibfnamefont {A.}~\bibnamefont
  {Santamaria}},\ }\href {\doibase 10.1016/0370-2693(94)00961-9} {\bibfield
  {journal} {\bibinfo  {journal} {Phys. Lett.}\ }\textbf {\bibinfo {volume} {B
  336}},\ \bibinfo {pages} {91} (\bibinfo {year} {1994})},\ \Eprint
  {http://arxiv.org/abs/hep-ph/9405427} {arXiv:hep-ph/9405427 [hep-ph]}
  \BibitemShut {NoStop}%
\bibitem [{\citenamefont {Masso}\ and\ \citenamefont
  {Toldra}(1994)}]{Masso:1994ww}%
  \BibitemOpen
  \bibfield  {author} {\bibinfo {author} {\bibfnamefont {E.}~\bibnamefont
  {Masso}}\ and\ \bibinfo {author} {\bibfnamefont {R.}~\bibnamefont {Toldra}},\
  }\href {\doibase 10.1016/0370-2693(94)91018-9} {\bibfield  {journal}
  {\bibinfo  {journal} {Phys. Lett.}\ }\textbf {\bibinfo {volume} {B 333}},\
  \bibinfo {pages} {132} (\bibinfo {year} {1994})},\ \Eprint
  {http://arxiv.org/abs/hep-ph/9404339} {arXiv:hep-ph/9404339 [hep-ph]}
  \BibitemShut {NoStop}%
\bibitem [{\citenamefont {Bilenky}\ and\ \citenamefont
  {Santamaria}(1999)}]{Bilenky:1999dn}%
  \BibitemOpen
  \bibfield  {author} {\bibinfo {author} {\bibfnamefont {M.~S.}\ \bibnamefont
  {Bilenky}}\ and\ \bibinfo {author} {\bibfnamefont {A.}~\bibnamefont
  {Santamaria}},\ }in\ \href@noop {} {\emph {\bibinfo {booktitle} {{Neutrino
  mixing. Festschrift in honour of Samoil Bilenky's 70th birthday. Proceedings,
  International Meeting, Turin, Italy, March 25-27, 1999}}}}\ (\bibinfo {year}
  {1999})\ pp.\ \bibinfo {pages} {50--61},\ \Eprint
  {http://arxiv.org/abs/hep-ph/9908272} {arXiv:hep-ph/9908272 [hep-ph]}
  \BibitemShut {NoStop}%
\bibitem [{\citenamefont {Das}\ \emph {et~al.}(2017)\citenamefont {Das},
  \citenamefont {Dighe},\ and\ \citenamefont {Sen}}]{Das:2017iuj}%
  \BibitemOpen
  \bibfield  {author} {\bibinfo {author} {\bibfnamefont {A.}~\bibnamefont
  {Das}}, \bibinfo {author} {\bibfnamefont {A.}~\bibnamefont {Dighe}}, \ and\
  \bibinfo {author} {\bibfnamefont {M.}~\bibnamefont {Sen}},\ }\href {\doibase
  10.1088/1475-7516/2017/05/051} {\bibfield  {journal} {\bibinfo  {journal}
  {JCAP}\ }\textbf {\bibinfo {volume} {1705}},\ \bibinfo {pages} {051}
  (\bibinfo {year} {2017})},\ \Eprint {http://arxiv.org/abs/1705.00468}
  {arXiv:1705.00468 [hep-ph]} \BibitemShut {NoStop}%
\bibitem [{\citenamefont {Dighe}\ and\ \citenamefont
  {Sen}(2018)}]{dighe2018nonstandard}%
  \BibitemOpen
  \bibfield  {author} {\bibinfo {author} {\bibfnamefont {A.}~\bibnamefont
  {Dighe}}\ and\ \bibinfo {author} {\bibfnamefont {M.}~\bibnamefont {Sen}},\
  }\href@noop {} {\bibfield  {journal} {\bibinfo  {journal} {Phys.\ Rev.\ D}\
  }\textbf {\bibinfo {volume} {97}},\ \bibinfo {pages} {043011} (\bibinfo
  {year} {2018})}\BibitemShut {NoStop}%
\bibitem [{\citenamefont {Gelmini}\ and\ \citenamefont
  {Roncadelli}(1981)}]{Gelmini:1980re}%
  \BibitemOpen
  \bibfield  {author} {\bibinfo {author} {\bibfnamefont {G.~B.}\ \bibnamefont
  {Gelmini}}\ and\ \bibinfo {author} {\bibfnamefont {M.}~\bibnamefont
  {Roncadelli}},\ }\href {\doibase 10.1016/0370-2693(81)90559-1} {\bibfield
  {journal} {\bibinfo  {journal} {Phys. Lett.}\ }\textbf {\bibinfo {volume} {B
  99}},\ \bibinfo {pages} {411} (\bibinfo {year} {1981})}\BibitemShut {NoStop}%
\bibitem [{\citenamefont {Gelmini}\ \emph {et~al.}(1982)\citenamefont
  {Gelmini}, \citenamefont {Nussinov},\ and\ \citenamefont
  {Roncadelli}}]{Gelmini:1982rr}%
  \BibitemOpen
  \bibfield  {author} {\bibinfo {author} {\bibfnamefont {G.~B.}\ \bibnamefont
  {Gelmini}}, \bibinfo {author} {\bibfnamefont {S.}~\bibnamefont {Nussinov}}, \
  and\ \bibinfo {author} {\bibfnamefont {M.}~\bibnamefont {Roncadelli}},\
  }\href {\doibase 10.1016/0550-3213(82)90107-9} {\bibfield  {journal}
  {\bibinfo  {journal} {Nucl. Phys.}\ }\textbf {\bibinfo {volume} {B 209}},\
  \bibinfo {pages} {157} (\bibinfo {year} {1982})}\BibitemShut {NoStop}%
\bibitem [{\citenamefont {Kolb}\ and\ \citenamefont
  {Turner}(1987)}]{Kolb:1987qy}%
  \BibitemOpen
  \bibfield  {author} {\bibinfo {author} {\bibfnamefont {E.~W.}\ \bibnamefont
  {Kolb}}\ and\ \bibinfo {author} {\bibfnamefont {M.~S.}\ \bibnamefont
  {Turner}},\ }\href {\doibase 10.1103/PhysRevD.36.2895} {\bibfield  {journal}
  {\bibinfo  {journal} {Phys. Rev.}\ }\textbf {\bibinfo {volume} {D 36}},\
  \bibinfo {pages} {2895} (\bibinfo {year} {1987})}\BibitemShut {NoStop}%
\bibitem [{\citenamefont {Chang}\ and\ \citenamefont
  {Choi}(1994)}]{Chang:1993yp}%
  \BibitemOpen
  \bibfield  {author} {\bibinfo {author} {\bibfnamefont {S.}~\bibnamefont
  {Chang}}\ and\ \bibinfo {author} {\bibfnamefont {K.}~\bibnamefont {Choi}},\
  }\href {\doibase 10.1103/PhysRevD.49.12} {\bibfield  {journal} {\bibinfo
  {journal} {Phys. Rev.}\ }\textbf {\bibinfo {volume} {D 49}},\ \bibinfo
  {pages} {12} (\bibinfo {year} {1994})},\ \Eprint
  {http://arxiv.org/abs/hep-ph/9303243} {arXiv:hep-ph/9303243 [hep-ph]}
  \BibitemShut {NoStop}%
\bibitem [{\citenamefont {Choi}\ \emph {et~al.}(1988)\citenamefont {Choi},
  \citenamefont {Kim}, \citenamefont {Kim},\ and\ \citenamefont
  {Lam}}]{Choi:1987sd}%
  \BibitemOpen
  \bibfield  {author} {\bibinfo {author} {\bibfnamefont {K.}~\bibnamefont
  {Choi}}, \bibinfo {author} {\bibfnamefont {C.~W.}\ \bibnamefont {Kim}},
  \bibinfo {author} {\bibfnamefont {J.}~\bibnamefont {Kim}}, \ and\ \bibinfo
  {author} {\bibfnamefont {W.~P.}\ \bibnamefont {Lam}},\ }\bibfield
  {booktitle} {\emph {\bibinfo {booktitle} {{3rd Asia Pacific Physics
  Conference Hong Kong, June 20-24, 1988}}},\ }\href {\doibase
  10.1103/PhysRevD.37.3225} {\bibfield  {journal} {\bibinfo  {journal} {Phys.
  Rev.}\ }\textbf {\bibinfo {volume} {D 37}},\ \bibinfo {pages} {3225}
  (\bibinfo {year} {1988})}\BibitemShut {NoStop}%
\bibitem [{\citenamefont {Kachelriess}\ \emph {et~al.}(2000)\citenamefont
  {Kachelriess}, \citenamefont {Tomas},\ and\ \citenamefont
  {Valle}}]{Kachelriess:2000qc}%
  \BibitemOpen
  \bibfield  {author} {\bibinfo {author} {\bibfnamefont {M.}~\bibnamefont
  {Kachelriess}}, \bibinfo {author} {\bibfnamefont {R.}~\bibnamefont {Tomas}},
  \ and\ \bibinfo {author} {\bibfnamefont {J.~W.~F.}\ \bibnamefont {Valle}},\
  }\href {\doibase 10.1103/PhysRevD.62.023004} {\bibfield  {journal} {\bibinfo
  {journal} {Phys. Rev.}\ }\textbf {\bibinfo {volume} {D 62}},\ \bibinfo
  {pages} {023004} (\bibinfo {year} {2000})},\ \Eprint
  {http://arxiv.org/abs/hep-ph/0001039} {arXiv:hep-ph/0001039 [hep-ph]}
  \BibitemShut {NoStop}%
\bibitem [{\citenamefont {Tomas}\ \emph {et~al.}(2001)\citenamefont {Tomas},
  \citenamefont {Pas},\ and\ \citenamefont {Valle}}]{Tomas:2001dh}%
  \BibitemOpen
  \bibfield  {author} {\bibinfo {author} {\bibfnamefont {R.}~\bibnamefont
  {Tomas}}, \bibinfo {author} {\bibfnamefont {H.}~\bibnamefont {Pas}}, \ and\
  \bibinfo {author} {\bibfnamefont {J.~W.~F.}\ \bibnamefont {Valle}},\ }\href
  {\doibase 10.1103/PhysRevD.64.095005} {\bibfield  {journal} {\bibinfo
  {journal} {Phys. Rev.}\ }\textbf {\bibinfo {volume} {D 64}},\ \bibinfo
  {pages} {095005} (\bibinfo {year} {2001})},\ \Eprint
  {http://arxiv.org/abs/hep-ph/0103017} {arXiv:hep-ph/0103017 [hep-ph]}
  \BibitemShut {NoStop}%
\bibitem [{\citenamefont {Farzan}(2003)}]{Farzan:2002wx}%
  \BibitemOpen
  \bibfield  {author} {\bibinfo {author} {\bibfnamefont {Y.}~\bibnamefont
  {Farzan}},\ }\href {\doibase 10.1103/PhysRevD.67.073015} {\bibfield
  {journal} {\bibinfo  {journal} {Phys. Rev.}\ }\textbf {\bibinfo {volume} {D
  67}},\ \bibinfo {pages} {073015} (\bibinfo {year} {2003})},\ \Eprint
  {http://arxiv.org/abs/hep-ph/0211375} {arXiv:hep-ph/0211375 [hep-ph]}
  \BibitemShut {NoStop}%
\bibitem [{\citenamefont {{Chakraborty}}\ \emph {et~al.}(2011)\citenamefont
  {{Chakraborty}}, \citenamefont {{Fischer}}, \citenamefont {{Mirizzi}},
  \citenamefont {{Saviano}},\ and\ \citenamefont
  {{Tom{\`a}s}}}]{2011PhRvL.107o1101C}%
  \BibitemOpen
  \bibfield  {author} {\bibinfo {author} {\bibfnamefont {S.}~\bibnamefont
  {{Chakraborty}}}, \bibinfo {author} {\bibfnamefont {T.}~\bibnamefont
  {{Fischer}}}, \bibinfo {author} {\bibfnamefont {A.}~\bibnamefont
  {{Mirizzi}}}, \bibinfo {author} {\bibfnamefont {N.}~\bibnamefont
  {{Saviano}}}, \ and\ \bibinfo {author} {\bibfnamefont {R.}~\bibnamefont
  {{Tom{\`a}s}}},\ }\href {\doibase 10.1103/PhysRevLett.107.151101} {\bibfield
  {journal} {\bibinfo  {journal} {Phys.\ Rev.\ Lett.}\ }\textbf {\bibinfo
  {volume} {107}},\ \bibinfo {eid} {151101} (\bibinfo {year} {2011})},\ \Eprint
  {http://arxiv.org/abs/1104.4031} {arXiv:1104.4031 [hep-ph]} \BibitemShut
  {NoStop}%
\bibitem [{\citenamefont {{Duan}}\ and\ \citenamefont
  {{Friedland}}(2011)}]{2011PhRvL.106i1101D}%
  \BibitemOpen
  \bibfield  {author} {\bibinfo {author} {\bibfnamefont {H.}~\bibnamefont
  {{Duan}}}\ and\ \bibinfo {author} {\bibfnamefont {A.}~\bibnamefont
  {{Friedland}}},\ }\href {\doibase 10.1103/PhysRevLett.106.091101} {\bibfield
  {journal} {\bibinfo  {journal} {Phys.\ Rev.\ Lett.}\ }\textbf {\bibinfo
  {volume} {106}},\ \bibinfo {eid} {091101} (\bibinfo {year} {2011})},\ \Eprint
  {http://arxiv.org/abs/1006.2359} {arXiv:1006.2359 [hep-ph]} \BibitemShut
  {NoStop}%
\bibitem [{\citenamefont {{Sawyer}}(2005)}]{2005PhRvD..72d5003S}%
  \BibitemOpen
  \bibfield  {author} {\bibinfo {author} {\bibfnamefont {R.~F.}\ \bibnamefont
  {{Sawyer}}},\ }\href {\doibase 10.1103/PhysRevD.72.045003} {\bibfield
  {journal} {\bibinfo  {journal} {Phys.\ Rev.\ D}\ }\textbf {\bibinfo {volume}
  {72}},\ \bibinfo {eid} {045003} (\bibinfo {year} {2005})},\ \Eprint
  {http://arxiv.org/abs/hep-ph/0503013} {hep-ph/0503013} \BibitemShut {NoStop}%
\bibitem [{\citenamefont {Chakraborty}\ \emph {et~al.}(2016)\citenamefont
  {Chakraborty}, \citenamefont {Hansen}, \citenamefont {Izaguirre},\ and\
  \citenamefont {Raffelt}}]{chakraborty2016self}%
  \BibitemOpen
  \bibfield  {author} {\bibinfo {author} {\bibfnamefont {S.}~\bibnamefont
  {Chakraborty}}, \bibinfo {author} {\bibfnamefont {R.~S.}\ \bibnamefont
  {Hansen}}, \bibinfo {author} {\bibfnamefont {I.}~\bibnamefont {Izaguirre}}, \
  and\ \bibinfo {author} {\bibfnamefont {G.}~\bibnamefont {Raffelt}},\
  }\href@noop {} {\bibfield  {journal} {\bibinfo  {journal} {Journal of
  Cosmology and Astroparticle Physics}\ }\textbf {\bibinfo {volume} {2016}},\
  \bibinfo {pages} {042} (\bibinfo {year} {2016})}\BibitemShut {NoStop}%
\bibitem [{\citenamefont {Sen}(2017)}]{sen2017supernova}%
  \BibitemOpen
  \bibfield  {author} {\bibinfo {author} {\bibfnamefont {M.}~\bibnamefont
  {Sen}},\ }\href@noop {} {\bibfield  {journal} {\bibinfo  {journal} {arXiv
  preprint arXiv:1702.06836}\ } (\bibinfo {year} {2017})}\BibitemShut {NoStop}%
\bibitem [{\citenamefont {Izaguirre}\ \emph {et~al.}(2017)\citenamefont
  {Izaguirre}, \citenamefont {Raffelt},\ and\ \citenamefont
  {Tamborra}}]{izaguirre2017fast}%
  \BibitemOpen
  \bibfield  {author} {\bibinfo {author} {\bibfnamefont {I.}~\bibnamefont
  {Izaguirre}}, \bibinfo {author} {\bibfnamefont {G.}~\bibnamefont {Raffelt}},
  \ and\ \bibinfo {author} {\bibfnamefont {I.}~\bibnamefont {Tamborra}},\
  }\href@noop {} {\bibfield  {journal} {\bibinfo  {journal} {Phys. Rev. Lett.}\
  }\textbf {\bibinfo {volume} {118}},\ \bibinfo {pages} {021101} (\bibinfo
  {year} {2017})}\BibitemShut {NoStop}%
\bibitem [{\citenamefont {Capozzi}\ \emph {et~al.}(2017)\citenamefont
  {Capozzi}, \citenamefont {Dasgupta}, \citenamefont {Lisi}, \citenamefont
  {Marrone},\ and\ \citenamefont {Mirizzi}}]{capozzi2017fast}%
  \BibitemOpen
  \bibfield  {author} {\bibinfo {author} {\bibfnamefont {F.}~\bibnamefont
  {Capozzi}}, \bibinfo {author} {\bibfnamefont {B.}~\bibnamefont {Dasgupta}},
  \bibinfo {author} {\bibfnamefont {E.}~\bibnamefont {Lisi}}, \bibinfo {author}
  {\bibfnamefont {A.}~\bibnamefont {Marrone}}, \ and\ \bibinfo {author}
  {\bibfnamefont {A.}~\bibnamefont {Mirizzi}},\ }\href@noop {} {\bibfield
  {journal} {\bibinfo  {journal} {Phys.\ Rev.\ D}\ }\textbf {\bibinfo {volume}
  {96}},\ \bibinfo {pages} {043016} (\bibinfo {year} {2017})}\BibitemShut
  {NoStop}%
\bibitem [{\citenamefont {Abbar}\ and\ \citenamefont
  {Duan}(2017)}]{abbar2017fast}%
  \BibitemOpen
  \bibfield  {author} {\bibinfo {author} {\bibfnamefont {S.}~\bibnamefont
  {Abbar}}\ and\ \bibinfo {author} {\bibfnamefont {H.}~\bibnamefont {Duan}},\
  }\href@noop {} {\bibfield  {journal} {\bibinfo  {journal} {arXiv preprint
  arXiv:1712.07013}\ } (\bibinfo {year} {2017})}\BibitemShut {NoStop}%
\bibitem [{\citenamefont {Dasgupta}\ and\ \citenamefont
  {Sen}(2018)}]{dasgupta2018fast}%
  \BibitemOpen
  \bibfield  {author} {\bibinfo {author} {\bibfnamefont {B.}~\bibnamefont
  {Dasgupta}}\ and\ \bibinfo {author} {\bibfnamefont {M.}~\bibnamefont {Sen}},\
  }\href@noop {} {\bibfield  {journal} {\bibinfo  {journal} {Phys.\ Rev.\ D}\
  }\textbf {\bibinfo {volume} {97}},\ \bibinfo {pages} {023017} (\bibinfo
  {year} {2018})}\BibitemShut {NoStop}%
\bibitem [{\citenamefont {Yang}\ and\ \citenamefont
  {Kneller}(2017)}]{Yang:2017asl}%
  \BibitemOpen
  \bibfield  {author} {\bibinfo {author} {\bibfnamefont {Y.}~\bibnamefont
  {Yang}}\ and\ \bibinfo {author} {\bibfnamefont {J.~P.}\ \bibnamefont
  {Kneller}},\ }\href {\doibase 10.1103/PhysRevD.96.023009} {\bibfield
  {journal} {\bibinfo  {journal} {Phys. Rev.}\ }\textbf {\bibinfo {volume} {D
  96}},\ \bibinfo {pages} {023009} (\bibinfo {year} {2017})},\ \Eprint
  {http://arxiv.org/abs/1705.09723} {arXiv:1705.09723 [astro-ph.HE]}
  \BibitemShut {NoStop}%
\bibitem [{\citenamefont {Patrignani}\ \emph {et~al.}(2016)\citenamefont
  {Patrignani} \emph {et~al.}}]{Olive:2016xmw}%
  \BibitemOpen
  \bibfield  {author} {\bibinfo {author} {\bibfnamefont {C.}~\bibnamefont
  {Patrignani}} \emph {et~al.} (\bibinfo {collaboration} {Particle Data
  Group}),\ }\href {\doibase 10.1088/1674-1137/40/10/100001} {\bibfield
  {journal} {\bibinfo  {journal} {Chin. Phys.}\ }\textbf {\bibinfo {volume}
  {C40}},\ \bibinfo {pages} {100001} (\bibinfo {year} {2016})}\BibitemShut
  {NoStop}%
\bibitem [{\citenamefont {Fischer}\ \emph {et~al.}(2010)\citenamefont
  {Fischer}, \citenamefont {Whitehouse}, \citenamefont {Mezzacappa},
  \citenamefont {Thielemann},\ and\ \citenamefont
  {Liebendorfer}}]{Fischer:2009af}%
  \BibitemOpen
  \bibfield  {author} {\bibinfo {author} {\bibfnamefont {T.}~\bibnamefont
  {Fischer}}, \bibinfo {author} {\bibfnamefont {S.~C.}\ \bibnamefont
  {Whitehouse}}, \bibinfo {author} {\bibfnamefont {A.}~\bibnamefont
  {Mezzacappa}}, \bibinfo {author} {\bibfnamefont {F.~K.}\ \bibnamefont
  {Thielemann}}, \ and\ \bibinfo {author} {\bibfnamefont {M.}~\bibnamefont
  {Liebendorfer}},\ }\href {\doibase 10.1051/0004-6361/200913106} {\bibfield
  {journal} {\bibinfo  {journal} {Astron. Astrophys.}\ }\textbf {\bibinfo
  {volume} {517}},\ \bibinfo {pages} {A80} (\bibinfo {year} {2010})},\ \Eprint
  {http://arxiv.org/abs/0908.1871} {arXiv:0908.1871 [astro-ph.HE]} \BibitemShut
  {NoStop}%
\bibitem [{\citenamefont {Keil}\ \emph {et~al.}(2003)\citenamefont {Keil},
  \citenamefont {Raffelt},\ and\ \citenamefont {Janka}}]{2003ApJ...590..971K}%
  \BibitemOpen
  \bibfield  {author} {\bibinfo {author} {\bibfnamefont {M.~T.}\ \bibnamefont
  {Keil}}, \bibinfo {author} {\bibfnamefont {G.~G.}\ \bibnamefont {Raffelt}}, \
  and\ \bibinfo {author} {\bibfnamefont {H.-T.}\ \bibnamefont {Janka}},\ }\href
  {\doibase 10.1086/375130} {\bibfield  {journal} {\bibinfo  {journal}
  {Astrophys. J.}\ }\textbf {\bibinfo {volume} {590}},\ \bibinfo {pages} {971}
  (\bibinfo {year} {2003})},\ \Eprint {http://arxiv.org/abs/astro-ph/0208035}
  {astro-ph/0208035} \BibitemShut {NoStop}%
\bibitem [{\citenamefont {Wu}\ \emph {et~al.}(2015)\citenamefont {Wu},
  \citenamefont {Qian}, \citenamefont {Mart\'{\i}nez-Pinedo}, \citenamefont
  {Fischer},\ and\ \citenamefont {Huther}}]{PhysRevD.91.065016}%
  \BibitemOpen
  \bibfield  {author} {\bibinfo {author} {\bibfnamefont {M.-R.}\ \bibnamefont
  {Wu}}, \bibinfo {author} {\bibfnamefont {Y.-Z.}\ \bibnamefont {Qian}},
  \bibinfo {author} {\bibfnamefont {G.}~\bibnamefont {Mart\'{\i}nez-Pinedo}},
  \bibinfo {author} {\bibfnamefont {T.}~\bibnamefont {Fischer}}, \ and\
  \bibinfo {author} {\bibfnamefont {L.}~\bibnamefont {Huther}},\ }\href
  {\doibase 10.1103/PhysRevD.91.065016} {\bibfield  {journal} {\bibinfo
  {journal} {Phys. Rev. D}\ }\textbf {\bibinfo {volume} {91}},\ \bibinfo
  {pages} {065016} (\bibinfo {year} {2015})}\BibitemShut {NoStop}%
\bibitem [{\citenamefont {Sarikas}\ \emph {et~al.}(2012)\citenamefont
  {Sarikas}, \citenamefont {de~Sousa~Seixas},\ and\ \citenamefont
  {Raffelt}}]{sarikas2012spurious}%
  \BibitemOpen
  \bibfield  {author} {\bibinfo {author} {\bibfnamefont {S.}~\bibnamefont
  {Sarikas}}, \bibinfo {author} {\bibfnamefont {D.}~\bibnamefont
  {de~Sousa~Seixas}}, \ and\ \bibinfo {author} {\bibfnamefont {G.}~\bibnamefont
  {Raffelt}},\ }\href@noop {} {\bibfield  {journal} {\bibinfo  {journal}
  {Phys.\ Rev.\ D}\ }\textbf {\bibinfo {volume} {86}},\ \bibinfo {pages}
  {125020} (\bibinfo {year} {2012})}\BibitemShut {NoStop}%
\bibitem [{\citenamefont {Pasquini}\ and\ \citenamefont
  {Peres}(2016)}]{Pasquini:2015fjv}%
  \BibitemOpen
  \bibfield  {author} {\bibinfo {author} {\bibfnamefont {P.~S.}\ \bibnamefont
  {Pasquini}}\ and\ \bibinfo {author} {\bibfnamefont {O.~L.~G.}\ \bibnamefont
  {Peres}},\ }\href {\doibase 10.1103/PhysRevD.93.053007,
  10.1103/PhysRevD.93.079902} {\bibfield  {journal} {\bibinfo  {journal} {Phys.
  Rev.}\ }\textbf {\bibinfo {volume} {D 93}},\ \bibinfo {pages} {053007}
  (\bibinfo {year} {2016})},\ \bibinfo {note} {[Erratum: Phys.
  Rev.D93,no.7,079902(2016)]},\ \Eprint {http://arxiv.org/abs/1511.01811}
  {arXiv:1511.01811 [hep-ph]} \BibitemShut {NoStop}%
\bibitem [{\citenamefont {Heurtier}\ and\ \citenamefont
  {Zhang}(2017)}]{Heurtier:2016otg}%
  \BibitemOpen
  \bibfield  {author} {\bibinfo {author} {\bibfnamefont {L.}~\bibnamefont
  {Heurtier}}\ and\ \bibinfo {author} {\bibfnamefont {Y.}~\bibnamefont
  {Zhang}},\ }\href {\doibase 10.1088/1475-7516/2017/02/042} {\bibfield
  {journal} {\bibinfo  {journal} {JCAP}\ }\textbf {\bibinfo {volume} {1702}},\
  \bibinfo {pages} {042} (\bibinfo {year} {2017})},\ \Eprint
  {http://arxiv.org/abs/1609.05882} {arXiv:1609.05882 [hep-ph]} \BibitemShut
  {NoStop}%
\bibitem [{\citenamefont {Bruus}\ and\ \citenamefont
  {Flensberg}(2004)}]{bruus2004many}%
  \BibitemOpen
  \bibfield  {author} {\bibinfo {author} {\bibfnamefont {H.}~\bibnamefont
  {Bruus}}\ and\ \bibinfo {author} {\bibfnamefont {K.}~\bibnamefont
  {Flensberg}},\ }\href@noop {} {\emph {\bibinfo {title} {Many-body quantum
  theory in condensed matter physics: an introduction}}}\ (\bibinfo
  {publisher} {Oxford University Press},\ \bibinfo {year} {2004})\ pp.\
  \bibinfo {pages} {69--71}\BibitemShut {NoStop}%
\bibitem [{\citenamefont {Bergmann}\ \emph {et~al.}(1999)\citenamefont
  {Bergmann}, \citenamefont {Grossman},\ and\ \citenamefont
  {Nardi}}]{Bergmann:1999rz}%
  \BibitemOpen
  \bibfield  {author} {\bibinfo {author} {\bibfnamefont {S.}~\bibnamefont
  {Bergmann}}, \bibinfo {author} {\bibfnamefont {Y.}~\bibnamefont {Grossman}},
  \ and\ \bibinfo {author} {\bibfnamefont {E.}~\bibnamefont {Nardi}},\ }\href
  {\doibase 10.1103/PhysRevD.60.093008} {\bibfield  {journal} {\bibinfo
  {journal} {Phys. Rev.}\ }\textbf {\bibinfo {volume} {D 60}},\ \bibinfo
  {pages} {093008} (\bibinfo {year} {1999})},\ \Eprint
  {http://arxiv.org/abs/hep-ph/9903517} {arXiv:hep-ph/9903517 [hep-ph]}
  \BibitemShut {NoStop}%
\bibitem [{\citenamefont {Giunti}\ and\ \citenamefont
  {Wook}(2007)}]{Giunti:1053706}%
  \BibitemOpen
  \bibfield  {author} {\bibinfo {author} {\bibfnamefont {C.}~\bibnamefont
  {Giunti}}\ and\ \bibinfo {author} {\bibfnamefont {K.~C.}\ \bibnamefont
  {Wook}},\ }\href {https://cds.cern.ch/record/1053706} {\emph {\bibinfo
  {title} {{Fundamentals of Neutrino Physics and Astrophysics}}}}\ (\bibinfo
  {publisher} {Oxford Univ.},\ \bibinfo {address} {Oxford},\ \bibinfo {year}
  {2007})\BibitemShut {NoStop}%
\end{thebibliography}%

\end{document}